\documentclass[10pt]{article}
\usepackage{graphicx}
\usepackage{amssymb}
\usepackage{epstopdf}
\usepackage{graphicx}
\usepackage{amssymb}
\usepackage{epstopdf}
\usepackage{dsfont}
\usepackage{xcolor}
\usepackage{amsmath} 
\usepackage{cite}
\usepackage{caption}
\usepackage{multirow}
\usepackage{subcaption}
\usepackage{tikz}
\usepackage{comment}
\usepackage{url}
\usepackage{booktabs}
\usepackage{float}

\begin{document}

\title{    Explainable Deep Learning for Price–Trade Dynamics: \\ From Black-Box Forecasts to Effective Parametric Models}

\author{
\begin{tabular}{c@{\hspace{2.5cm}}c}
{\Large Manuel~Naviglio}\thanks{Email: \texttt{manuel.naviglio@sns.it}} 
&
{\Large Fabrizio~Lillo}\thanks{Email: \texttt{fabrizio.lillo@sns.it}}
\end{tabular}\\[0.4cm]
{\large Scuola Normale Superiore, Pisa, Italy}
}

\maketitle

\begin{abstract}
Understanding the joint dynamics of prices and trades is central to market microstructure, where returns and order flow interact through nonlinear and state-dependent mechanisms. Linear models are interpretable but may miss these effects, while deep neural networks improve forecasting at the cost of transparency. We use neural networks as tools for structural discovery rather than only for prediction. A deep feed-forward network is trained on high-frequency returns and signed volumes for large- and small-tick stocks and compared with a linear VAR benchmark. The neural network improves predictive performance, especially for returns, revealing nonlinear dependencies beyond the linear specification.

Using Shapley-based explainability, we show that the dominant contributions are concentrated at the most recent lags. Model-implied responses are consistent with conditional averages reconstructed from the data. Unlike empirical averages, however, the neural-network decomposition isolates individual regressor contributions to the aggregate dependence. Lagged signed volume generates sign-preserving and saturating effects, consistent with nonlinear price impact and order-flow persistence. Lagged returns act as state variables: when the previous trade does not move the price, the model predicts continuation in the direction of past order flow, whereas non-zero returns generate attenuation or reversal.

Building on these findings, we introduce a parsimonious SHAP-inspired nonlinear parametric model. It reproduces the main return--volume dependencies, outperforms the linear VAR benchmark, and achieves performance comparable to the neural network. A multi-lag extension captures residual longer-memory effects while preserving interpretability. Overall, explainability offers a route from black-box prediction to economically meaningful parametric models of price and trade dynamics.
\end{abstract}

\section{Introduction}\label{sec:introduction}

The dynamics of trading volumes and prices are intrinsically linked through the microstructure of financial markets. In modern electronic markets, this interaction is mediated by the limit order book (LOB), where buy and sell orders are continuously submitted, canceled, and executed. Price changes arise from the interaction between incoming market orders and available liquidity, making order flow a primary driver of short-term price dynamics. This mechanism underlies several well-documented stylized facts, including persistent order flow, transient price impact, and short-term return reversals~\cite{bouchaud2003fluctuations,lillo2004long,toth2011anomalous}.

Modeling the joint dynamics of prices and trades is therefore central to market microstructure. On the one hand, accurate forecasts of returns and trading activity are essential for applications such as optimal execution, risk management, and algorithmic trading~\cite{almgren2001optimal}. On the other hand, models that faithfully reproduce the statistical properties of the data are crucial for realistic market simulations, which are widely used to study liquidity provision and market impact. In particular, market impact—the price response induced by trading—plays a fundamental role in determining transaction costs, especially for metaorders executed over time~\cite{bouchaud2003fluctuations,bucci2019are}. Capturing these mechanisms requires models that account for the complex and nonlinear interaction between order flow and price formation.

A large body of literature has addressed this problem using linear parametric models. A prominent example is the vector autoregression (VAR) framework introduced by Hasbrouck~\cite{hasbrouck1991measuring}, as well as its variants such as the Transient Impact Model~\cite{bouchaud2003fluctuations}. These models describe price and trade dynamics through linear dependencies on past observations and have been widely adopted due to their interpretability and tractability. However, their ability to capture the full complexity of high-frequency financial data is inherently limited, as they impose restrictive assumptions on the functional form of the underlying relationships.

More recently, machine learning methods have emerged as flexible alternatives capable of capturing nonlinear and high-dimensional dependencies directly from data. In particular, deep learning models applied to limit order book data have demonstrated strong predictive performance~\cite{sirignano2019deep,zhang2019deeplob}. Despite these advances, a key limitation remains: such models are typically treated as black boxes, making it difficult to understand which features drive their predictions and whether the extracted patterns correspond to economically meaningful mechanisms.

In this work, we adopt a different perspective. Rather than using nonlinear
models solely as forecasting devices, we employ them as tools for structural
discovery. The idea is to exploit the flexibility of neural networks in a first
stage, allowing the model to learn nonlinear dependencies directly from the
data, and then to use explainability techniques to identify and interpret the
functional forms that drive the predictions. By combining neural networks with
Shapley-based explainability methods~\cite{lundberg2017unified}, we decompose
the fitted predictions into feature-level contributions and reconstruct the
nonlinear dependence of returns and signed volumes on their lagged values.

The central hypothesis of this work is that, although nonlinear models are
complex at the parametric level, the relationships they learn need not be
arbitrary or uninterpretable. On the contrary, if the model is capturing stable
features of the data-generating process, its predictions should be organized
around structured and economically meaningful mechanisms. In this sense,
explainability is used not merely as a diagnostic tool for understanding a
black-box predictor, but as a way to extract candidate structural relations
from the model.

Our analysis shows that the neural network learns a set of robust nonlinear
patterns that are consistent with well-known mechanisms of market
microstructure. Lagged signed volume generates sign-preserving and saturating
responses, which can be interpreted as nonlinear price impact in the return
equation and order-flow persistence in the signed-volume equation. Lagged
returns, instead, mainly act as state variables: they modulate the strength of
the order-flow signal and generate short-term reversal or attenuation effects,
depending on the magnitude of the previous price move. These relationships are
stable across assets with some differences between large-tick and small-tick
stocks, indicating that the model captures features that are sensitive to the
underlying microstructure regime.

We further validate these findings by comparing the model-implied predictive
surfaces with analogous conditional averages computed directly from the data.
The close qualitative agreement between the two provides an important
consistency check: the nonlinear structures revealed by the neural network are
not artifacts of the architecture or of the explainability method, but reflect
empirical regularities already present in the data.

Furthermore, the advantage of the explainability analysis is that it allows
these regularities to be decomposed into the separate contributions of lagged
returns, lagged signed volumes, and the remaining past information. This
provides access to the internal structure of the nonlinear predictor and makes
it possible to isolate the functional role of individual regressors. In
particular, the Shapley decomposition shows that the lagged-volume contribution
has a clean sign-like and saturating structure, with no substantial spurious
dependence on lagged returns. The return contribution is instead more
state-dependent and is strongly organized by the sign of previous order flow:
when the previous trade does not move the price, the model tends to preserve
the sign of lagged signed volume, whereas a non-zero previous return generates
attenuation or reversal.

This decomposition allows us to design simpler parametric models. In
particular, we introduce a SHAP-inspired nonlinear parametric specification that
incorporates the main functional forms identified by the neural network: a
saturating dependence on lagged signed volume and a state-dependent correction
driven by the magnitude of lagged returns and by the sign of previous order
flow.

A key result of the paper is that this reduced specification improves upon the linear VAR benchmark and achieves predictive performance comparable to that of the neural network, while being more stable with respect to the choice of lag order. The model therefore retains a large fraction of the predictive power of the flexible nonlinear architecture, but in a parsimonious and economically interpretable form.

Overall, our results demonstrate that explainability provides a powerful
framework to uncover and interpret nonlinear structure in financial data. It
offers a principled way to move from black-box prediction to interpretable model
construction: neural networks can be used as discovery devices, while Shapley
decompositions and aggregated conditional response analysis can translate the learned
relationships into economically meaningful mechanisms and parsimonious
parametric specifications. In this way, explainability helps bridge flexible
machine-learning models and traditional econometric modeling, and provides a
way to assess whether the learned nonlinear relationships reflect genuine market
mechanisms rather than spurious statistical patterns.

The remainder of the paper is organized as follows. 
Section~\ref{sec:models} introduces the modeling framework and presents both linear and nonlinear specifications. 
Section~\ref{sec:data} describes the data and preprocessing steps. 
Section~\ref{sec:forecasting} compares the forecasting performance of the models. 
Section~\ref{sec:explainability} develops the explainability analysis and extracts the nonlinear dependencies, decomposing the contributions of the most relevant regressors. 
%Section~\ref{sec:data_nonlinear} validates these findings directly on the data. 
%Section~\ref{sec:var_nonlinear} shows how simple nonlinear transformations can be incorporated into linear VAR-type models. 
Section~\ref{sec:shap_parametric_model}
introduces and calibrates a SHAP-inspired parametric nonlinear model.
Section~\ref{sec:conclusions} concludes.

\section{Price and trades models}\label{sec:models}
A price and trade model describes the joint dynamics of prices and trading activity, capturing the relationships between order flow, price formation, and market impact. These models aim to characterize how trades influence prices and how market participants adjust their behavior in response to price movements. They are essential for forecasting market dynamics, optimizing execution strategies, and simulating realistic market conditions. In this section, we first provide a general definition of price and trade models, outlining their key components. We then present two classes of models considered in this study: parametric autoregressive linear models and nonparametric nonlinear models.

Price and trade models in discrete time assume a recursive dependence on lagged variables and a cross-dependence between price and trade volume. They can be formulated in the following general form:  

\begin{equation}\label{eq:general_model}
\begin{split}  
& p_t = {\mathcal F}_1(p_{t-1}, p_{t-2}, \dots, v_{t-1}, v_{t-2}, \dots), \\  
& v_t = {\mathcal F}_2(p_{t-1}, p_{t-2}, \dots, v_{t-1}, v_{t-2}, \dots),  
\end{split}
\end{equation}  

where \( t \in \mathbb{N} \) represents the trade time, \( p_t \) is the midpoint price just before the trade, and \( v_t \) is the signed volume of the trade at time \( t \)\footnote{As usual, \( v_t > 0 \) (\( v_t < 0 \)) for buyer- (seller-) initiated trades. We will also refer to \( v_t \) as the order flow.}.  

In this formulation, the functions \( \mathcal{F}_1 \) and \( \mathcal{F}_2 \) define the model, describing the relationship between the present values of \( p_t \) and \( v_t \) with their respective lagged values. The first equation expresses the dependence of the midpoint price on its past values and on the current and past trade volumes, while the second equation models the order flow as a function of past price and volume values. This structure captures both the time series properties of each variable and their mutual influence.

To ensure stationarity, the first equation is rewritten in terms of midpoint price increments, $\Delta p_t = p_{t+1} - p_{t}$, or price returns. Stationarity is a fundamental property in time series modeling, as it ensures that statistical properties such as mean and variance do not change over time, making the model more reliable for both estimation and forecasting. Non-stationary models can lead to spurious correlations and unstable predictions, limiting their applicability in real-world financial settings.  

In this work, we consider logarithmic returns, i.e., $r_t = \log(p_{t+1}/p_t)$. Thus, the models are always expressed in the form:

\begin{equation}\label{eq:general_model_returns}
\begin{split}
    & r_t = \mathcal{G}_1(r_{t-1}, r_{t-2}, \dots, v_{t-1}, v_{t-2}, \dots), \\
    & v_t = \mathcal{G}_2(r_{t-1}, r_{t-2}, \dots, v_{t-1}, v_{t-2}, \dots).
\end{split}
\end{equation}

We focus here on lagged dependencies only. In contrast, several traditional market-impact models, such as propagator models and Hasbrouck-type specifications, also allow for a contemporaneous effect of order flow on returns. In our framework, the contemporaneous contribution of $v_t$ is not included explicitly when predicting $r_t$, but is instead absorbed into the residual component, which may therefore retain part of the instantaneous dependence between returns and volumes. We return to this point later in the paper. This choice also keeps all time series aligned in length and preserves a consistent structure across different datasets.

As before, the functions $\mathcal{G}_1$ and $\mathcal{G}_2$ define the model structure and can be specified using different approaches. When building ${\mathcal G}_1$ and $\mathcal{G}_2$, a key distinction lies between \textit{parametric} and \textit{nonparametric} models. In a \textit{parametric approach}, the functional forms of ${\mathcal G}_1$ and ${\mathcal G}_2$ are explicitly defined based on theoretical principles or empirical observations. This requires specifying a limited set of parameters that characterize the model, which are then estimated from the data. Parametric models offer interpretability and computational efficiency but may struggle to capture complex, nonlinear dependencies in financial time series. 

In contrast, a \textit{nonparametric approach} does not impose a predefined functional form. Instead, the model is inferred directly from historical data, allowing for greater flexibility in capturing intricate patterns and nonlinear relationships. Machine learning techniques, such as neural networks, exemplify this approach by learning the mapping between inputs and outputs without assuming a specific mathematical structure. While nonparametric models can achieve higher predictive accuracy, they may lack interpretability and require larger datasets for robust estimation. The choice between parametric and nonparametric models depends on the trade-off between interpretability and predictive power, as well as the availability of domain knowledge to guide the modeling process.

In this work, we first compare a linear parametric benchmark, implemented as a VAR model, with a nonlinear nonparametric benchmark, implemented as a deep neural network. Since the latter achieves higher predictive accuracy, especially for returns, we use explainability techniques to identify the nonlinear dependencies learned by the network. The final objective is to translate these dependencies into an interpretable nonlinear parametric model with competitive predictive performance.

\subsection{Parametric Linear models}

We consider a linear version of the general model in Eq.~\eqref{eq:general_model_returns}, based on the VAR framework defined as:
\begin{equation}\label{hasbrouckmodel}
\begin{split}
r_t = \sum_{i=1}^{p} a_i r_{t-i} + \sum_{i=1}^{p} b_i v_{t-i} + u_{t}^r, \\
v_t = \sum_{i=1}^{p} c_i r_{t-i} + \sum_{i=1}^{p} d_i v_{t-i} + u_{t}^v,
\end{split}
\end{equation}
where \( r_t \) is the midpoint price return, and \( v_t \) is the signed volume.

The terms \( u_{t}^r \) and \( u_{t}^v \) are white noise processes, which are generally correlated, thereby capturing the residual contemporaneous dependence between returns and order flow. The parameter \( p \) represents the lag, i.e., the number of past observations used for prediction. Although the two equations appear decoupled, they are linked through these residual terms, which encode information about the instantaneous interaction between price changes and trading activity. As we discuss later, this contributions can be decoupled recovering the Hasbrouck model~\cite{hasbrouck1991measuring} and, as a consequence, the impact models such as the Transient Impact Model~\cite{bouchaud2003fluctuations}.

Linear models of this type have been widely used to describe price formation and order flow dynamics because of their tractability and interpretability. They provide a natural framework to model transient price impact and order flow persistence, and have been extensively employed in empirical market microstructure studies. At the same time, their linear structure imposes strong restrictions on the form of the dependencies between returns and order flow. This is particularly relevant in the context of market impact, where empirical evidence points to nonlinear and state-dependent effects. More generally, models calibrated only on public trade and quote data may fail to reproduce the price trajectories observed during real metaorder executions. In particular, recent evidence shows that both linear and nonlinear models calibrated on public market data can generate almost linear impact profiles with limited post-trade reversion, in contrast with the concave execution profiles and subsequent reversion typically observed for real metaorders~\cite{Naviglio03042026}. This motivates the use of more flexible nonlinear specifications, while also highlighting the need to interpret their predictions with care.

\begin{comment}
By defining \( y_t = (r_t, v_t)' \), the model can be rewritten compactly as:  

\begin{equation}\label{svarP}
y_t = \sum_{i=1}^{p} \tilde{A}_i y_{t-i} + \tilde{u}_t,
\end{equation}  

where \( \tilde{A}_i \) are coefficient matrices, and \( \tilde{u}_t \) represents correlated noise terms. The SVAR model parameters can be efficiently estimated from data using ordinary least squares (OLS), given the clear causal structure in which trades impact prices contemporaneously, but not vice versa. We refer to this model as H(\( p \)).

A special case of the H(\( p \)) model is the Transient Impact Model, which is given by:  

\begin{equation}\label{transientimpact}
\begin{split}
r_t = \sum_{i=0}^{p} b_i v_{t-i} + u_{1,t}, \\
v_t = \sum_{i=1}^{p} d_i v_{t-i} + u_{2,t},
\end{split}
\end{equation}  

and can again be written in the compact form of Eq.~\eqref{svarP}. In this case, we adopt a linear model as an ansatz, which is then calibrated on data. Thus, the functional form is explicitly provided as an input to our study.  

It is also possible to evaluate such models independently by considering autoregressive models where each variable depends only on its own lagged values. In the following sections, we compare these approaches to assess which choices are preferable.
\end{comment}

\subsection{Non parametric Non Linear models}
\begin{figure}[!htb]
\centering
  \includegraphics[width=0.85\textwidth]{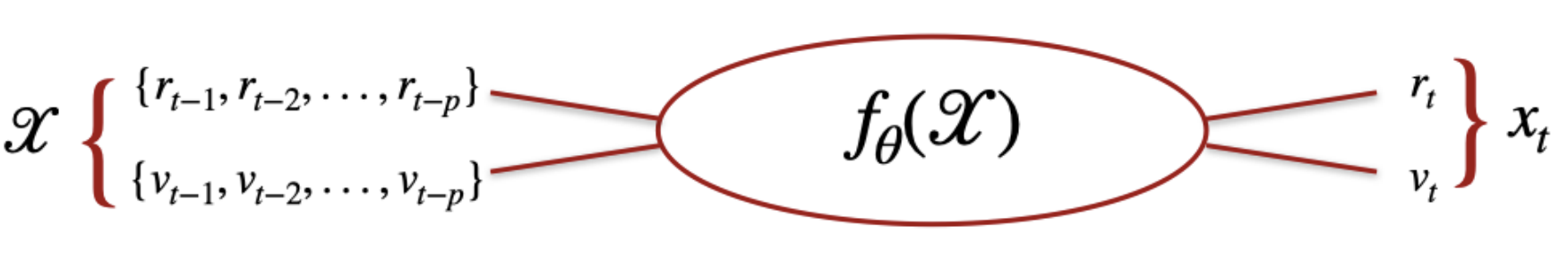}
    \caption{Schematic representation of non parametric model. Input data are processed through multiple hidden layers, each applying a nonlinear transformation, to produce the final output. A single function \( f_\theta \) maps the input data to both return and volume forecasts, allowing the network to learn joint representations and capture complex nonlinear dependencies between the two variables.}
\label{fig:NN_pict}
\end{figure} 

\begin{comment}
\begin{figure}[!htb]
\centering
  \includegraphics[width=0.85\textwidth]{figs/Model_NoCont.pdf}
    \caption{Schematic representation of the \textit{Separated Forecast Model} (SFM). Two independent functions, \( \mathcal{G}_1 \) and \( \mathcal{G}_2 \), separately predict returns and volumes. This structure provides a more interpretable model and aligns with the traditional VAR framework, at the cost of potentially losing joint learning advantages.}
\label{fig:NN_pict_Sep}
\end{figure} 
\end{comment}

In contrast to linear parametric models, nonparametric models do not impose a predefined functional form, but instead learn the relationship between inputs and outputs directly from the data. In this work, we adopt neural networks (NNs) to construct such models. Their use is motivated by the fact that feedforward neural networks provide flexible function approximators and, under suitable conditions, can approximate a broad class of nonlinear mappings~\cite{cybenko1989approximation,hornik1989multilayer,goodfellow2016deep}.

Neural networks define nonlinear mappings obtained by composing parametrized transformations across multiple layers. Given an input sequence \( \mathcal{X}_t \) of length \( p \), the network produces a prediction through a function \( f_{\theta}(\mathcal{X}_t) \), where \( \theta \) denotes the collection of trainable parameters. As in the VAR setting, the parameter \( p \) determines the maximum lookback horizon and will therefore be referred to as the lag. The function \( f_{\theta} \) is not specified ex ante, but is learned from the data by minimizing a loss function measuring the discrepancy between predictions and realized values, through gradient-based optimization~\cite{goodfellow2016deep}. Thus, in this case, we have the following relation between lagged quantities and predictions
\begin{equation}
    x_t = f_\theta(\mathcal{X}_t) + u_t,
\end{equation}
where, in complete analogy with the linear model, $x_t = (r_t, v_t)$  are the outputs of the model, $u_t = (u_t^r, u_t^v)$ are the correlated residuals, while $\mathcal{X}_t = (r_{t-1},..., r_{t-p}; v_{t-1}, ..., v_{t-p}) $ are the lagged inputs. In analogy with the VAR specification, the lag index \(\ell\) denotes the temporal distance from the predicted variable. Thus, the \(\ell\)-th lag of the DNN corresponds to the input pair (\(r_{t-\ell},v_{t-\ell}\)), with \(\ell=1,\ldots,p\). Unlike in the VAR case, however, these lagged variables enter as coordinates of the nonlinear map \(f_\theta\), so that their effect on the prediction is not restricted to be linear or additive.

Figure~\ref{fig:NN_pict} illustrates the neural network framework used in this work. In contrast to the linear setting, where the dynamics of returns and volumes are described by separate equations and contemporaneous dependence is reflected in correlated residuals, here both components are captured jointly by a single nonlinear mapping.

Several neural-network architectures can be used to parameterize the function
\(f_{\theta}\), including convolutional and recurrent architectures designed to
exploit local or long-range temporal dependencies in sequential data
~\cite{lecun1998gradient,hochreiter1997long,goodfellow2016deep}. In this work,
we use a deep feed-forward fully connected neural network with hidden layers as a
flexible nonlinear benchmark. For a lag order \(p\), the input vector is
\(\mathcal{X}_t\) so that the input dimension is \(2p\). The network has two outputs,
corresponding to the one-step-ahead predictions of \(r_t\) and \(v_t\).

The architecture consists of three hidden dense layers with \(64\), \(64\), and
\(32\) neurons, respectively. Each hidden layer uses a hyperbolic tangent
activation function, while the output layer is linear, as appropriate for a
regression task. All models are trained by minimizing the mean squared error
(MSE) loss using the Adam optimizer~\cite{kingma2014adam}.

This choice is motivated by two considerations. First, deep feed-forward neural
networks with hidden layers provide flexible nonlinear function approximators
~\cite{cybenko1989approximation,hornik1989multilayer}. Second, compared with
more complex sequence-specific architectures, this architecture defines a more
transparent and easily manipulable mapping from lagged inputs to predictions.
This makes it particularly suitable for the aggregated dependence and
Shapley-based analyses developed below
~\cite{friedman2001greedy,lundberg2017unified}. The same architecture is used
across stocks and lag specifications, so that differences in performance and
explainability patterns are not driven by changes in the neural-network design.
While we focus on this architecture for interpretability and reproducibility,
the approach could in principle be extended to other nonlinear architectures.

\subsection{Residual Contemporaneous Dependence}
\label{sec:residual_structural_decomposition}

Both the linear VAR model in Eq.~\eqref{hasbrouckmodel} and the nonlinear
models introduced above use lagged information only. In particular, the return
equation does not include the contemporaneous signed volume \(v_t\) as a direct
regressor. At first sight, this may seem restrictive from a market
microstructure perspective, since trades are expected to have an immediate
effect on prices. The point, however, is that contemporaneous order-flow effects
are not ignored. Since the models estimate reduced-form conditional
expectations based on past information, any instantaneous relation between
unexpected order flow and unexpected price changes remains encoded in the
residuals.

Let \(\mathcal{X}_t\) denote the lagged information set used by the model. For a
generic reduced-form predictor we write
\begin{equation}
\begin{split}
r_t &= f_r(\mathcal{X}_t) + u_t^r,\\
v_t &= f_v(\mathcal{X}_t) + u_t^v,
\end{split}
\label{eq:reduced_form_residuals}
\end{equation}
where \(u_t^r\) and \(u_t^v\) are the components of returns and signed volumes
that cannot be predicted from lagged variables alone. If trades have an
immediate effect on prices, these two residuals should remain contemporaneously
dependent even after the lagged dynamics have been removed.

This distinction is important because directly including \(v_t\) in the return
equation may lead to a misleading interpretation. Indeed, \(v_t\) contains both
a component predictable from past information and a genuinely contemporaneous
innovation:
\[
v_t
=
\mathbb{E}[v_t\mid \mathcal{X}_t]
+
u_t^v.
\]
Therefore, a model of the form \(r_t=G(\mathcal{X}_t,v_t)+\eta_t\) receives as
input not only the unexpected order-flow shock \(u_t^v\), but also the
predictable part of order flow, which is already determined by lagged
variables. In nonlinear models, these two effects are generally mixed inside
the same function \(G\), making it difficult to distinguish lagged
predictability from instantaneous impact. For this reason, we first estimate
the reduced-form model using only lagged information and then analyze the
residual dependence.

Following a Hasbrouck-type recursive interpretation, we separate the predictable component of the dynamics from the contemporaneous innovation. We assume that, in trade time, the innovation in order flow is contemporaneously prior to the quote revision. 
This motivates the triangular structural representation
\begin{equation}
u_t^v=\epsilon_t^v,
\qquad
u_t^r=\phi(u_t^v)+\epsilon_t^r.
\label{eq:nonlinear_residual_structural}
\end{equation}
Here, \(\epsilon_t^v\) is the structural order-flow shock, while \(\epsilon_t^r\) is the return innovation that remains after removing the contemporaneous impact of order flow. 
The function \(\phi\) therefore captures the instantaneous price impact of the unexpected component of signed volume. 
In the linear Hasbrouck model this reduces to \(\phi(u)=b_0u\), whereas in our nonlinear setting we allow \(\phi\) to be a nonparametric function.

This decomposition separates two objects. The reduced-form forecast,
\[
\widehat r_t^{\,\mathrm{lagged}}= f_r(\mathcal{X}_t),
\]
uses only information available up to time \(t-1\) and is therefore the object
used in the forecasting comparisons. By contrast, once the contemporaneous
signed volume is observed, one may form the structural conditional prediction
\begin{equation}
\widehat r_t^{\,\mathrm{struct}}
=
 f_r(\mathcal{X}_t)
+
\hat\phi\!\left(v_t-f_v(\mathcal{X}_t)\right).
\label{eq:structural_prediction}
\end{equation}
This second quantity is not a purely lagged forecast; it is conditional on the
realized order-flow innovation.

In the empirical analysis below, we therefore proceed in two steps. First, we
evaluate the forecasting performance of the reduced-form models using lagged
information only. Second, we estimate the residual contemporaneous projection defined in Equation~\eqref{eq:nonlinear_residual_structural}.
\begin{comment}
and verify whether the resulting structural shocks are approximately
orthogonal:
\[
\operatorname{Corr}(\epsilon_t^r,\epsilon_t^v)\simeq 0,
\qquad
\epsilon_t^v=u_t^v.
\]
This condition is not a proof of full statistical independence, since nonlinear
or higher-order dependence may still remain. Nevertheless, it provides a useful
diagnostic that the main contemporaneous dependence between return and
order-flow innovations has been removed. 
\end{comment}
A more detailed discussion of the
identification logic and of the difference between direct contemporaneous
conditioning and residual decomposition is provided in
Appendix~\ref{app:residual_structural_identification}.

\section{Data}\label{sec:data}
In this work, we use data obtained from LOBSTER\footnote{\url{lobsterdata.com}}, containing observations from NASDAQ-listed stocks. The dataset consists of two files: the message file and the order book file. The message file records all transactions that occur within the limit order book (LOB), while the order book file provides the state of the LOB immediately after each transaction.  

In our analysis, we focus exclusively on ``Execution of a visible limit order" events, excluding hidden orders. When adopting such a selection criterion, it is necessary to define a consistent price measure. We define the price as the mid-price of the LOB immediately before the execution of a visible order. Thus, we compute it based on the LOB configuration just before the execution of the visible order. The LOB then evolves due to subsequent events until another visible order is executed, at which point we again record the state of the LOB just before the event to define the price at that time. This approach follows the TIM convention (see~\cite{Naviglio03042026} for further details).  

%{\bf FAB: Forse si puo' tagliare} To implement this methodology, we first removed the first row of the message file and the last row of the order book file, ensuring that the volumes in the message file correspond to the preceding row in the order book file. We then filtered out all events except visible orders. Finally, we computed the returns between two subsequent prices, aligning them with the volume of the first price. In other words, we removed the last volume entry from our dataset. To prevent price jumps and, consequently, outliers in returns data, we applied this procedure separately to each daily file. In fact, price jumps can occur between the closing of one trading day and the opening of the next.  

Subsequently, we aggregated orders occurring at the same timestamp with the same sign. To mitigate potential distortions due to auction effects, we also removed the first and last 30 minutes of trading data. Finally, to reduce the influence of extreme observations and potential outliers, we applied a quantile filter by removing observations outside the \([0.005,0.995]\) quantile range. Furthermore, the training of the DNN has been obtained by normalizing data.

For all the models presented, the first 50\% of the sample is used for training,
while the remaining 50\% is used for out-of-sample testing.

In the analysis carried out in this work we consider June 2024 for 5 small tick stocks and 5 large tick stocks. As reported in Table~\ref{tab:spread_ticks}, large tick stocks are characterized by bid--ask spreads that remain close to one tick for most of the time, whereas small tick stocks typically display spreads of several ticks (see~\cite{eisler2012price} for details).

\begin{table}[ht]
\centering
\begin{tabular}{|c|c|c|}
\hline
 & \textbf{ticker} & \textbf{mean spread (ticks)} \\
\hline
\multirow{5}{*}{\rotatebox[origin=c]{90}{large tick}}
& BAC   &  1.16\\
& INTC  &  1.16\\
& CSCO  &  1.18\\
& CMCSA &  1.19\\
& PFE   &  1.20\\
\hline
\multirow{5}{*}{\rotatebox[origin=c]{90}{small tick}}
& AMZN  &  1.70\\
& AAPL  &  1.82\\
& GILD  &  2.03\\
& TSLA  &  3.69\\
& NVDA  &  6.27\\
\hline
\end{tabular}
\caption{Classification of the stocks considered in the analysis, together with their mean spread measured in ticks.}
\label{tab:spread_ticks}
\end{table}

\section{Forecasting Performance of the Models}\label{sec:forecasting}

In this section, we evaluate the forecasting performance of the considered models, 
comparing parametric linear models with nonparametric nonlinear models. Performance 
is measured using the coefficient of determination \(R^2\), which quantifies how well 
the predicted values explain the variance of the actual data.

{\bf Lag choice.} Before comparing the different model classes, we discuss the choice of the lag order
used in the analysis. A detailed study of the dependence of forecasting performance
on the lag length is reported in Appendix~\ref{app:lag_selection}. For the VAR
benchmark, predictive performance initially improves as the lag order increases,
but the out-of-sample \(R^2\) rapidly reaches a stable level and does not display
systematic gains at longer lags. This suggests that most of the useful predictive
information captured by the linear model is concentrated in the recent past.

The neural-network model leads to the same qualitative conclusion, although its
lag-dependence is less regular. After the first lags, the test performance fluctuates
around a comparable range of values rather than increasing monotonically with the
lag order. As discussed in Appendix~\ref{app:lag_selection}, these fluctuations are
consistent with the fact that changing the lag order also changes the input
representation and the nonlinear mapping learned by the network.

Overall, both analyses indicate that very long lag structures are not required for
the forecasting task considered here. We therefore set \(p=20\) in the following
analysis. This choice is sufficiently large to include the relevant short-memory
dependencies observed in the data, while avoiding unnecessarily long input vectors
and excessive parameter proliferation.

\begin{table}[ht]
\centering
\begin{tabular}{|c|c|c|c|c|c|}
\hline
 & \textbf{ticker} 
 & \(\mathbf{R^2_{\mathrm{out}}(r)}\) \textbf{VAR}
 & \(\mathbf{R^2_{\mathrm{out}}(r)}\) \textbf{DNN}
 & \(\mathbf{R^2_{\mathrm{out}}(v)}\) \textbf{VAR}
 & \(\mathbf{R^2_{\mathrm{out}}(v)}\) \textbf{DNN} \\
\hline
\multirow{5}{*}{\rotatebox[origin=c]{90}{large tick}}
& BAC   & 0.0494 & 0.0591 & 0.0299 & 0.0215 \\
& INTC  & 0.0814 & 0.1217 & 0.0198 & 0.0288 \\
& CSCO  & 0.0764 & 0.0982 & 0.0328 & 0.0273 \\
& CMCSA & 0.0604 & 0.0859 & 0.0350 & 0.0486 \\
& PFE   & 0.0356 & 0.0367 & 0.0183 & 0.0091 \\
\hline
\multirow{5}{*}{\rotatebox[origin=c]{90}{small tick}}
& AMZN  & 0.0246 & 0.0341 & 0.0685 & 0.0705 \\
& AAPL  & 0.0643 & 0.0951 & 0.1147 & 0.1372 \\
& GILD  & 0.0499 & 0.0429 & 0.1055 & 0.1173 \\
& TSLA  & 0.0433 & 0.0906 & 0.1115 & 0.1475 \\
& NVDA  & 0.0166 & 0.0454 & 0.0872 & 0.1068 \\
\hline
\end{tabular}
\caption{Out-of-sample \(R^2\) for the VAR and DNN models at lag \(p=20\), separately for returns \(r\) and signed volumes \(v\). Stocks are grouped according to the large-tick/small-tick classification used in Table~\ref{tab:spread_ticks}.}
\label{tab:var_mlp_r2_out}
\end{table}

{\bf Performances.} The empirical performances of the VAR and DNN models are summarized in
Table~\ref{tab:var_mlp_r2_out}, which reports the out-of-sample \(R^2\) obtained
at lag order \(p=20\) for both returns and signed volumes. The comparison shows
that the DNN generally improves upon the VAR benchmark, although the strength of
the improvement depends on the predicted variable and on the tick-size regime.

For returns, the improvement is rather systematic: the DNN achieves a higher
out-of-sample \(R^2\) for almost all stocks in the sample. This indicates that the
nonlinear model is able to extract predictive structure that is not captured by
the linear VAR specification. For signed volumes, the evidence is more
heterogeneous. The DNN improves upon the VAR for all small-tick stocks, whereas
for large-tick stocks the two models are closer and the VAR remains competitive in
some cases. Overall, this suggests that nonlinear dependencies contribute to
forecasting performance beyond the linear benchmark, but that the incremental gain
is not uniform across variables and market regimes.

An interesting pattern also emerges when comparing the relative predictability of
returns and signed volumes across tick-size regimes. For large-tick stocks,
returns are more predictable than signed volumes, both under the VAR and under the
DNN. Conversely, for small-tick stocks, signed volumes are more predictable than
returns. This difference is consistent with the distinct microstructural nature of
the two regimes. In large-tick assets, the spread is typically constrained close
to one tick, so price changes are strongly affected by discreteness and queue
depletion at the best quotes. Recent return--volume states can therefore contain
relevant information about the direction of the next price move. Signed volumes,
instead, may remain noisier, since they also reflect liquidity provision,
queue-positioning decisions and trading motives not fully summarized by the
variables included in the model.

In small-tick assets, price changes are less mechanically constrained by the
minimum tick size and may depend on a richer liquidity state across several price
levels. Returns are therefore harder to forecast from the recent history of
returns and signed volumes alone. Signed order flow, by contrast, typically
displays stronger persistence: buyer-initiated trades tend to be followed by
further buyer-initiated trades, and similarly for seller-initiated trades. This
persistence makes signed volumes more predictable than returns in the small-tick
regime.

Overall, the out-of-sample results indicate that the DNN provides a useful
improvement over the VAR benchmark, especially for return prediction and for
signed-volume prediction in small-tick stocks. At the same time, the comparison
between returns and signed volumes shows that predictive performance is itself
informative about market microstructure: large-tick stocks display a stronger
link between recent order-flow pressure and discrete price changes, while
small-tick stocks reveal a clearer persistence in the direction of trading
pressure.

These results motivate the analysis of the next section. Since the DNN generally 
outperforms the linear VAR benchmark, we ask which additional dependencies are 
captured by the nonlinear model and how they contribute to the improvement in 
forecasting performance. The comparison of predictive accuracy therefore leads 
naturally to an explainability problem: understanding what the neural network has 
learned about the joint dynamics of returns and signed volumes.

In the following, we use explainability tools to characterize the functional 
relationships learned by the DNN. In particular, we investigate whether the network 
mainly reproduces and refines the linear effects already present in the VAR, or 
whether it captures genuinely nonlinear dependencies that are absent from the 
linear benchmark.

\section{Explaining the Neural Network}\label{sec:explainability}

\begin{comment}
\begin{figure}[!htb]
\centering
\begin{subfigure}{0.48\textwidth}
  \centering
  \includegraphics[width=\linewidth]{figs/r_v.png}
  \caption{.}
  \label{fig:r_v}
\end{subfigure}
\hfill
\begin{subfigure}{0.48\textwidth}
  \centering
  \includegraphics[width=\linewidth]{figs/r_r.png}
  \caption{}
  \label{fig:r_r}
\end{subfigure}
\caption{\textit{.}}
\label{fig:Explain}
\end{figure}
\end{comment}

As shown in the previous section, nonlinear models tend to improve predictive
performance relative to their linear counterparts. This suggests that the data may
contain dependencies that are not fully captured by a purely linear specification.
At the same time, the flexibility of neural networks makes their predictions less
immediately interpretable, since the functional form learned from the data is not
specified a priori.

For this reason, we now analyze the trained neural network itself. The goal is to
understand which dependencies the model has learned and to assess whether the
improvement over the VAR benchmark is associated with genuinely nonlinear effects
in the joint dynamics of returns and signed volumes.

The key point of the present section is that, once a nonlinear neural network is
trained on the data, explainability methods allow us to decompose its prediction
into the contributions of the individual input variables. This provides
information that is not directly accessible from empirical conditional averages
alone. The data can reveal that a nonlinear dependence exists between returns
and signed volumes, but the trained model, through its attribution structure,
allows us to separate the role of lagged signed volume, lagged returns, and the
remaining past history in the construction of the prediction.

In this sense, the neural network is used as a structural discovery device. If
the fitted model captures stable patterns that are also consistent with direct
empirical evidence, its explanations can be interpreted as a guide to the
minimal nonlinear mechanisms present in the data. As we show below, this makes
it possible to move from a flexible nonparametric predictor to a parsimonious
parametric model whose functional form is designed to reproduce the patterns
identified by the neural network.

As defined above, the nonlinear predictor is a map
\[
 f:\mathbb{R}^{2p}\to\mathbb{R}^2,
\qquad
 f(\mathcal{X}_t)=\big(\hat r_t,\hat v_t\big),
\]
which returns the model-based forecasts of next-period return and signed
volume. In what follows, we focus on each component individually and indicate them with $f_r(\mathcal{X}_t)$ and $f_v(\mathcal{X}_t)$, for the return and volume components, respectively.

Thus, in this section we first extract a set of diagnostic observations that allow us to assess whether the model behaves consistently with the financial mechanisms we expect. We begin by computing Shapley values, which enable us to identify interpretable and economically meaningful patterns in the model predictions. We then construct empirical response surfaces relating the relevant variables, both from the data and from the trained model, and compare them directly. This comparison provides an additional benchmark to verify that the model reproduces the main structures present in the data.

Once these checks indicate that the model is capturing the expected financial relationships, we can use the Shapley decomposition with greater confidence to isolate and interpret the marginal contribution of each input variable to the model output, while keeping in mind that these contributions should be understood as model-based explanations rather than causal effects.

%one scalar output at atime. We therefore write
%\[
%\hat f_\theta(x_t)
%\]
%for a generic scalar prediction extracted from the vector-valued model, so that, depending on the context,
%\[
%\hat f_\theta(x_t)=\hat r_t
%\qquad\text{or}\qquad
%\hat f_\theta(x_t)=\hat v_t.
%\]

\subsection{The Shapley decomposition}
For a given scalar output $f_a(\mathcal{X}_t)$, with $a\in \{r,v\}$ the Shapley additive explanations (SHAP)~\cite{lundberg2017unified} express the prediction as the sum of a baseline term and the contributions of all input features. Denoting by $\mathcal{J}$ the full set of input regressors, we write
\begin{equation}
f_a(\mathcal{X}_t)
=
\mathbb{E}[ f_{a}(\mathcal{X}_t)]
+
\sum_{j\in \mathcal{J}} \phi_j^{a}(\mathcal{X}_t).
\label{eq:shap_decomposition_full}
\end{equation}
In our setting, the input vector consists of lagged returns and lagged signed volumes, so that the decomposition can be written explicitly as
\begin{equation}
 f_a(\mathcal{X}_t)
=
\mathbb{E}[ f_a(\mathcal{X}_t)]
+
\sum_{\ell=1}^{p} \phi_{r_{t-\ell}}^{a}(\mathcal{X}_t)
+
\sum_{\ell=1}^{p} \phi_{v_{t-\ell}}^{a}(\mathcal{X}_t).
\label{eq:shap_decomposition_lags}
\end{equation}
Here:
\begin{itemize}
    \item $\mathbb{E}[ f_a (\mathcal{X}_t)]$ is the baseline prediction, i.e.\ the average output of the model over the reference sample;
    \item $\phi_{r_{t-\ell}}^{a}(\mathcal{X}_t)$ is the Shapley contribution of the lagged return $r_{t-\ell}$ on the predicted variable $a$;
    \item $\phi_{v_{t-\ell}}^{a}(\mathcal{X}_t)$ is the Shapley contribution of the lagged signed volume $v_{t-\ell}$ on the predicted variable $a$.
\end{itemize}

This representation makes it possible to quantify the marginal contribution of each lagged input separately. In particular, it allows us to distinguish the role of recent lags, such as $r_{t-1}$ and $v_{t-1}$, from the aggregate effect of more distant past values.

For a scalar output \( f_a\), given the decomposition~\eqref{eq:shap_decomposition_lags}, the Shapley value of feature \(j\) at point \(x_t\) can be written as
\begin{equation}
\phi_j^{a}(x_t)
=
\sum_{S\subseteq \mathcal{J}\setminus\{j\}}
\frac{|S|!\,\big(|\mathcal{J}|-|S|-1\big)!}{|\mathcal{J}|!}
\left[
v_a(S\cup\{j\};x_t)-v_a(S;x_t)
\right],
\label{eq:shapley_definition}
\end{equation}
where \(v_a(S;x_t)\) denotes the value of the model prediction on the variable $a$ when only the subset of features \(S\) is observed. In practice, this quantity is defined by marginalizing or replacing the missing features with respect to a reference distribution, usually represented by a background sample. Thus, Eq.~\eqref{eq:shapley_definition} expresses the contribution of feature \(j\) as the average change in the model prediction obtained when \(j\) is added to all possible subsets of the remaining features.

It is useful to keep in mind what this decomposition implies in the linear
benchmark defined in Eq.~\eqref{hasbrouckmodel}. Under the background formulation adopted here, since the VAR
predictor is linear in the lagged regressors, the Shapley contribution of each
input reduces to its linear coefficient times the deviation of that input from
its background mean. For
instance, for the return equation one obtains 
\begin{equation}
    \phi_{r_{t-i}}^{r}(\mathcal{X}_t)
    =
    a_i
    \left(
    r_{t-i}
    -
    \mathbb E[r_{t-i}]
    \right),
    \qquad
    \phi_{v_{t-i}}^{r}(\mathcal{X}_t)
    =
    b_i
    \left(
    v_{t-i}
    -
    \mathbb E[v_{t-i}]
    \right),
    \label{eq:var_shap_return}
\end{equation}
while for the signed-volume equation
\begin{equation}
    \phi_{r_{t-i}}^{v}(\mathcal{X}_t)
    =
    c_i
    \left(
    r_{t-i}
    -
    \mathbb E[r_{t-i}]
    \right),
    \qquad
    \phi_{v_{t-i}}^{v}(\mathcal{X}_t)
    =
    d_i
    \left(
    v_{t-i}
    -
    \mathbb E[v_{t-i}]
    \right).
    \label{eq:var_shap_volume}
\end{equation}
Thus, in the VAR case, the Shapley decomposition recovers the usual linear contribution of each lagged
regressor, centered with respect to the reference sample. This observation provides
a useful benchmark for the nonlinear case. For the neural network, the same
decomposition summarizes a nonparametric prediction in terms of feature-wise
contributions, and deviations from the linear pattern above can reveal dependencies
that are not captured by the VAR specification.

The exact computation of Eq.~\eqref{eq:shapley_definition} is infeasible in our setting, since it requires averaging over all possible subsets of the input variables. With \(2p\) input features, corresponding to \(p\) lagged returns and \(p\) lagged signed volumes, the number of coalitions grows exponentially with the lag length. For this reason, we compute Shapley-based attributions using the \texttt{DeepExplainer} method implemented in the SHAP library~\cite{lundberg2017unified}. This method provides an efficient approximation of Shapley values for neural networks by combining ideas from DeepLIFT-style backpropagation rules with an expectation over a background reference sample~\cite{shrikumar2017learning,lundberg2017unified}. In this way, the contribution of each feature is evaluated relative to the distribution of background observations, rather than by explicitly enumerating all feature coalitions.

Therefore, throughout the paper we refer to these quantities as SHAP values. The resulting attributions preserve the additive interpretation in Eq.~\eqref{eq:shap_decomposition_full} and allow us to decompose the fitted nonlinear prediction into feature-level contributions.

\subsection{Shapley values}

In the following we want to understand whether the improvement observed in the previous section is
associated with economically meaningful dependencies in the data. In other words,
we would like to verify that the model is not only exploiting its greater
flexibility to improve forecasting accuracy, but is also learning patterns that are
consistent with plausible financial and microstructural mechanisms.

A first diagnostic in this direction is provided by the raw Shapley values. Since
Shapley values decompose each prediction into feature-wise contributions, they
allow us to inspect which lagged returns and signed volumes are used by the
network, and with which sign and intensity. Systematic patterns in these
contributions can therefore provide preliminary evidence that the model is
extracting stable structure from the return--volume dynamics, rather than relying
on purely noisy fluctuations.

\begin{figure}[H]
    \centering
    \begin{subfigure}{0.48\textwidth}
        \includegraphics[width=\textwidth]{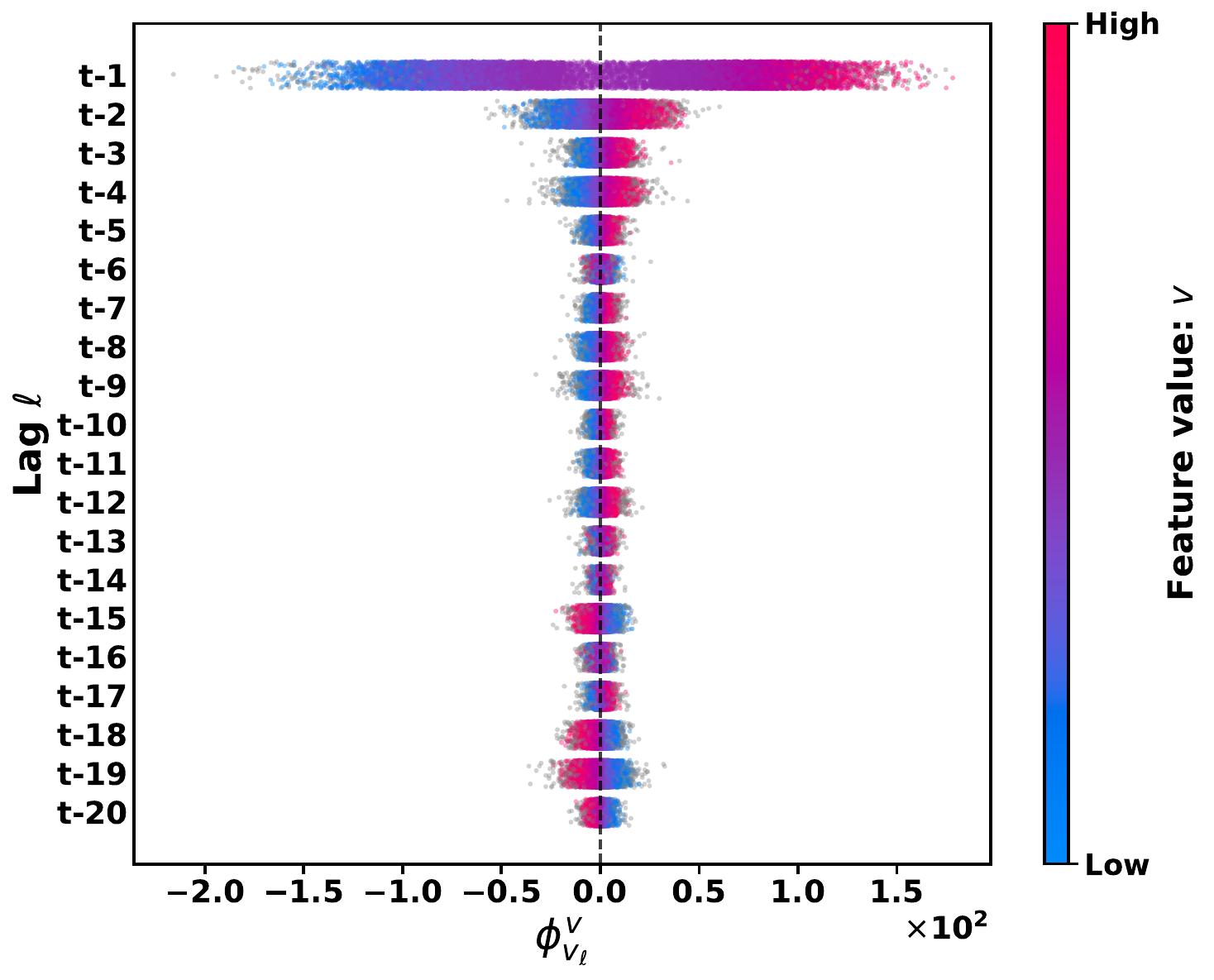}
        \caption{Effect of lagged signed volume $v_{t-\ell}$ on predicted volume $v_t$. Positive (negative) volumes generate positive (negative) contributions, reflecting persistence of order flow.}
    \end{subfigure}
    \begin{subfigure}{0.48\textwidth}
        \includegraphics[width=\textwidth]{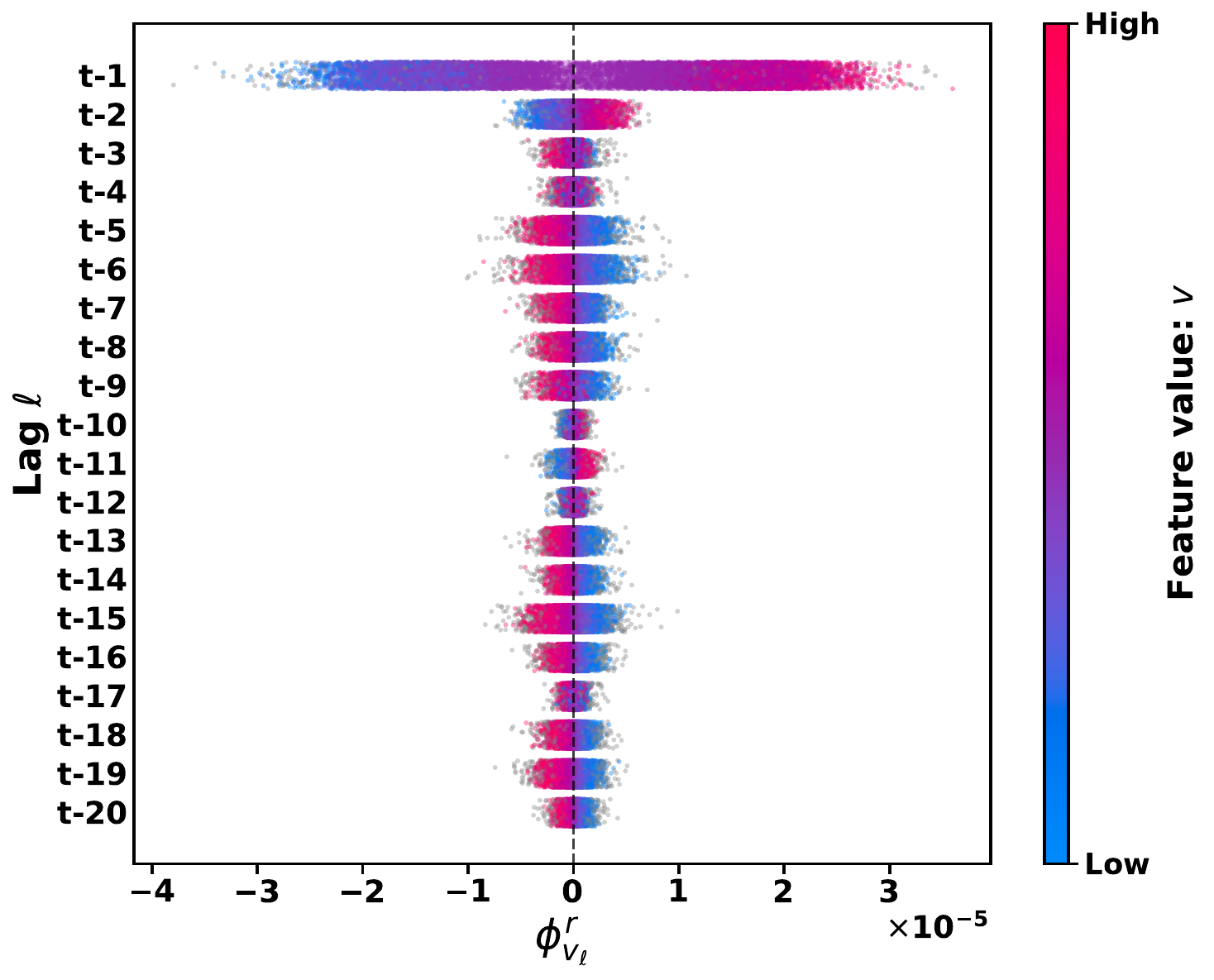}
        \caption{Effect of lagged signed volume $v_{t-\ell}$ on predicted return $r_t$. Positive order flow induces positive contributions, consistent with price impact.}
    \end{subfigure}
    
    \begin{subfigure}{0.48\textwidth}
        \includegraphics[width=\textwidth]{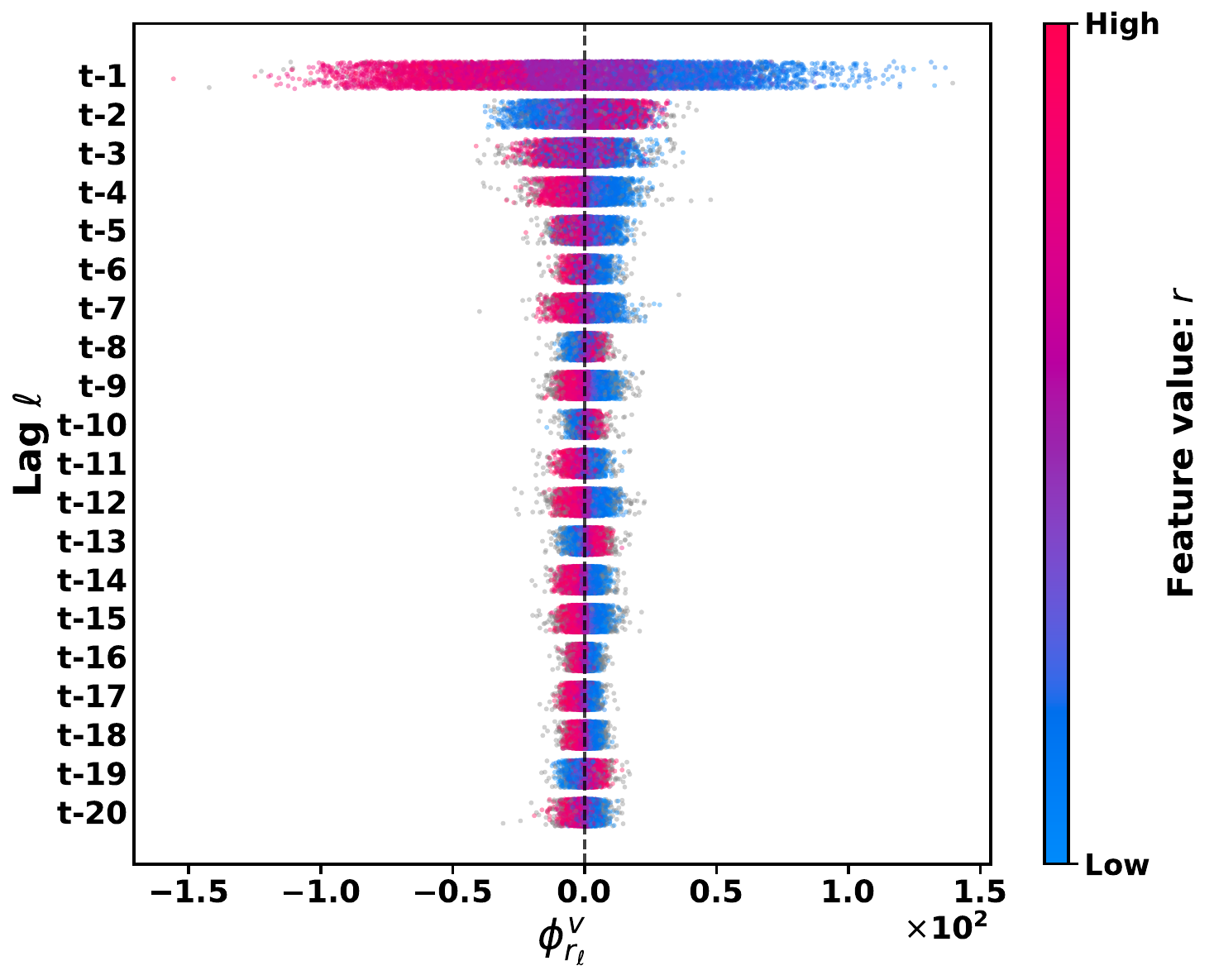}
        \caption{Effect of lagged return $r_{t-\ell}$ on predicted volume $v_t$. Large price moves increase expected trading activity, consistent with the volatility--volume relationship.}
    \end{subfigure}
    \begin{subfigure}{0.48\textwidth}
        \includegraphics[width=\textwidth]{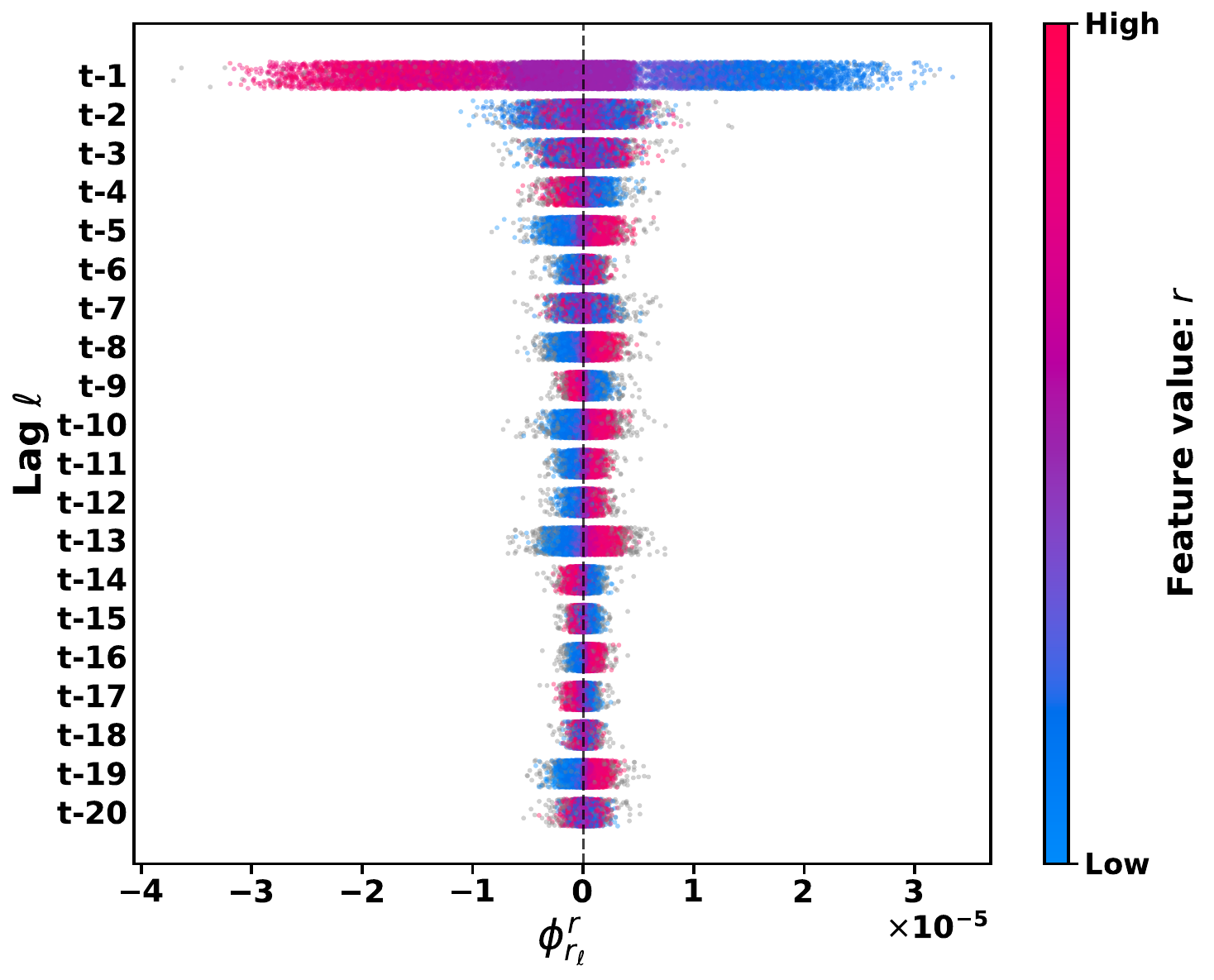}
        \caption{Effect of lagged return $r_{t-\ell}$ on predicted return $r_t$. Positive past returns generate negative contributions, revealing short-term mean reversion.}
    \end{subfigure}
    
    \caption{Shapley decomposition for AAPL. Colors indicate feature values (blue: low, red: high).}
    \label{fig:Shapley}
\end{figure}

In Figure~\ref{fig:Shapley}, we show the Shapley values for a representative stock,
namely AAPL. The results presented below are qualitatively robust across all
considered tickers, both large- and small-tick, and AAPL can therefore be viewed as
a benchmark case. We consider four configurations, corresponding to the effect of
lagged signed volumes and lagged returns on both predicted returns and predicted
volumes. Each panel reports the distribution of Shapley values across observations
for each lag \(\ell = 1,\dots,p\), with color indicating the realized value of the
corresponding feature: blue for low values and red for high values.

A first important feature is the strong concentration of Shapley magnitude at
the first lag. Across all panels, the largest contributions are associated with
\(\ell=1\), while the explanatory power rapidly decays for larger lags. This
suggests that the neural network mainly relies on the most recent
return--volume pair when forming its predictions.

The sign structure of the Shapley values is also economically meaningful.
Lagged signed volumes generate sign-preserving contributions: positive order
flow pushes both the predicted volume and the predicted return upward, while
negative order flow pushes them downward. This captures, respectively,
order-flow persistence in the volume equation and directional price impact in
the return equation.

Lagged returns display a different pattern. In the return prediction, positive
past returns tend to generate negative contributions and negative past returns
positive contributions, revealing a short-term reversal effect. In the volume
prediction, the contribution of lagged returns is more symmetric and mainly
related to the magnitude of past price changes, consistently with the link
between volatility and trading activity.

Overall, the SHAP profiles indicate that the neural network has learned
interpretable short-horizon mechanisms: order-flow persistence, directional
price impact, and return reversal. A more detailed analysis of the distribution
of Shapley importance across lags is reported in
Appendix~\ref{app:shap_lag_importance}.

As already suggested by
Figure~\ref{fig:Shapley}, Shapley importance is strongly concentrated at the
first lag across all prediction blocks and both tick-size regimes. This
motivates the focus on the first-lag coordinates
\((r_{t-1},v_{t-1})\) in the predictive-surface analysis below.

\subsection{Model-Implied Predictive Surface}\label{sec:Surfaces}

As justified by the previous observations, we focus on the most recent return--volume observations. Importantly, we do not re-estimate the model using only the first lag. The predictor is still fitted using the full \(p\)-lag input vector \(\mathcal{X}_t\); we only examine its dependence on the most recent pair \((r_{t-1},v_{t-1})\), while the remaining lagged coordinates retain the values observed in each empirical realization. This choice allows us to isolate and visualize the dominant short-term dependence without discarding the additional information contained in deeper lags. Re-fitting a model using only \(t-1\) would instead define a different forecasting specification and would therefore confound the analysis of the fitted nonlinear response with a change in the information set available to the model.

At the pointwise level, this is simply the map
\[
\mathcal{X}_t \mapsto f(\mathcal{X}_t),
\]
viewed as a function of the pair \((r_{t-1},v_{t-1})\), where \(f\) denotes the fitted model predictor. In practice, when we plot the cloud of points
\[
\big(r_{t-1},\,v_{t-1},\,f(\mathcal{X}_t)\big),
\]
we are visualizing the model response on the empirical support of the data. This representation is useful because it shows directly how the fitted nonlinear predictor behaves in regions of the state space that are actually visited by the market. Moreover, this object has a natural empirical counterpart obtained by replacing the model prediction \(f(\mathcal{X}_t)\) with the realized target variable. Comparing the model-implied response with the corresponding empirical cloud therefore allows us to assess whether the dependencies reconstructed by the neural network are genuinely present in the data, or whether they are artifacts introduced by the flexibility of the model.

To obtain more readable objects, we consider bin-averaged versions of the predictive surface. The construction, however, differs depending on whether the horizontal axis is given by \(v_{t-1}\) or by \(r_{t-1}\).

When the response is represented as a function of \(v_{t-1}\), we partition the support of \(v_{t-1}\) into uniform bins, denoted by \(\{B_u^v\}_u\). We then compute conditional averages of the model prediction within each bin, further splitting the sample according to the regime of \(r_{t-1}\), namely \(r_{t-1}>0\), \(r_{t-1}<0\), and \(r_{t-1}=0\). The corresponding model-implied response is therefore
\begin{equation}
\hat m_{a|v}(u,k_r)
:=
\mathbb{E}\!\left[f_a(\mathcal{X}_t)\mid v_{t-1}\in B_u^v,\ \mathrm{sign}(r_{t-1})=k_r\right],
\label{eq:model_surface_v_binned}
\end{equation}
where \(k_r\in\{-1,0,+1\}\) indexes the sign regime of \(r_{t-1}\) and $a \in \{r,v\}$.

When the response is represented instead as a function of \(r_{t-1}\), uniform binning is less appropriate, since log-returns are concentrated around a finite set of local clusters induced by price discreteness. For this reason, we identify the relevant regions of the support of \(r_{t-1}\) through a Gaussian kernel density estimator, and define clusters by locating the local maxima of the estimated density and assigning each observation to the interval delimited by the midpoints between adjacent peaks. Denoting these data-driven clusters by \(\{B_u^r\}_u\), we define
\begin{equation}
\hat m_{a|r}(u,k_v)
:=
\mathbb{E}\!\left[f_a(\mathcal{X}_t)\mid r_{t-1}\in B_u^r,\ \operatorname{sign}(v_{t-1})=k_v\right],
\label{eq:model_surface_r_binned}
\end{equation}
where \(k_v\in\{-1,+1\}\) denotes the sign of \(v_{t-1}\). Clearly, volumes are non vanishing and cannot have zero sign.

Similarly, we can define the same quantities on the Shapley functions

\begin{equation}
\begin{aligned}
s_v^{a}(u,k_r)
&:=
\mathbb{E}\!\left[
\phi_{v_{t-1}}^{a}(\mathcal{X}_t)
\mid
v_{t-1}\in B_u^v,\ \operatorname{sign}(r_{t-1})=k_r
\right],\\
s_{r}^a(u,k_v)
&:=
\mathbb{E}\!\left[
\phi_{r_{t-1}}^a(\mathcal{X}_t)
\mid
r_{t-1}\in B_u^r,\ \operatorname{sign}(v_{t-1})=k_v
\right],\\
s_{r}^{*,a}(u,k_r)
&:=
\mathbb{E}\!\left[
\phi_{r_{t-1}}^{a}(\mathcal{X}_t)
\mid
v_{t-1}\in B^v_u,\ \operatorname{sign}(r_{t-1})=k_r
\right], \\
s_{v}^{*,a}(u,k_v)
&:=
\mathbb{E}\!\left[
\phi_{v_{t-1}}^{a}(\mathcal{X}_t)
\mid
r_{t-1}\in B^r_u,\ \operatorname{sign}(v_{t-1})=k_v
\right], \\
s_{\mathrm{rest},v}^{a}(u,k_r)
&:=
\mathbb{E}\!\left[
\phi_{\mathrm{rest}}^{a}(\mathcal{X}_t)
\mid
v_{t-1}\in B^v_u,\ \operatorname{sign}(r_{t-1})=k_r
\right],\\
s_{\mathrm{rest},r}^{a}(u,k_v)
&:=
\mathbb{E}\!\left[
\phi_{\mathrm{rest}}^{a}(\mathcal{X}_t)
\mid
r_{t-1}\in B^r_u,\ \operatorname{sign}(v_{t-1})=k_v
\right].
\end{aligned}
\label{eq:shapley_profiles_definitions}
\end{equation}

Here, we have defined 
\begin{equation}
    \phi_{\text{rest}}^a = \sum_{\ell = 2}^p \big[ \phi_{r_{t-\ell}}^a(\mathcal{X}_t) + \phi_{v_{t-\ell}}^a(\mathcal{X}_t) \big].
\end{equation}

Using these definitions, we can recover Equation~\eqref{eq:shap_decomposition_full} in conditional form as
\begin{equation}
\begin{split}
\hat m_{a|v}(u,k_r)
&=
\mathbb{E}[f_a(\mathcal{X}_t)]
+
s_v^{a}(u,k_r)
+
s_r^{*,a}(u,k_r)
+
s_{\mathrm{rest},v}^{a}(u,k_r),\\
\hat m_{a|r}(u,k_v)
&=
\mathbb{E}[f_a(\mathcal{X}_t)]
+
s_r^{a}(u,k_v)
+
s_v^{*,a}(u,k_v)
+
s_{\mathrm{rest},r}^{a}(u,k_v).
\end{split}
\label{eq:binned_shap_decomposition}
\end{equation}

The quantities  defined above provide a Shapley-based decomposition of the binned predictive surfaces. The non-starred terms,
\[
s_v^{a}(u,k_r)
\qquad\text{and}\qquad
s_r^{a}(u,k_v),
\]
measure the average Shapley contribution of the variable displayed on the horizontal axis of the corresponding aggregated conditional response surfaces plot. More precisely, \(s_v^{a}(u,k_r)\) captures the contribution of \(v_{t-1}\) when the model response is studied as a function of \(v_{t-1}\), conditional on the regime of \(r_{t-1}\). Analogously, \(s_r^{a}(u,k_v)\) captures the contribution of \(r_{t-1}\) when the model response is studied as a function of \(r_{t-1}\), conditional on the sign of \(v_{t-1}\).

The starred quantities,
\[
s_r^{*,a}(u,k_r)
\qquad\text{and}\qquad
s_v^{*,a}(u,k_v),
\]
measure the average Shapley contribution of the conditioning variable, rather than of the variable shown on the horizontal axis. Therefore, they are useful to understand how the separation between the different conditional curves is modulated by the second covariate. For example, when the response is plotted as a function of \(v_{t-1}\), the quantity \(s_r^{*,a}(u,k_r)\) measures the contribution of \(r_{t-1}\), which is the variable defining the conditioning regimes. Similarly, when the response is plotted as a function of \(r_{t-1}\), the quantity \(s_v^{*,a}(u,k_v)\) measures the contribution of \(v_{t-1}\), which determines the conditioning regimes.

For the linear VAR benchmark, these quantities have a simple expected structure. 
Since the fitted response is affine in the lagged variables, the non-starred profiles are expected to be linear functions of the variable displayed on the horizontal axis. 
By contrast, the starred profiles measure the contribution of the conditioning variable and should therefore be approximately flat within each conditioning regime, up to residual effects induced by empirical correlations between \(r_{t-1}\) and \(v_{t-1}\) and by the binning procedure.

\subsubsection{Aggregated conditional surfaces plots results}\label{sec:aggregated} 

In this section, we first consider the aggregated surfaces defined on the left-hand side of Equations~\eqref{eq:binned_shap_decomposition}. These surfaces provide a useful diagnostic tool to assess whether the model learns relationships that are meaningful from a financial point of view. Moreover, as anticipated, they can be directly benchmarked against the data whose results are reported in Appendix~\ref{sec:data_nonlinear}.

\begin{figure}[!htb]
\centering
\begin{subfigure}{0.45\textwidth}
  \includegraphics[width=\linewidth]{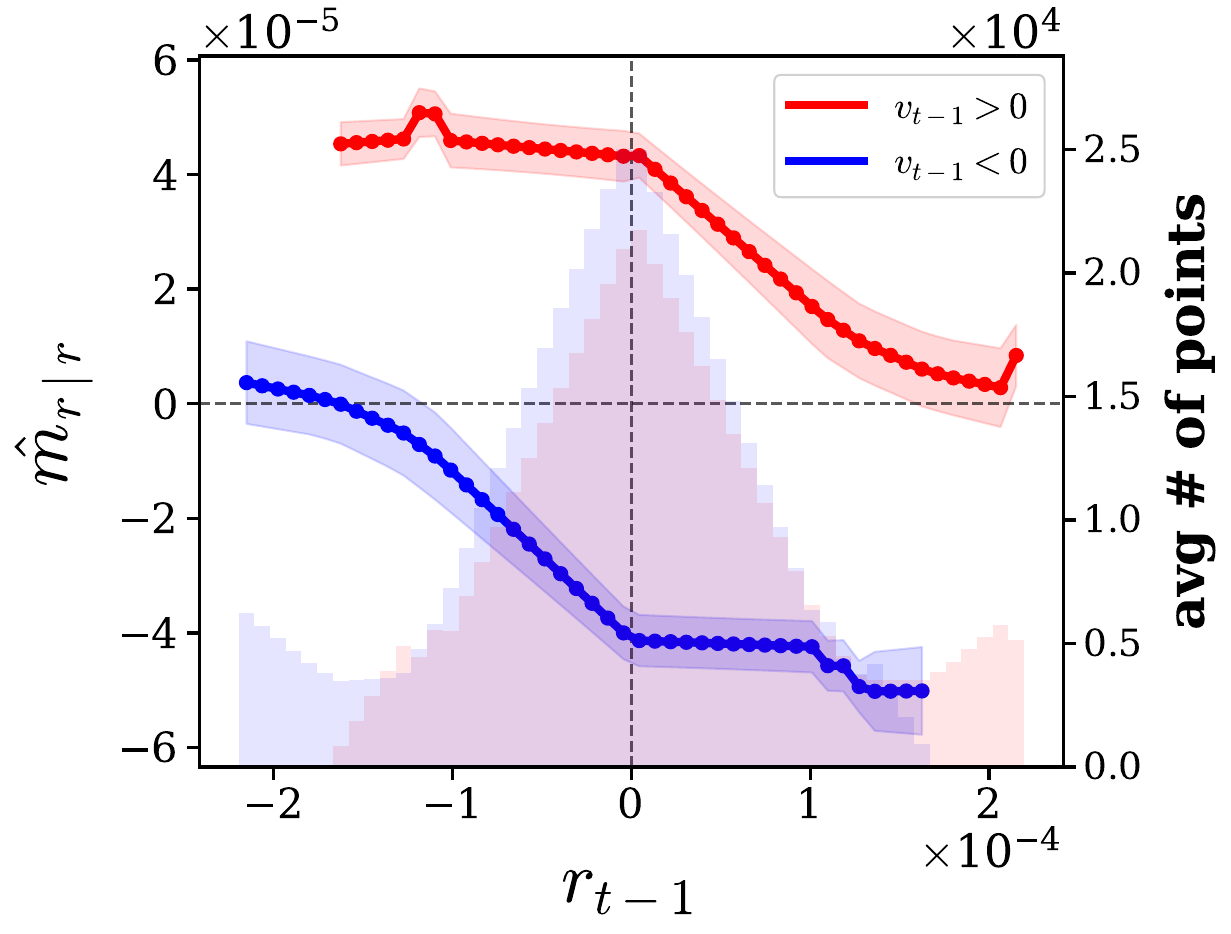}
\end{subfigure}
\begin{subfigure}{0.45\textwidth}
  \includegraphics[width=\linewidth]{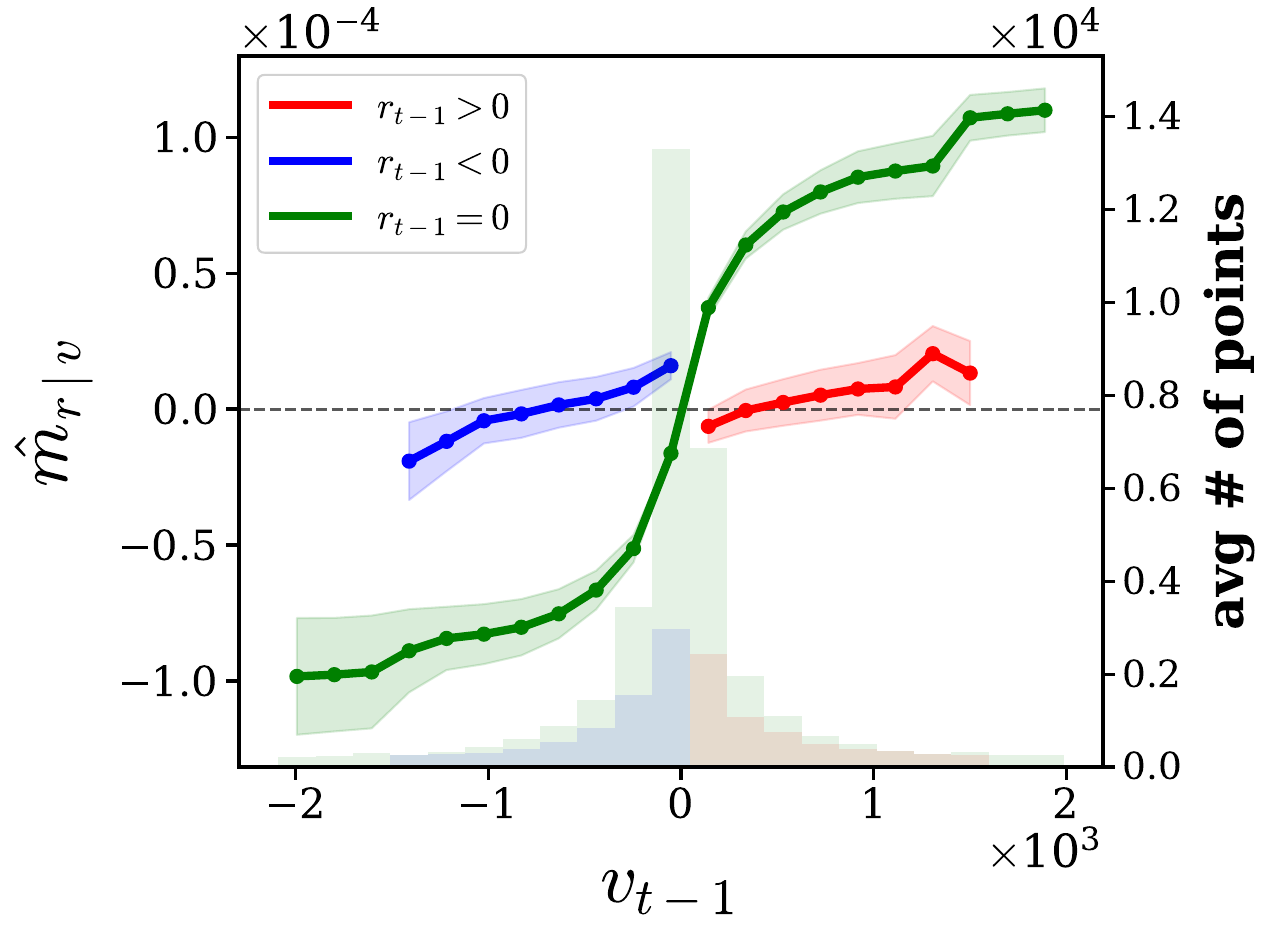}
\end{subfigure}

\vspace{0.3cm}

\begin{subfigure}{0.45\textwidth}
  \includegraphics[width=\linewidth]{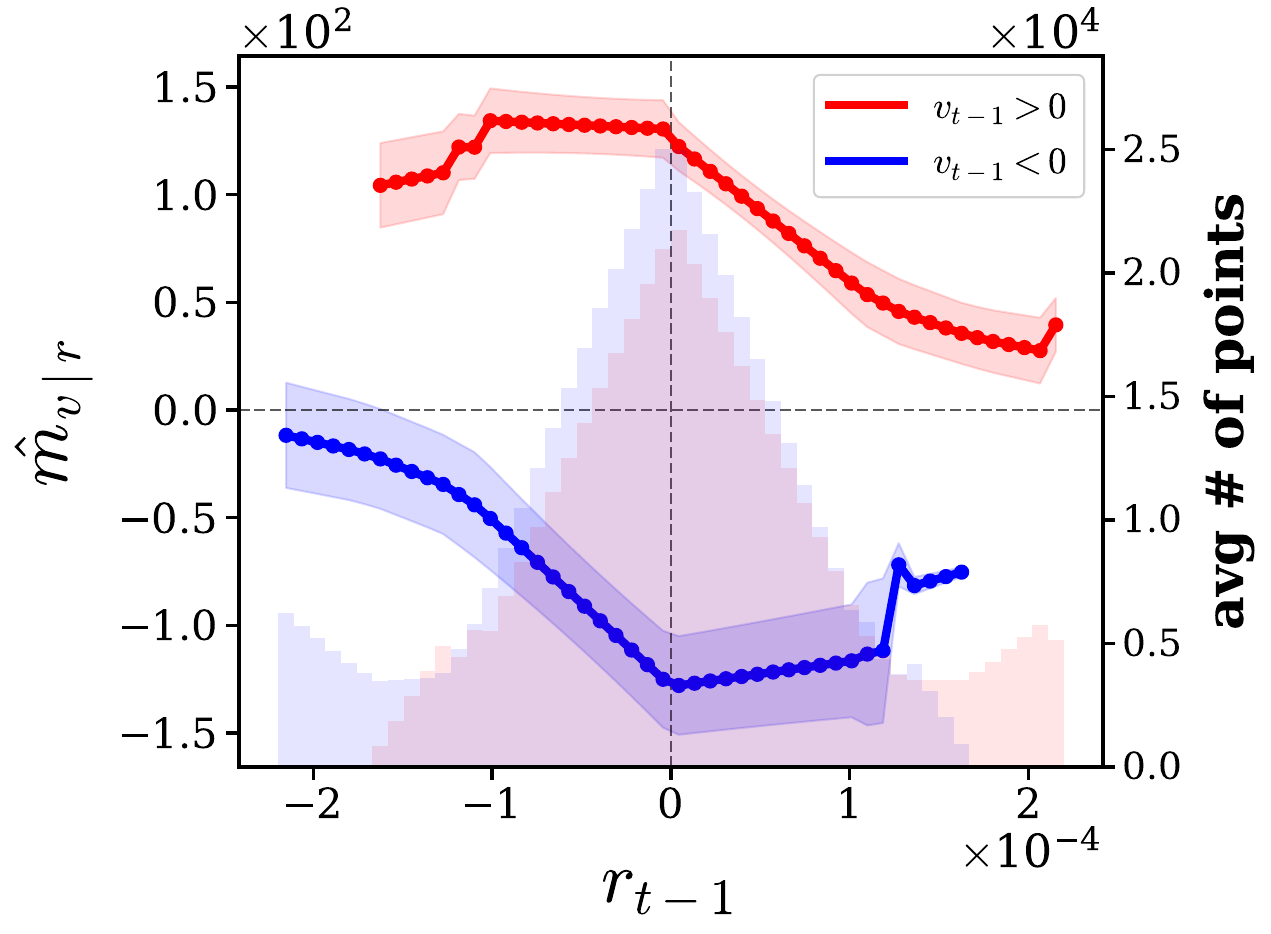}
\end{subfigure}
\begin{subfigure}{0.45\textwidth}
  \includegraphics[width=\linewidth]{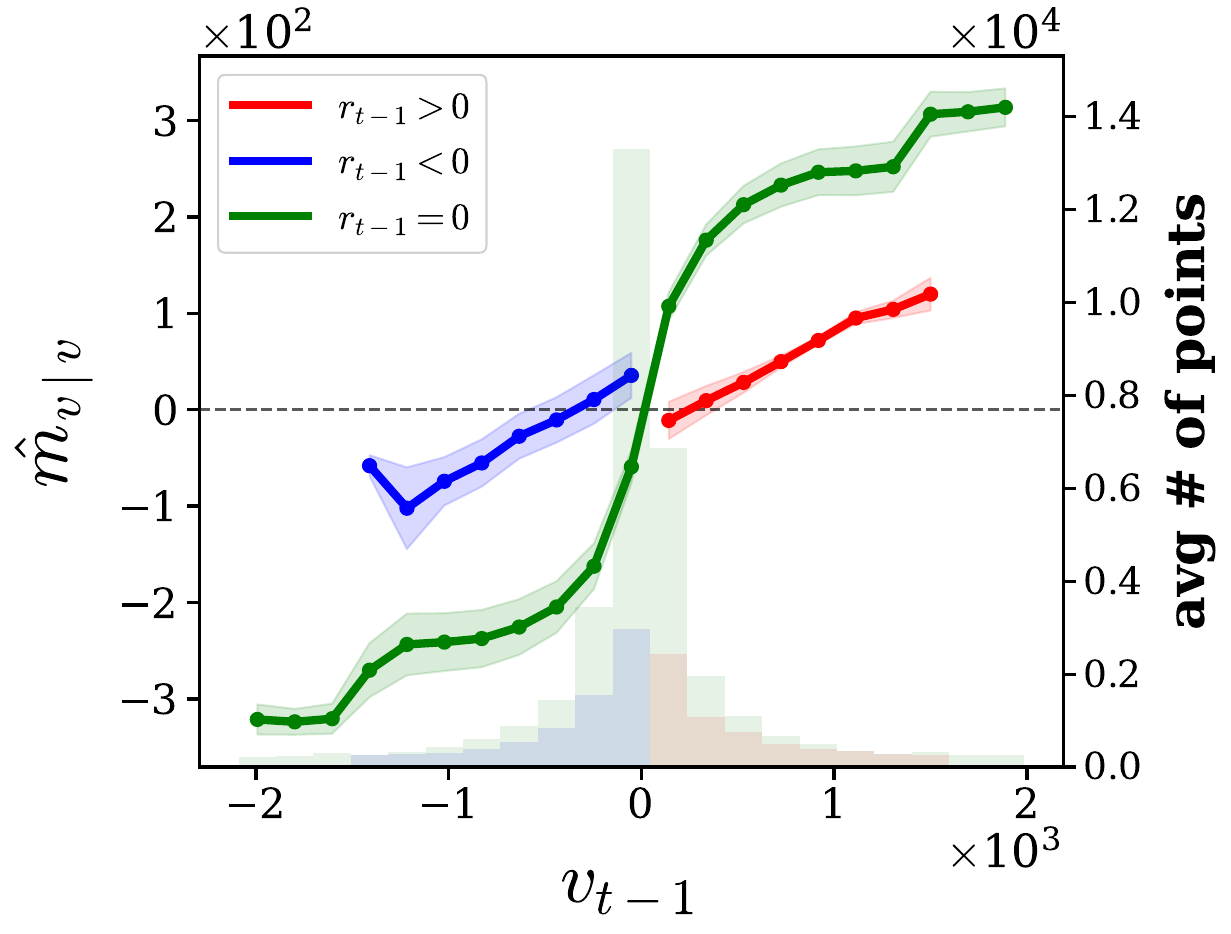}
\end{subfigure}

\caption{Results for Equations~\eqref{eq:model_surface_v_binned} and~\eqref{eq:model_surface_r_binned} for large tick (LT) stocks.}

\label{fig:PDP_LT}
\end{figure}

In figures~\ref{fig:PDP_ST} and~\ref{fig:PDP_LT} we report the resulting plots for the left hand side of Equations~\eqref{eq:binned_shap_decomposition}, namely for Eq.s~\eqref{eq:model_surface_v_binned}-\eqref{eq:model_surface_r_binned} for the small tick (ST) and large tick (LT) buckets, starting from ticker-specific grouped aggregated conditional response surfaces computed individually for each stock. 

These results are obtained starting from the four grouped aggregated conditional response surfaces computed for each stock: the effect of \(r_{t-1}\) on \(v_t\) conditional on the sign of \(v_{t-1}\), the effect of \(r_{t-1}\) on \(r_t\) conditional on the sign of \(v_{t-1}\), the effect of \(v_{t-1}\) on \(v_t\) conditional on the regime of \(r_{t-1}\), and the effect of \(v_{t-1}\) on \(r_t\) conditional on the regime of \(r_{t-1}\). Then an average across stocks is performed. The histograms reported in transparency represent the number of observations in each bin, thus providing a measure of empirical support. In Appendix~\ref{sec:data_nonlinear}, we report a more detailed discussion of the procedure used to obtain these results. 

\begin{figure}[!htb]
\centering
\begin{subfigure}{0.45\textwidth}
  \includegraphics[width=\linewidth]{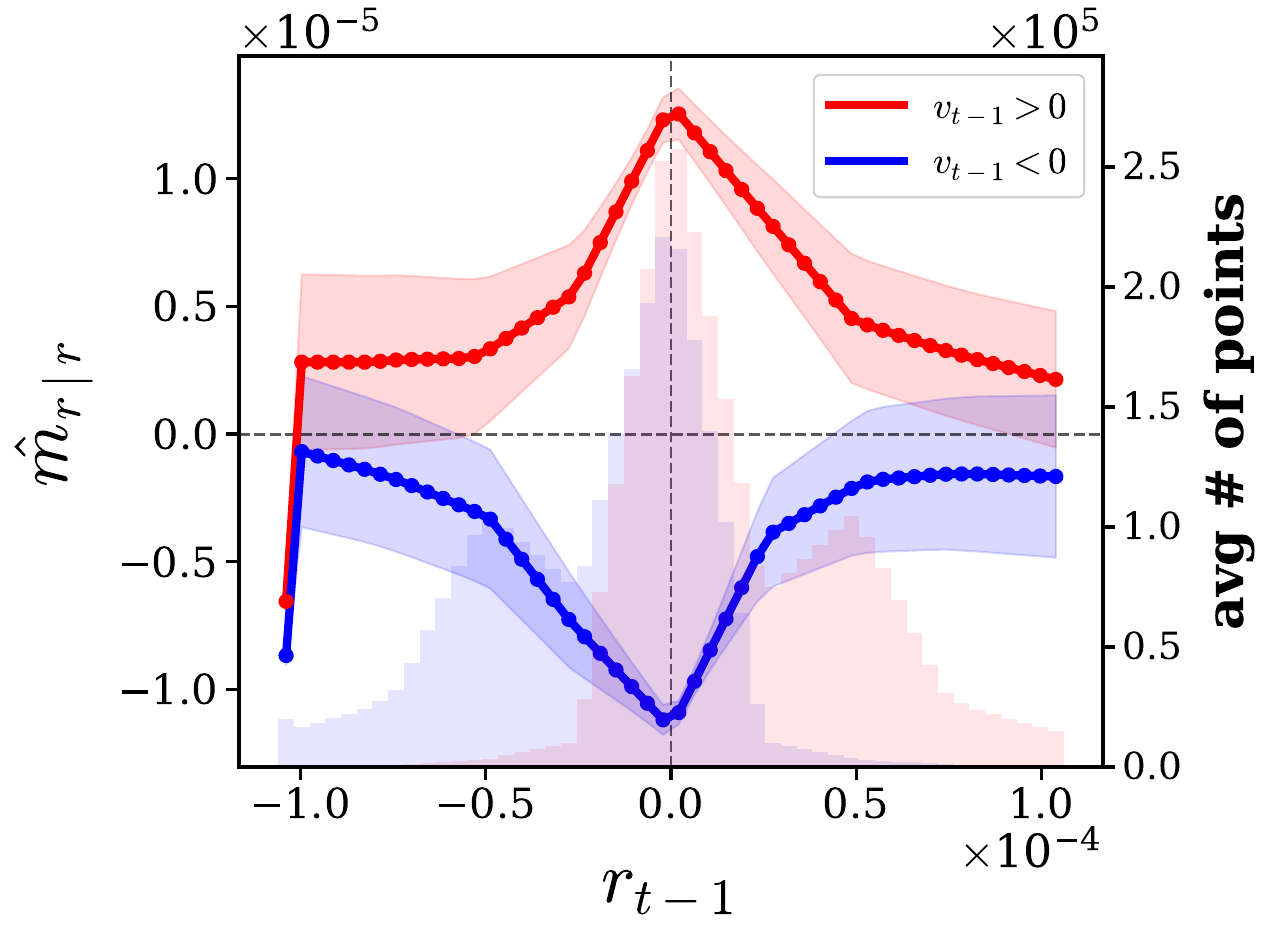}
\end{subfigure}
\begin{subfigure}{0.45\textwidth}
  \includegraphics[width=\linewidth]{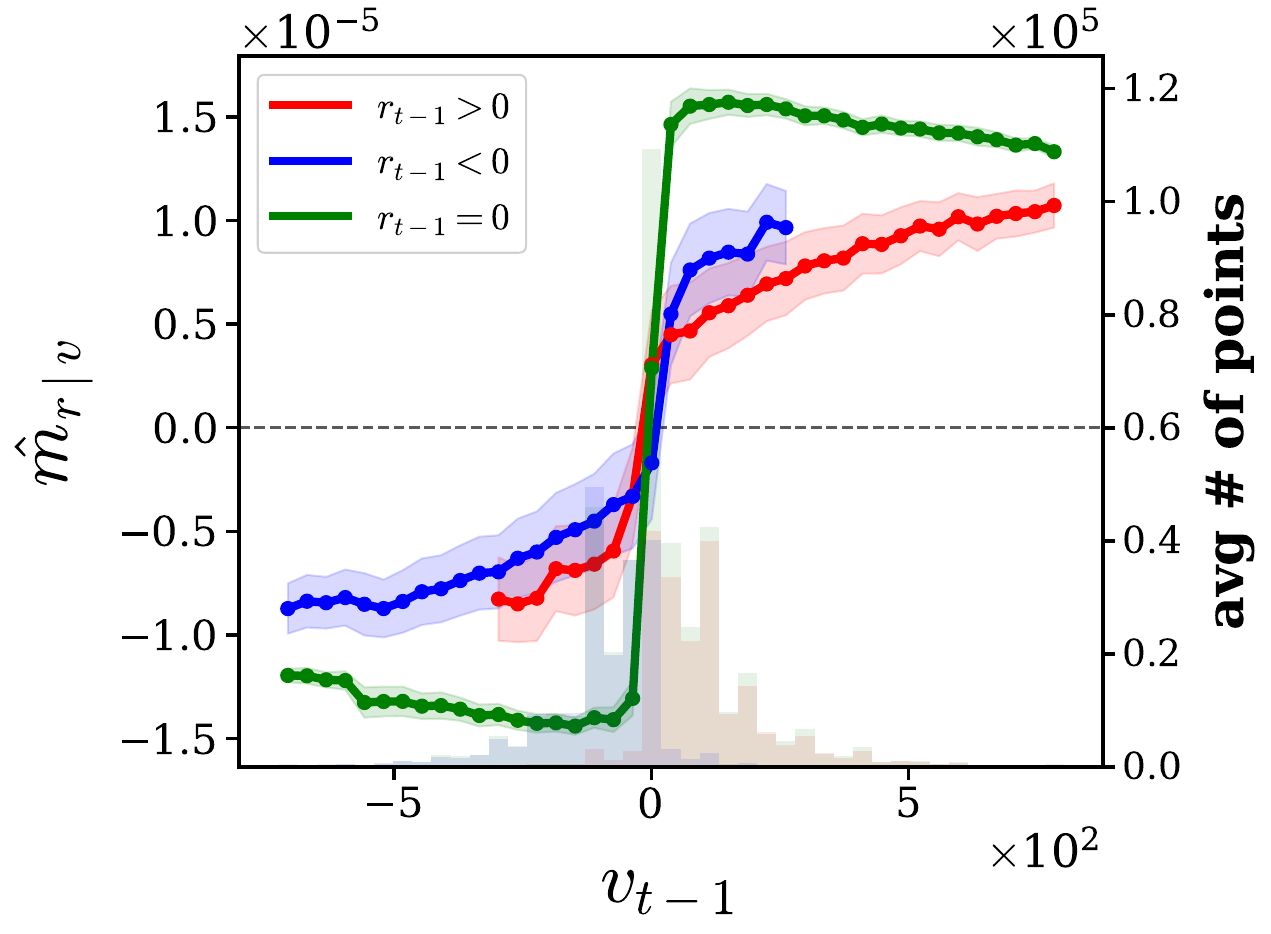}
\end{subfigure}

\vspace{0.3cm}

\begin{subfigure}{0.45\textwidth}
  \includegraphics[width=\linewidth]{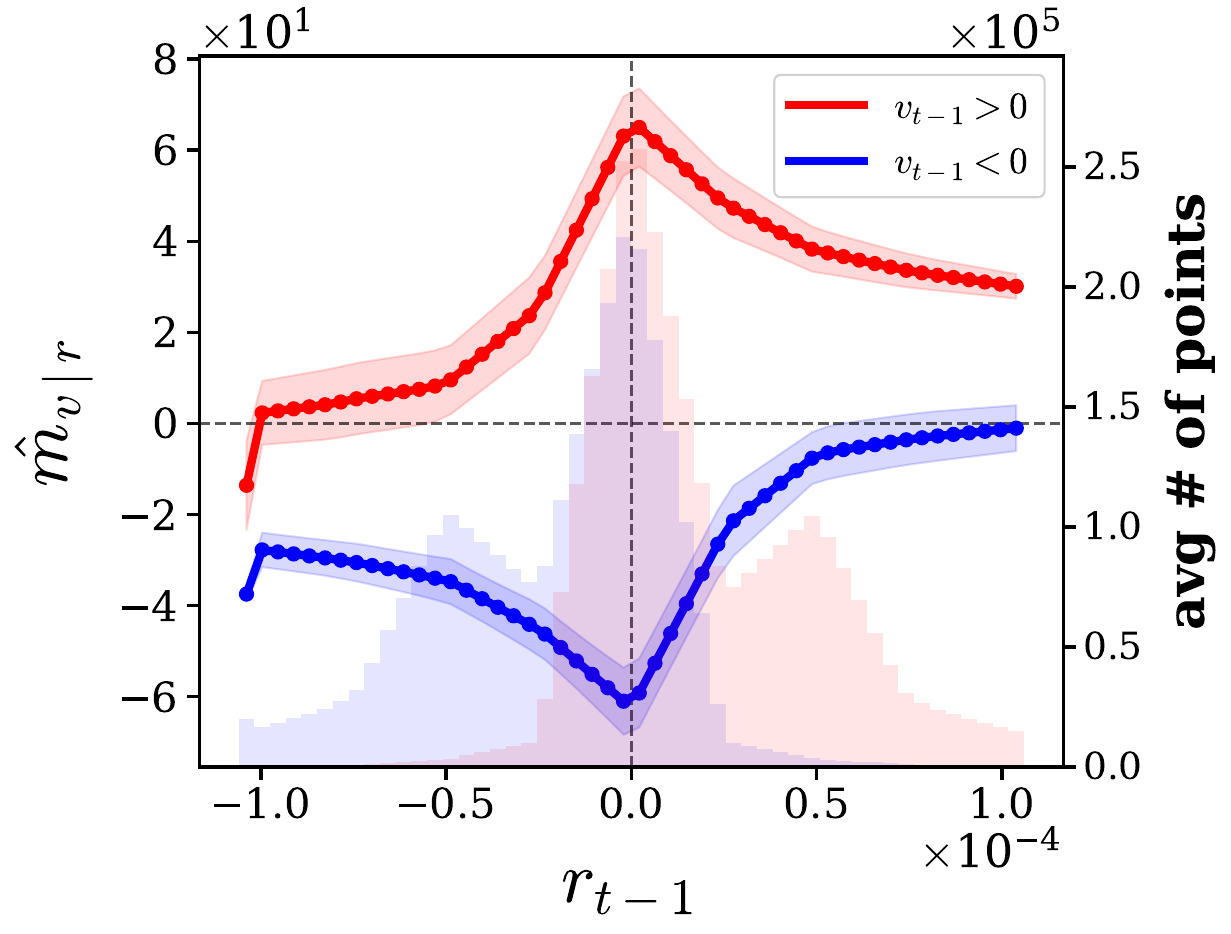}
\end{subfigure}
\begin{subfigure}{0.45\textwidth}
  \includegraphics[width=\linewidth]{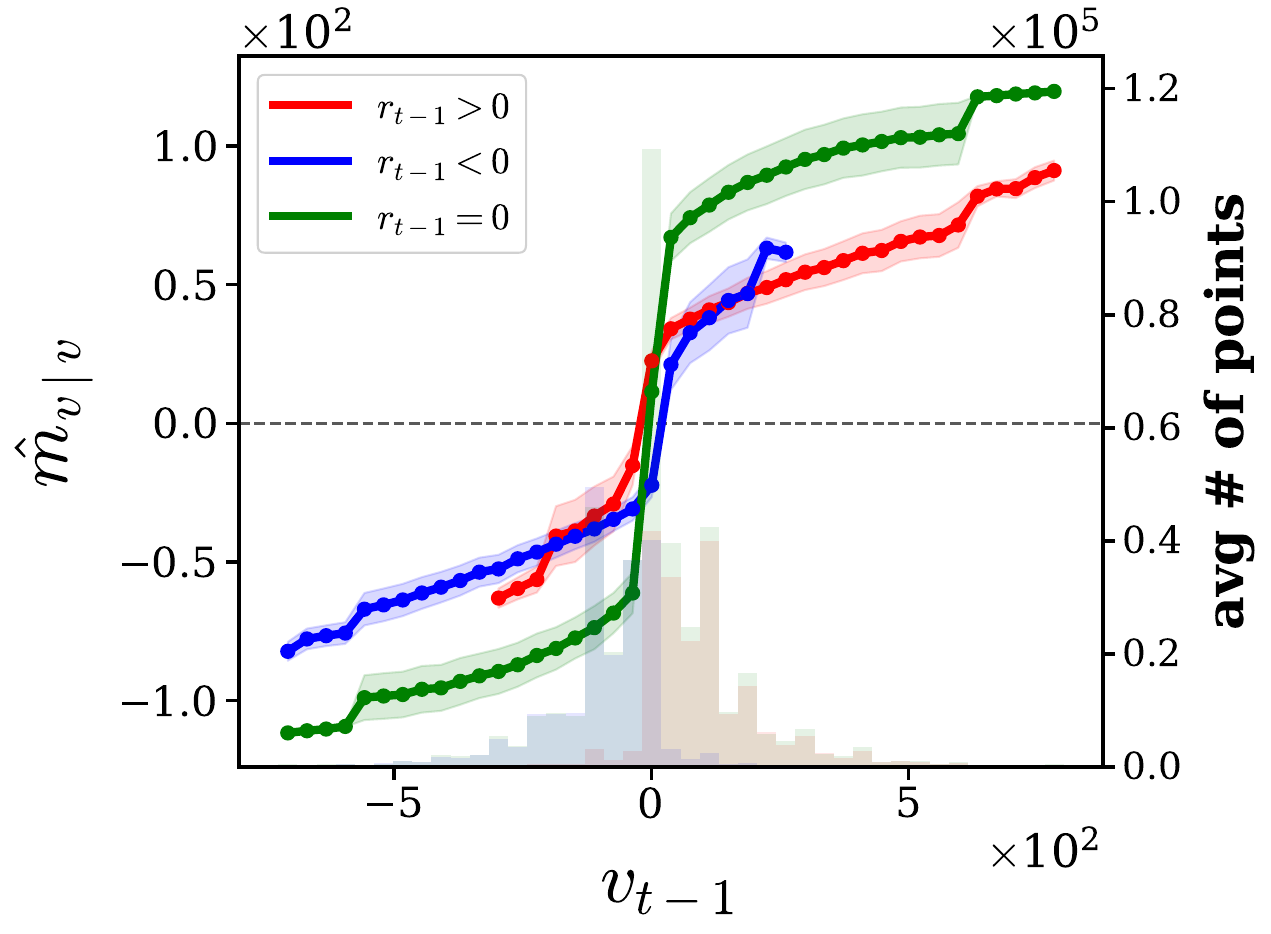}
\end{subfigure}

\caption{Results for Equations~\eqref{eq:model_surface_v_binned} and~\eqref{eq:model_surface_r_binned} for small tick (ST) stocks.}
\label{fig:PDP_ST}
\end{figure}

From an interpretative viewpoint, the figures reveal that the model has learned clear nonlinear and state-dependent relationships between returns and signed volumes. These effects are more pronounced for ST stocks. The most direct structure appears in the panels where the horizontal axis is \(v_{t-1}\). Both the predicted return, \(\hat m_{r|v}\), and the predicted signed volume, \(\hat m_{v|v}\), display a large discontinuity around the origin.

For returns, the sign-like response to \(v_{t-1}\) can be interpreted as a nonlinear price-impact relation: buy-initiated volume predicts positive price changes, while sell-initiated volume predicts negative price changes. The response is not purely linear, but tends to flatten for large absolute values of \(v_{t-1}\), consistently with empirical evidence showing that price impact is nonlinear and often concave in traded volume~\cite{lillo2003master,bouchaud2009price}. For signed volumes, the same structure reflects the persistence of order flow: positive lagged signed volume is followed by positive predicted signed volume, and negative lagged signed volume by negative predicted signed volume. In this sense, the model recovers the nonlinear analogue of the order-flow autocorrelation structure documented in the empirical literature~\cite{bouchaud2004fluctuations,toth2015equity}. It is particularly interesting to compare the curves obtained by conditioning on the regime of \(r_{t-1}\), namely \(r_{t-1}<0\), \(r_{t-1}=0\), and \(r_{t-1}>0\). The three regimes generate visibly distinct predictive responses, indicating that the effect of lagged signed volume depends on the recent price state. Especially in the ST case, a sharp transition around \(v_{t-1}=0\), although less pronounced than in the \(r_{t-1}=0\) regime, remains clearly visible for both \(r_{t-1}<0\) and \(r_{t-1}>0\). This suggests that, in ST stocks, the model captures a sharper change between sell- and buy-pressure regimes, whereas in LT stocks the response to lagged signed volume is smoother once prices have just moved.

The panels where the horizontal axis is \(r_{t-1}\) convey a different message. In both \(\hat m_{r|r}\) and \(\hat m_{v|r}\), the dominant feature is not a simple monotonic dependence on past returns, but rather a strong separation between the curves associated with \(v_{t-1}>0\) and \(v_{t-1}<0\). This indicates that, even when the model response is projected onto lagged returns, the sign of lagged order flow remains the main state variable organizing the prediction. More specifically, conditional on \(v_{t-1}>0\), the model tends to predict positive values for both \(r_t\) and \(v_t\) over most of the support of \(r_{t-1}\), whereas conditional on \(v_{t-1}<0\), the predicted values remain negative. Thus, both positive and negative lagged returns mainly affect the intensity of the response, while the direction of the prediction is primarily governed by the sign of order flow.

This suggests that, at this horizon, \(r_{t-1}\) does not act as an autonomous directional predictor. Rather, it works as a state variable modulating the strength of the continuation of order-flow pressure. A positive lagged signed volume reflects recent buy pressure and therefore tends to predict both positive future signed volume and positive short-term price changes; analogously, a negative lagged signed volume reflects sell pressure and leads to negative predicted responses. This behavior is consistent with evidence that short-horizon price changes are strongly related to order-flow imbalance and to the state of liquidity at the best quotes~\cite{cont2014price}. In large tick assets, where price changes are constrained by the tick size, part of the information contained in trading pressure may therefore be expressed through order-flow and queue dynamics rather than through smooth price adjustment. This interpretation is also consistent with the distinction between large- and small-tick regimes discussed in Eisler, Bouchaud and Kockelkoren~\cite{eisler2012price}.

Thus, rather than determining the sign of the prediction, \(r_{t-1}\) primarily controls its intensity. The directional effect is strongest when
\(r_{t-1}\) is close to zero. As \(|r_{t-1}|\) increases, the predicted response
moves back toward zero. This suggests that large previous price changes weaken
the continuation of the order-flow signal. Economically, this can be interpreted as an attenuation or reversal mechanism.
If the previous return is small, the order-flow pressure may not yet have been
fully incorporated into prices, so the model reinforces the prevailing
buy/sell-pressure regime. If the previous return is large in absolute value,
part of the order-flow pressure may already have been absorbed by the price
move, or liquidity provision may induce a short-term reversal. In that case, the
model reduces the expected continuation of the same directional signal.

Note that small tick stocks exhibit much stronger nonlinearities and sharper transitions, while large tick stocks display smoother and more constrained dynamics, consistent with the role of tick size in limiting price adjustments. 

In Appendix~\ref{sec:data_nonlinear}, we report the results obtained by applying
the same aggregation procedure directly to the data. The only difference is that,
instead of averaging the model prediction, we average the realized target
variable. Overall, the comparison shows that the model reproduces fairly well,
at least qualitatively, the main dependencies observed in the empirical
conditional surfaces. The agreement is not perfect, as expected. In particular, especially for
large-tick stocks, the model-implied profiles \(\hat m_{a|r}\) display a more
pronounced asymmetry with respect to the vertical axis than the corresponding
empirical surfaces. This suggests that the neural network may overemphasize
some state-dependent asymmetries or decay patterns in this direction. These differences are natural, since the model provides a fitted and smoothed
representation of empirical conditional averages, which are themselves affected
by finite-sample fluctuations and uneven empirical support across bins.
Nevertheless, the comparison remains an important consistency check: the main
nonlinear structures learned by the model are also visible when the same
procedure is applied directly to the data.

Finally, in Appendix~\ref{app:var_predictive_surfaces} we also report the corresponding aggregated predictive surfaces obtained from the VAR benchmark. 
As expected, the comparison reveals clear differences with respect to both the empirical surfaces and the neural-network ones. 
In the VAR case, the aggregated dependencies are essentially linear, with only mild deviations induced by the empirical correlations between \(v_{t-1}\) and \(r_{t-1}\). 
This shows that such correlations, once processed through a linear model, are not sufficient to reproduce the nonlinear aggregate dependencies observed in the data. 
These dependencies must therefore be explicitly captured by the predictive model, as happens in the nonlinear neural-network specification.

\begin{figure}[!htb]
\centering
\begin{subfigure}{0.45\textwidth}
  \includegraphics[width=\linewidth]{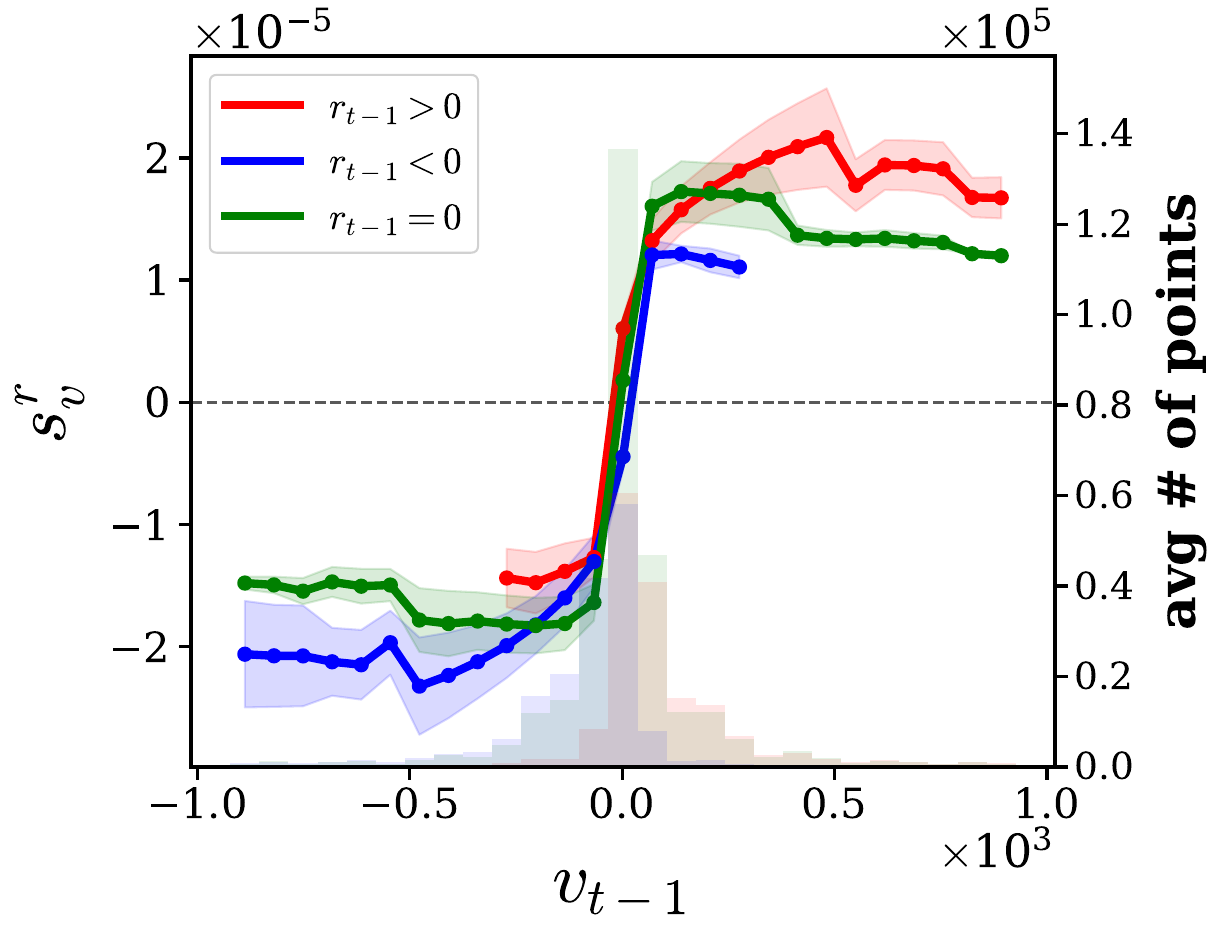}
\end{subfigure}
\begin{subfigure}{0.45\textwidth}
  \includegraphics[width=\linewidth]{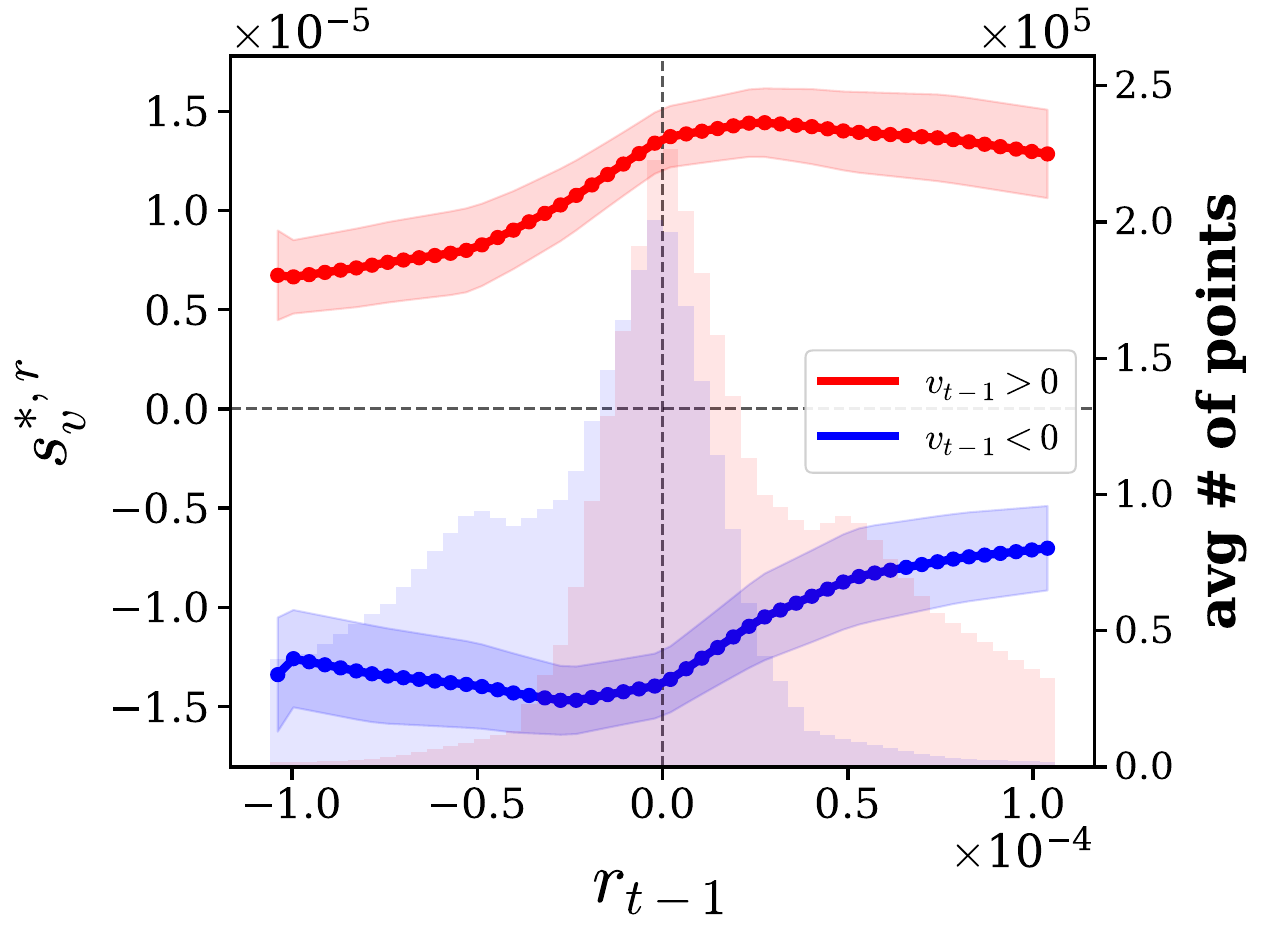}
\end{subfigure}

\vspace{0.3cm}
\begin{subfigure}{0.45\textwidth}
  \includegraphics[width=\linewidth]{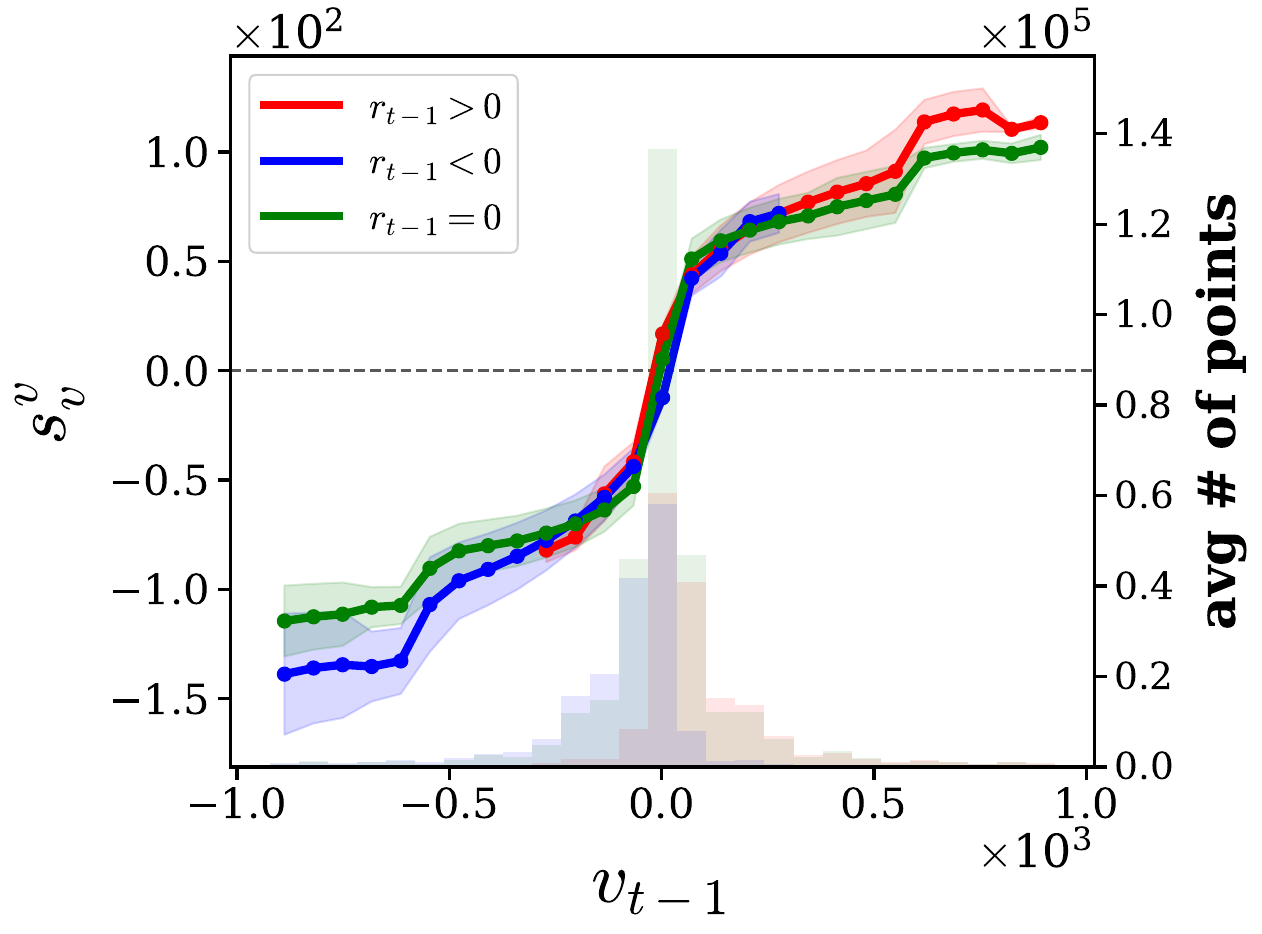}
\end{subfigure}
\begin{subfigure}{0.45\textwidth}
  \includegraphics[width=\linewidth]{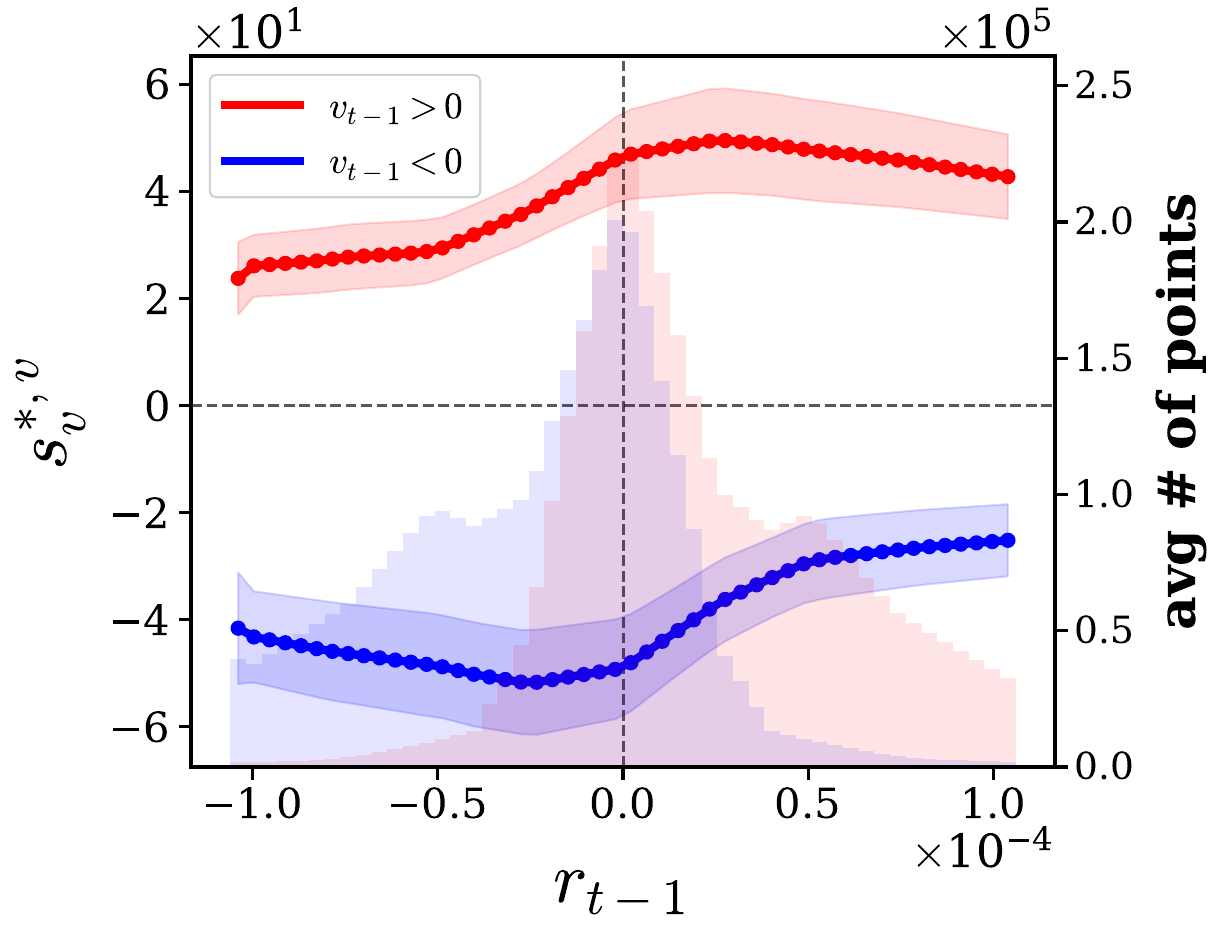}
\end{subfigure}

\caption{Average binned Shapley contributions for small-tick (ST) stocks for the volume's term.}
\label{fig:SHAP_ST_v}
\end{figure}

\begin{figure}[!htb]
\centering
\begin{subfigure}{0.45\textwidth}
  \includegraphics[width=\linewidth]{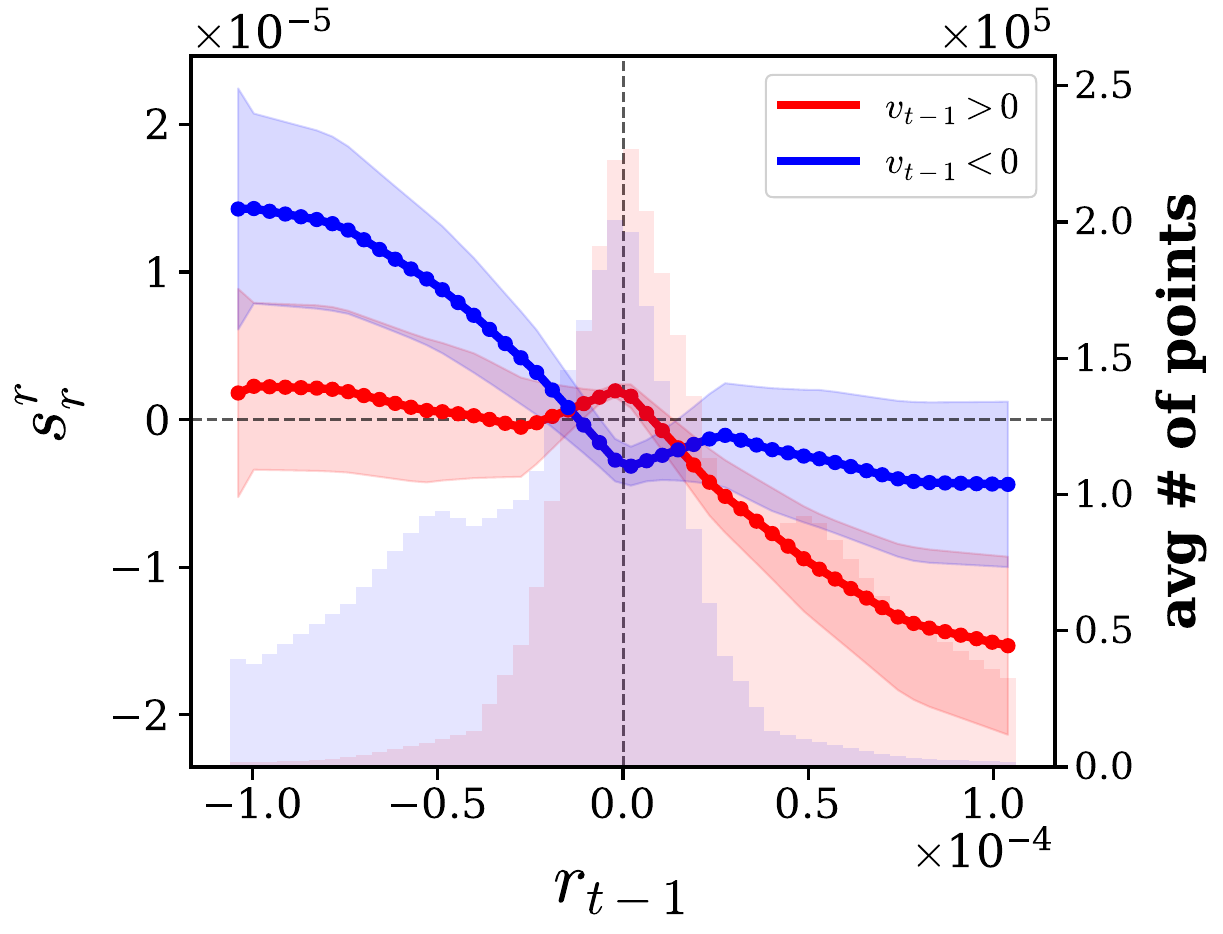}
\end{subfigure}
\begin{subfigure}{0.45\textwidth}
  \includegraphics[width=\linewidth]{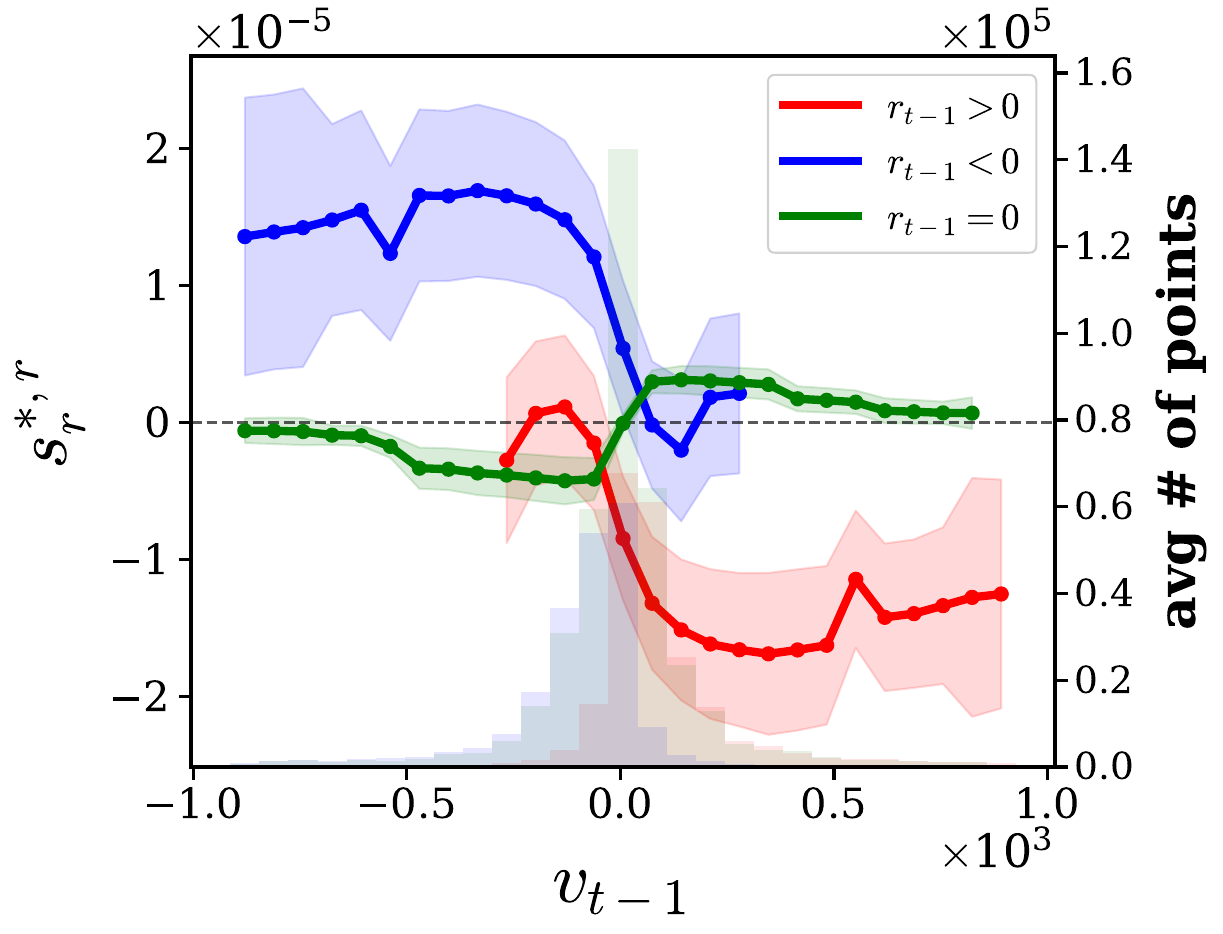}
\end{subfigure}

\vspace{0.3cm}

\begin{subfigure}{0.45\textwidth}
  \includegraphics[width=\linewidth]{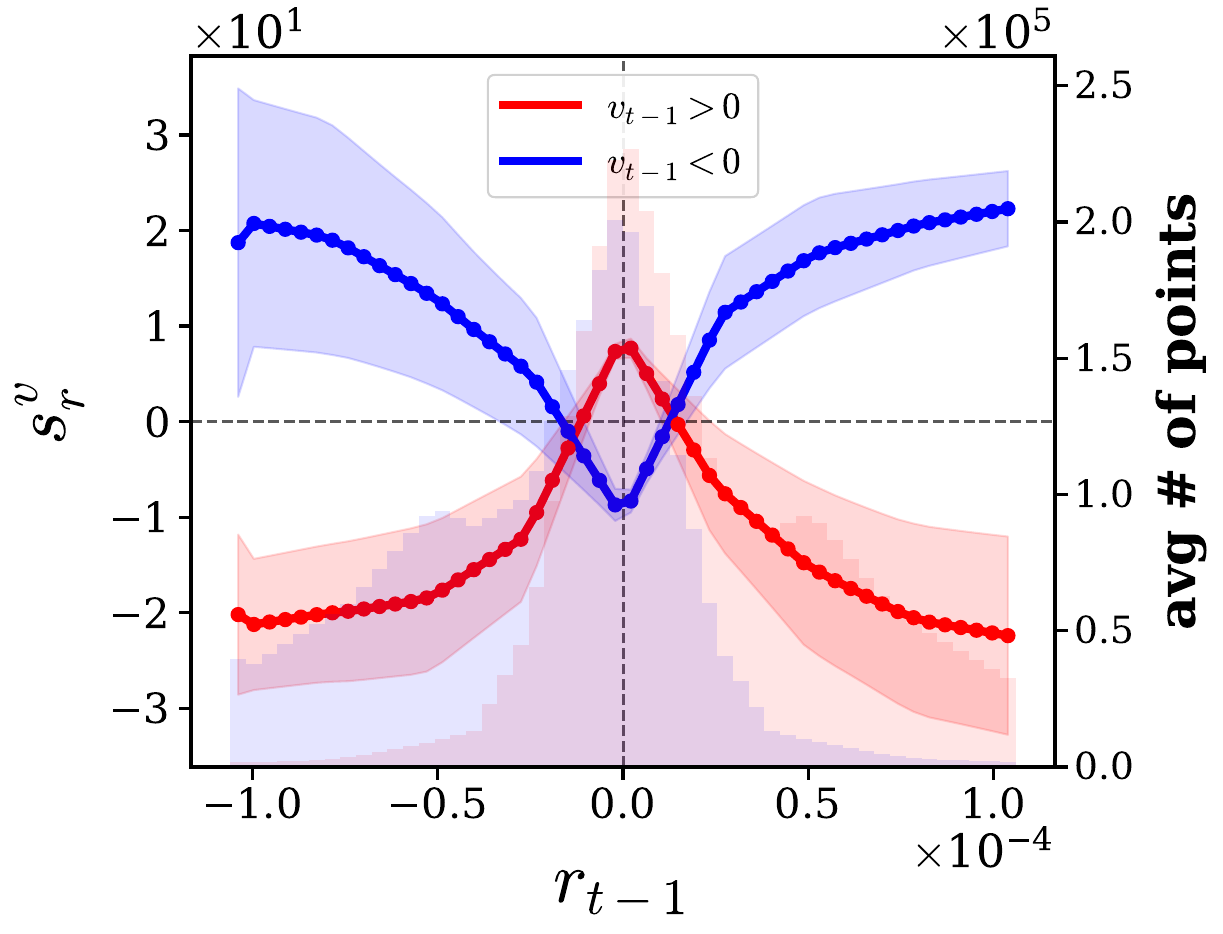}
\end{subfigure}
\begin{subfigure}{0.45\textwidth}
  \includegraphics[width=\linewidth]{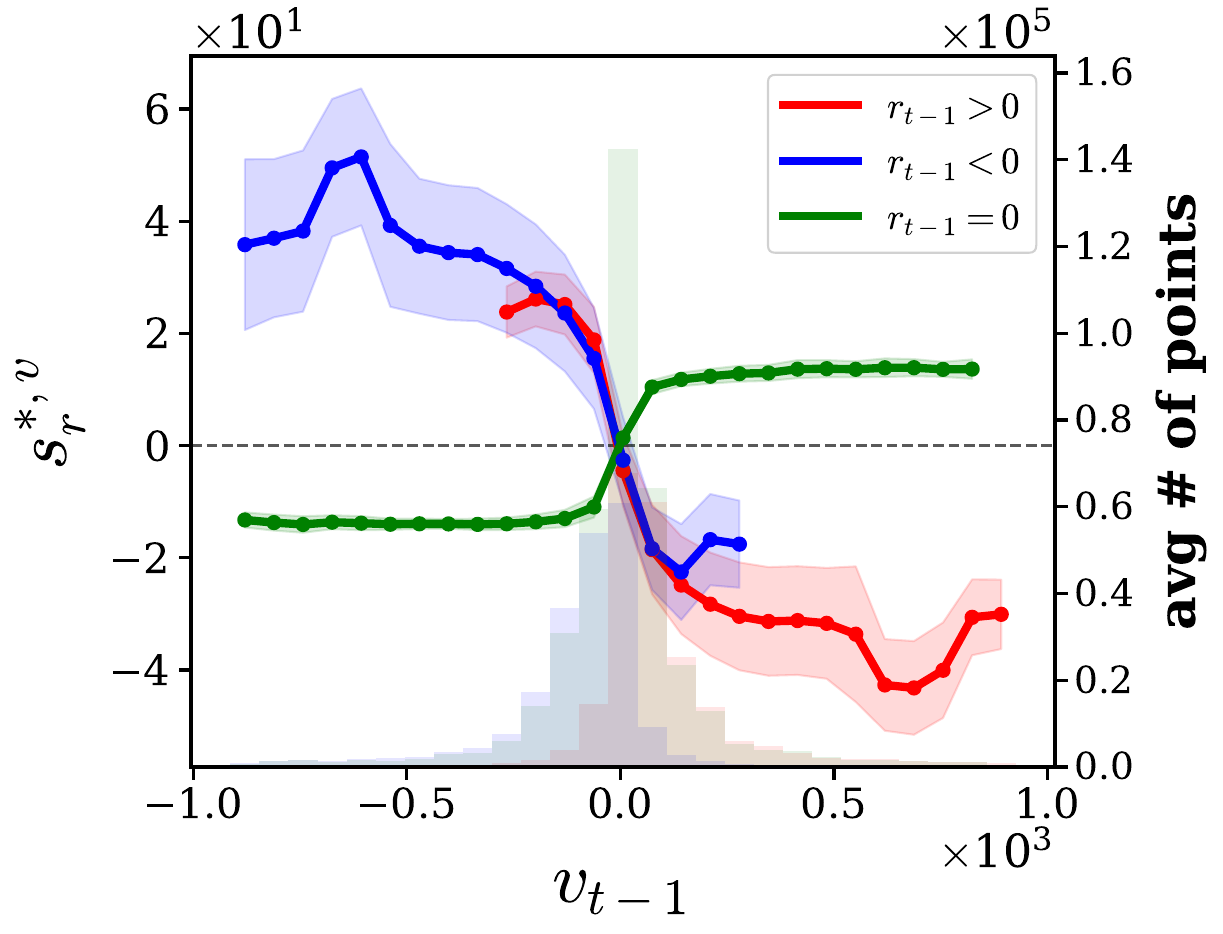}
\end{subfigure}

\caption{Average binned Shapley contributions for ST stocks for the return's term.}
\label{fig:SHAP_ST_r}
\end{figure}

\subsubsection{Shap values results}

We now turn to the Shapley-based decomposition introduced in
Equation~\eqref{eq:binned_shap_decomposition}. The goal of this section is
to move from the analysis of the aggregate predictive response of the model
to the interpretation of the isolated contributions of the individual
regressors. In the previous sections, we showed that the fitted model is able
to reproduce meaningful dependence patterns between lagged returns and lagged
signed volumes. This was assessed both by looking at the raw Shapley values and also by looking at the model-based response surfaces
compared with the corresponding empirical surfaces
extracted from the data. These consistency checks suggest that the nonlinear
structures learned by the model are not merely artifacts of the estimation
procedure, but reflect relevant patterns already present in the data. We can
therefore use the Shapley decomposition with greater confidence to study how
each input variable contributes to the model prediction.

In particular, Equation~\eqref{eq:binned_shap_decomposition} separates two types of quantities. 
The non-starred components, \(s_v^a\) and \(s_r^a\), measure the average Shapley contribution of the variable along which the profile is constructed: \(s_v^a\) for profiles in \(v_{t-1}\), and \(s_r^a\) for profiles in \(r_{t-1}\). 
They therefore describe how the contribution assigned to a variable changes as that same variable varies. The starred components, \(s_v^{*,a}\) and \(s_r^{*,a}\), instead project the contribution of one variable along the direction of the other. 
Thus, \(s_v^{*,a}\) measures how the contribution of \(v_{t-1}\) varies across \(r_{t-1}\), while \(s_r^{*,a}\) measures how the contribution of \(r_{t-1}\) varies across \(v_{t-1}\). 
These profiles are therefore useful to detect residual cross-dependencies or interaction effects between the two regressors.

This distinction is useful also as a benchmark against the linear VAR case.
In a purely linear model without interaction terms, the contribution of each
regressor depends only on that regressor itself. Consequently, the
non-starred profiles reproduce the corresponding linear dependence, up to an
additive constant, while the starred profiles are flat. A non-flat starred
component therefore signals that, in the nonlinear model, the contribution
assigned to one variable is modulated by the value of the other variable.
This provides a way to identify nonlinear cross-dependencies learned by the
model and to distinguish them from purely additive linear effects.

In the following, we first analyse the volume component by considering the
profiles \(s_v^a\) and \(s_v^{*,a}\). We then apply the same analysis to the
return component, studying \(s_r^a\) and \(s_r^{*,a}\). For reasons of space,
in this section we report only the results for small-tick stocks. The
corresponding results for large-tick stocks, which are qualitatively similar,
are provided in Appendix~\ref{app:LT_Shaps}.

Let us first consider the results in Figure~\ref{fig:SHAP_ST_v} obtained for the volume term. 

We observe that, while in the aggregated case in the right panels of Figures~\ref{fig:PDP_ST} and~\ref{fig:PDP_LT} the curves conditional on \(r_{t-1}>0\), \(r_{t-1}<0\), and \(r_{t-1}=0\) are visibly separated, in this case the three curves in the left panels become much closer to each other. This is confirmed also by the right panels showing the spurious dependence of the volume term as a function of $r_{t-1}$. This dependence is, to a first approximation, quite flat, meaning that there is
no relevant dependence on $r_{t-1}$ in the contribution associated with
$v_{t-1}$. Thus, the separation previously observed in the aggregated profiles as a function of \(v_{t-1}\), which is not fully accounted for by the direct \(v_{t-1}\) contribution, is likely captured by a residual contribution in the \(r_{t-1}\) term.

Let us now consider the profiles constructed along the \(r_{t-1}\) direction. 
In the non-starred component \(s_r^v\), we observe a change in the role played by past returns depending on their magnitude. 
For small price changes, the contribution of \(r_{t-1}\) tends to preserve the sign of the subsequent signed volume. 
By contrast, for larger absolute values of \(r_{t-1}\), the return contribution changes sign, generating a mean-reverting effect on future order flow. 
This suggests that, for small-tick stocks, zero or very small price changes are associated with persistence in trading pressure, whereas larger price changes induce a reversal contribution to \(v_t\). Furthermore, although the separation between the two curves is reduced with respect to the aggregated profiles, it is still present in the Shapley profiles when conditioning on the sign of \(v_{t-1}\). 
This suggests that the return contribution also contains a residual component associated with lagged signed volume. 

The starred components \(s_r^{*,a}\) provide complementary information on the separation observed in the full predictive surfaces. 
In fact, these two panels reveal a clear and interpretable residual dependence of the return contribution \(\phi_r^v\) on \(v_{t-1}\). When the previous signed volume does not move the price, namely when \(r_{t-1}\) is close to zero, the model attributes to the return term a contribution that reinforces the persistence of order flow and, as a consequence, maintains the sign of the previous trade. When, instead, the previous order flow is accompanied by a non-zero price change, the return contribution becomes mean-reverting and tends to reverse the sign of future signed volume and return.

In Appendix~\ref{app:LT_Shaps}, we report the corresponding results for LT stocks. 
Although some differences are present, the qualitative picture remains similar. 
This is particularly true for the starred components, which again show persistence of the order-flow sign when there is no price movement, and a sign reversal when the lagged return is non-zero. 
The same mechanism is also visible in the non-starred profiles: when \(r_{t-1}=0\), the return contribution remains consistent with the sign of lagged signed volume, whereas for non-zero returns it tends to reverse it.

At the same time, clear differences between ST and LT stocks emerge, especially in the non-starred components. 
For LT stocks, the profiles display more pronounced asymmetries and a more evident mean-reverting behavior. 
Moreover, their shape is closer to a linear, almost straight-line dependence than in the ST case.

Overall, this Shapley-based decomposition provides a more detailed interpretation of the aggregate dependence plots. 
While the latter can also be constructed directly from the data, they do not separate the different contributions entering the fitted response. 
The Shapley decomposition, instead, allows us to decompose the aggregate predictive patterns into the individual contributions associated with the different regressors.

Before constructing the reduced parametric model, we first extract the residual
contemporaneous component associated with the unexpected part of signed volume.
Since the models considered so far use only lagged information, the
instantaneous contribution of \(v_t\) remains encoded in the residual dependence
between return and signed-volume innovations. As shown below, this residual
decomposition reveals interpretable nonlinear patterns, similar to those
identified in the lagged predictive structure.

\subsection{The structural shocks}
\label{sec:structural_shocks}

We now use the fitted nonlinear model to extract structural shocks defined in Equation~\eqref{eq:nonlinear_residual_structural} from the reduced-form residuals. In this way, we aim to extract the contemporaneous effect of order flow on returns, not included in the previous model, which remains encoded in the residual dependence between return and signed-volume innovations.

Following the triangular structural interpretation in Equation~\eqref{eq:nonlinear_residual_structural}, the resulting structural shocks are estimated as
\begin{equation}
\hat\epsilon_t^v=u_t^v,
\qquad
\hat\epsilon_t^r
=
u_t^r-\hat\phi(u_t^v).
\label{eq:estimated_structural_shocks_section}
\end{equation}

We estimate \(\phi\) using a nonparametric binned estimator,
\begin{equation}
\phi_{\mathrm{bin}}(u_j)
=
\mathbb{E}
\left[
u_t^r
\mid
u_t^v\in B_j
\right],
\label{eq:phi_binned_structural}
\end{equation}
which provides an agnostic estimate of the conditional mean
\[
\phi(u)=\mathbb{E}[u_t^r\mid u_t^v=u].
\]
This allows us to capture possible nonlinearities in the residual contemporaneous impact. 
Moreover this choice implies
\[
\mathbb{E}[\epsilon_t^r\mid u_t^v]=0,
\]
and therefore, in population,
\[
\operatorname{Cov}(\epsilon_t^r,u_t^v)=0.
\]

To quantify the relevance of the residual impact map, we compute the fraction of variation in the return residual accounted for by the contemporaneous order-flow projection,
\begin{equation}
R_\phi^2
=
1-
\frac{
\sum_t\left(u_t^r-\hat\phi(u_t^v)\right)^2
}{
\sum_t\left(u_t^r-\overline{u^r}\right)^2
}.
\label{eq:r2_phi}
\end{equation}
Here, \(u_t^r\) is the component of the return that remains unexplained by lagged returns and lagged signed volumes, while \(\hat\phi(u_t^v)\) is the component of this residual return accounted for by the contemporaneous innovation in signed volume. Thus, \(R_\phi^2\) measures how much of the lag-unexplained return variation is captured by the residual contemporaneous order-flow relation.

If \(R_{\mathrm{lag}}^2\) denotes the fraction of total return variation explained by the lagged predictor alone, then for zero-mean return residuals,
\[
\overline{u^r}=0,
\]
the corresponding gain in total explanatory power is
\begin{equation}
\Delta R_{\mathrm{total}}^2
=
\left(1-R_{\mathrm{lag}}^2\right)R_\phi^2.
\label{eq:r2_total_gain}
\end{equation}
This zero-mean condition holds in population when the reduced-form predictor coincides with the conditional mean, and is expected to hold approximately in finite samples. Hence, \(R_\phi^2\) should be interpreted as the fraction of the remaining return variation explained after the lagged predictive component has been removed.

%We also report correlations between absolute and squared structural shocks to check whether residual dependence remains in the magnitudes of the innovations.

The analysis is performed separately for LT and ST stocks. This is consistent with the evidence presented above, which shows that the two buckets differ in their predictive surfaces, Shapley profiles, and relative predictability of returns and signed volumes. Since \(\phi\) captures the instantaneous impact of unexpected order flow on unexpected returns, its shape and explanatory power may also depend on the tick-size regime.

\paragraph{Empirical results.}

Table~\ref{tab:structural_shocks_ticker} reports the results of the binned structural decomposition for each ticker, together with equal-weight averages within the LT and ST buckets. 

\begin{table}[htbp]
\centering
\caption{Structural shock decomposition by ticker.}
\label{tab:structural_shocks_ticker}
\begin{tabular}{llc}
\toprule
Bucket 
& Ticker
& \(R_\phi^2\) \\
\midrule
LT & BAC   & 0.163 \\
LT & CMCSA & 0.117 \\
LT & CSCO  & 0.148 \\
LT & INTC  & 0.178 \\
LT & PFE   & 0.205 \\
\midrule
LT & Average & 0.162 \\
\midrule
ST & AAPL  & 0.143 \\
ST & AMZN  & 0.266 \\
ST & GILD  & 0.222 \\
ST & NVDA  & 0.146 \\
ST & TSLA  & 0.116 \\
\midrule
ST & Average & 0.179 \\
\bottomrule
\end{tabular}
\end{table}
We observe that the explanatory power of the residual impact function is economically meaningful. The average value of \(R_\phi^2\) is \(0.162\) for LT stocks and \(0.179\) for ST stocks. Thus, around \(16\%\)--\(18\%\) of the variance of the reduced-form return innovation can be explained by the contemporaneous innovation in signed volume. This fraction is sizeable, especially because the lagged nonlinear model has already removed the predictable component based on past returns and volumes. The result therefore confirms that a relevant part of price formation occurs through an instantaneous order-flow channel. At the ticker level, \(R_\phi^2\) ranges from \(0.117\) to \(0.205\) in the LT bucket and from \(0.116\) to \(0.266\) in the ST bucket, showing that the residual contemporaneous impact is present across all assets, although with heterogeneous intensity.

\begin{figure}[!htb]
\centering
\includegraphics[width=0.48\textwidth]{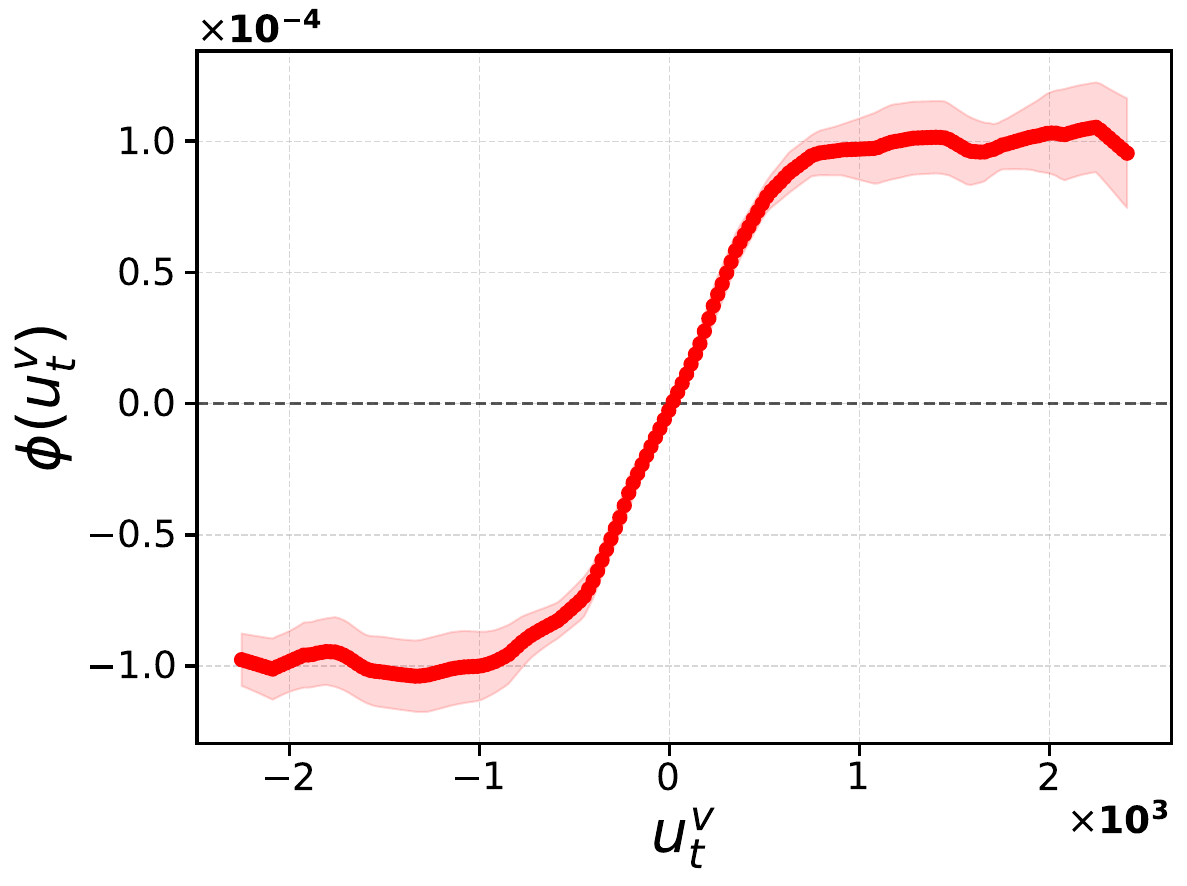}
\includegraphics[width=0.48\textwidth]{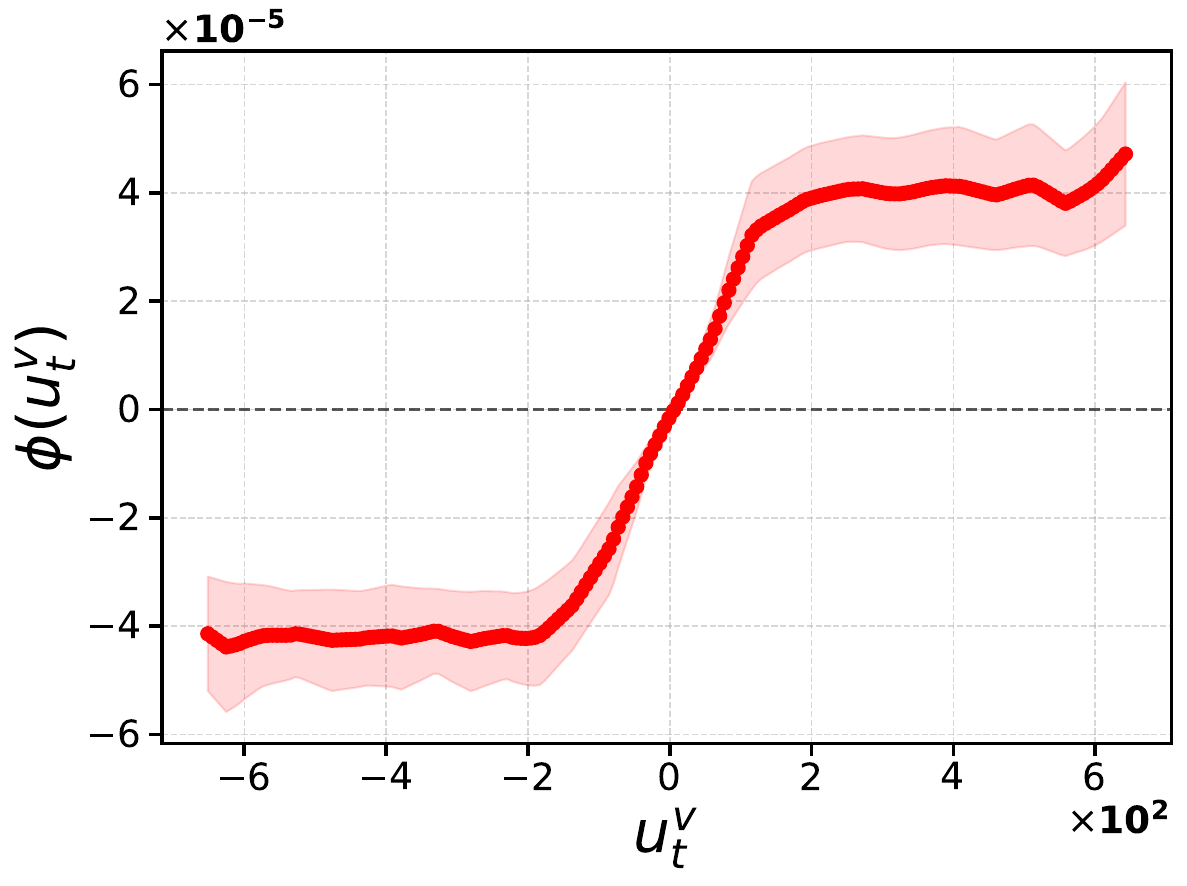}
\caption{Average binned residual impact function \(\hat\phi(u_t^v)\) for LT stocks (left) and ST stocks (right). The function maps order-flow innovations into return innovations after removing the lagged predictable component.}
\label{fig:phi_average_LT_ST}
\end{figure}

The shape of the estimated residual impact functions is shown in Figure~\ref{fig:phi_average_LT_ST}. In both LT and ST stocks, the function \(\hat\phi\) is increasing and sign-preserving: negative order-flow innovations generate negative return innovations, while positive order-flow innovations generate positive return innovations. The response is not purely linear. It is steeper around the origin and tends to flatten in the tails, suggesting a nonlinear and saturating instantaneous impact. This behavior is consistent with the interpretation of \(\phi\) as a residual price-impact function: after removing the lagged predictable component, unexpected buy pressure pushes the return innovation upward, while unexpected sell pressure pushes it downward.

The two buckets display similar qualitative shapes, but with different scales. The LT residual impact curve is defined over a wider range of order-flow innovations and reaches a larger return-residual magnitude. The ST curve is more concentrated around the origin and displays a smoother saturation pattern. This suggests that the contemporaneous residual impact channel is present in both regimes, but its scale and effective support are affected by tick-size constraints.

\section{From Explanations to a Parametric Model}\label{sec:shap_parametric_model}

The analysis developed in the previous sections suggests that the nonlinear neural network learns economically interpretable dependencies. 

In particular, the aggregated conditional response surfaces describe the
model-implied response in selected regions of the state space and are
qualitatively validated by the corresponding empirical surfaces reconstructed
directly from the data. The Shapley decomposition provides an additional layer
of information by separating this aggregate response into the contributions of
the individual input variables. We use these insights to introduce a simpler,
calibratable, and interpretable parametric model.

The idea is therefore to construct a reduced-form model based on the explanations developed in the previous section. In this way, the model keeps the interpretability and tractability of a VAR-type specification, but incorporates the nonlinear mechanisms revealed by the neural network.

\subsection{One-lag model construction}

The SHAP analysis showed that a large part of the nonlinear structure learned by the neural network is concentrated in the most recent return--volume coordinates, \((r_{t-1},v_{t-1})\). This does not imply that higher-order lags are negligible: they still contain useful predictive information, in particular through the persistence of order flow. Rather, the evidence suggests that the strongest nonlinear effects are already visible at the first lag. For this reason, we first ask how much explanatory power is lost when replacing a linear model with several lags by a nonlinear Markovian specification based only on the most recent return and signed-volume observations. In this way, the analysis compares two different sources of predictability: the use of a richer lag structure in a linear model and the use of nonlinear interactions in a low-dimensional, one-lag representation.

\paragraph{Volume terms.} 
We first consider the volume-related contributions. For small-tick stocks, these profiles are reported in Figure~\ref{fig:SHAP_ST_v}, while the corresponding large-tick profiles are reported in Appendix Figure~\ref{fig:SHAP_star_LT}. The figure shows the profiles \(s_v^r\) and \(s_v^v\), which describe the average Shapley contribution of the lagged signed-volume term as a function of the lagged signed volume itself, respectively for the return and signed-volume predictions. We also report the corresponding starred profiles, \(s_v^{*,r}\) and \(s_v^{*,v}\). These quantities measure how the remaining Shapley contribution varies when projected on the same signed-volume coordinate, and are therefore useful to detect possible spurious interaction effects attributed to the other variables.

As widely discussed, this term appears to have no spurious dependencies on $r_{t-1}$ but only a sign-like dependence on the signed volume. Given this, we can guess the term \(\Phi_{v}^a\), where $a \in \{r,v\}$, to be an odd, sign-preserving, saturating function:
\begin{equation}
\Phi_{v}^a(v)
=
A_{v}^a\tanh(\beta_{v}^av).
\label{eq:F_rv}
\end{equation}
This preserves the sign-like directional response observed in the projections of Shapley term $\phi_v^a$, while avoiding a discontinuous jump at zero. The steepness parameter allows the same specification to capture smoother LT responses and sharper ST responses. This term appears in the same way in the return equation, $a=r$, and in the volume equation $a=v$.

\paragraph{Return terms.}
We now consider the return-related contributions, reported in Figure~\ref{fig:SHAP_ST_r} for small-tick stocks. The analogous profiles for large-tick stocks are reported in Appendix Figure~\ref{fig:SHAP_star_LT}. The interpretation is more delicate than for the volume terms, because a substantial part of the structure associated with the return contribution is organized by lagged signed volume.

A relevant structure is present in the starred components. These profiles show how the remaining Shapley contribution varies when projected on the signed-volume coordinate. They indicate that the sign of \(v_{t-1}\) plays a central role in determining the direction of the next-step contribution. In particular, when the previous trade did not induce a price change, \(r_{t-1}=0\), the contribution tends to preserve the sign of \(v_{t-1}\): positive lagged signed volume produces a positive contribution, while negative lagged signed volume produces a negative contribution, both for the next return and for the next signed volume. Conversely, when a price change occurred, \(r_{t-1}\neq 0\), the contribution tends to generate an opposite pressure, favouring a sign reversal in the next-step prediction. This means that the return-related term contains information about the joint return--volume state, rather than a purely return-driven effect.

The same mechanism is visible directly in the non-starred return profiles. Around \(r_{t-1}=0\), the sign of the contribution is mainly determined by the sign of \(v_{t-1}\): the two conditional curves lie on opposite sides of zero, consistently with the sign-preserving behavior observed in the starred components. Thus, even when the profile is plotted as a function of \(r_{t-1}\), the dominant separation between curves is induced by the lagged signed-volume state.

For non-zero lagged returns, the dependence on \(r_{t-1}\) becomes more subtle and differs between tick-size regimes. In the small-tick case, shown in Figure~\ref{fig:SHAP_ST_r}, the dependence is approximately symmetric with respect to the sign of \(r_{t-1}\). As \(|r_{t-1}|\) increases, the sign-preserving contribution associated with \(v_{t-1}\) is progressively weakened and may become mean-reverting. This motivates the following parsimonious reduced-form specification:
\begin{equation}
\Phi_{r}^a(r, v)
=
q\left[
A_{r}^a
-
B_{r}^a\tanh\left(\lambda_{r}^a|r|\right)
\right],
\qquad
q=\operatorname{sign}(v_{t-1}).
\label{eq:H_vr}
\end{equation}
This functional form captures the dominant symmetric component of the empirical profiles: for small lagged returns, the contribution is mainly aligned with the sign of the previous signed volume, while for larger \(|r_{t-1}|\) this contribution is attenuated through the saturating hyperbolic tangent term.

The large-tick profiles, reported in Appendix Figure~\ref{fig:SHAP_star_LT}, suggest that this symmetry is only approximate. In particular, away from \(r_{t-1}=0\), the dependence on positive and negative lagged returns appears partially asymmetric, indicating a possible symmetry-breaking correction. Such a correction could be introduced, for instance, through an additional term depending on the signed return \(r_{t-1}\), or more generally on the relative sign variable \(q r_{t-1}\). We do not include this correction in the baseline specification, since our goal is to construct a parsimonious reduced model that captures the dominant and most stable empirical pattern across stocks. So we keep Equation~\eqref{eq:H_vr} as a first-order reduced form for the symmetric component of the return contribution, while the asymmetric correction is left as a natural extension to be tested separately.

Note that taking \(B_{r}^a>A_{r}^a\), the function \(\Phi_{r}^a\) changes sign at a finite threshold in \(|r|\). Indeed, the term in brackets,
\[
A_{r}^a-B_{r}^a\tanh(\lambda_{r}^a|r|),
\]
is positive for sufficiently small \(|r|\), since \(\tanh(\lambda_{r}^a|r|)\simeq 0\) when \(r\) is close to zero. In this region,
\[
\Phi_{r}^a(r, v)\simeq qA_{r}^a,
\]
so the contribution has the same sign as the lagged signed volume. Therefore, when \(v_{t-1}>0\), the term contributes positively to the prediction of \(r_t\), while when \(v_{t-1}<0\), it contributes negatively. In this regime, the previous return is too small to indicate that the order-flow pressure has already been incorporated into prices, and the model therefore reinforces the directional signal carried by \(v_{t-1}\).

As \(|r|\) increases, the quantity \(\tanh(\lambda_{r}^a|r|)\) increases from zero to one. If \(B_{r}^a>A_{r}^a\), there exists a threshold
\[
|r|^\star
=
\frac{1}{\lambda_{r}^a}
\operatorname{arctanh}\!\left(\frac{A_{r}^a}{B_{r}^a}\right)
\]
such that
\[
A_{r}^a-B_{r}^a\tanh(\lambda_{r}^a|r|)>0
\quad \text{for } |r|<|r|^\star,
\]
while
\[
A_{r}^a-B_{r}^a\tanh(\lambda_{r}^a|r|)<0
\quad \text{for } |r|>|r|^\star.
\]
Beyond this threshold, the sign of \(\Phi_{r}^a\) becomes opposite to the sign of \(q=\operatorname{sign}(v_{t-1})\). Hence, for large absolute returns, the term no longer reinforces the order-flow regime, but instead attenuates or reverses it. This behavior is consistent with the combined evidence from the aggregated surfaces and the SHAP decomposition: the return-state contribution reinforces the order-flow regime when the previous price move is small, while it attenuates or reverses it when the previous return is large in magnitude.

\paragraph{Summary of the parametric model.}

Given these specifications we can, thus, summarize the two equations as
\begin{equation}
\begin{split}
r_t
&= \alpha_r + \Phi_v^r(v) + \Phi_r^r(r,v) =
\alpha_r
+
A_{v}^r\tanh(\beta_{v}^rv_{t-1})
+
\operatorname{sign}(v_{t-1})
\left[
A_{r}^r
-
B_{r}^r\tanh(\lambda_{r}^r|r_{t-1}|)
\right]
+
u_t^r,\\
v_t
&=  \alpha_v + \Phi_v^v(v) + \Phi_r^v(r,v) = 
\alpha_v
+
A_{v}^v\tanh(\beta_{v}^vv_{t-1})
+
\operatorname{sign}(v_{t-1})
\left[
A_{r}^v
-
B_{r}^v\tanh(\lambda_{r}^v|r_{t-1}|)
\right]
+
u_t^v.
\end{split}
\label{eq:parametric_model_summary}
\end{equation}

The first nonlinear term in each equation captures the direct contribution of lagged signed volume: directional order-flow impact in the return equation and order-flow persistence in the volume equation. The second term captures the correction induced by the previous return state. For small \(|r_{t-1}|\), this term reinforces the sign of \(v_{t-1}\); for sufficiently large \(|r_{t-1}|\), it attenuates or reverses the order-flow signal. Thus, the model provides a compact parametric representation of the main mechanisms isolated by the aggregated surfaces and SHAP analyses.

\paragraph{Recovering the SHAP profiles.} The projected quantities in Equation~\eqref{eq:shapley_profiles_definitions} and the relations in Equation~\eqref{eq:binned_shap_decomposition} can be recovered by conditioning Equations~\eqref{eq:parametric_model_summary} in the directions \{$r_{t-1}, v_{t-1}$\}.

If we compute the response of the reduced model equation as a function of \(v_{t-1}\), conditioning on the sign \(k_r=\mathrm{sign}(r_{t-1})\), we obtain approximately
\begin{equation}
m_{a|v}^{\mathrm{par}}(u,k)
\simeq
\alpha_a
+
\Phi_{v}^a(v_u)
+
\mathbb{E}\!\left[
\Phi_{r}^a(r,v)
\mid
v_{t-1}\in B_u^v,\ \mathrm{sign}(r_{t-1})=k_r
\right]
 = \alpha_a + s_{v, \text{par}}^a + s_{r, \text{par}}^{*,a}.
\label{eq:new_pdp_rv}
\end{equation}
In fact, in this decomposition, \(\Phi_{v}^a(v_u)\) reproduces the observed non-starred profile \(s_v^a\), while the conditional expectation of \(\Phi_{r}^a\) reproduces the regime-dependent shift associated with the observed \(s_r^{*,a}\). Here, we observe that $s_{r,\text{par}}^{*,a}$ reproduces the $s_r^{*,a}$ shown in the right panels of Figures~\ref{fig:SHAP_ST_r} and~\ref{fig:SHAP_star_ST}. In fact, for $|r|$ sufficiently large, we have that
\[
     s_{r, \text{par}}^{*,a}  \simeq - C \operatorname{sign}(v_{t-1}), \quad C=B_{r}^a\tanh(\lambda_{r}^a|r|) - A_{r}^a>0, 
\]
which reproduces the blue and red curves in the top left panels in Figures~\ref{fig:SHAP_star_LT} and~\ref{fig:SHAP_star_ST}. Differently, for $|r| \sim 0$ we have that 
\[
     s_{r, \text{par}}^{*,a}  \simeq A_{r}^a \operatorname{sign}(v_{t-1}), \quad A_{r}^a>0, 
\]
which reproduces the green curve.
Hence, the separation between the three curves in \(\hat m_{a|v}\) is not attributed to different direct volume effects, but to the return-state correction induced by \(\Phi_{r}^a\), exactly as observed in the explainability of the model.

Conversely, if we compute the response as a function of \(r_{t-1}\), conditioning on \(q=\operatorname{sign}(v_{t-1})\), we obtain
\begin{equation}
m_{a|r}^{\mathrm{par}}(u,q)
\simeq
\alpha_a
+
\mathbb{E}\!\left[
\Phi_{v}^a(v_{t-1})
\mid
r_{t-1}\in B_u^r,\ \operatorname{sign}(v_{t-1})=q
\right]
+
\Phi_{r}^a(r_u) = \alpha_a + s_{v, \text{par}}^{*,a} + s_{r,\text{par}}^a.
\label{eq:new_pdp_rr}
\end{equation}
Also in this case, $s_{v,\text{par}}^{*,a}$ reproduces what observed in the right panels of Figures~\ref{fig:SHAP_ST_v} and~\ref{fig:SHAP_star_LT}, being just a constant as a function of $r_{t-1}$ but whose sign depends on the sign of $v_{t-1}$. Hence, again, the expectation of \(\Phi_{v}^a\) generates the vertical separation between the \(v_{t-1}>0\) and \(v_{t-1}<0\) curves, corresponding to the starred contribution \(s_v^{*,a}\), while \(\Phi_{r}^a(r_u)\) describes the shape of the response along the \(r_{t-1}\) axis, corresponding to the non-starred profile \(s_r^a\).

In the next section, we calibrate the reduced model on data and test whether the proposed specifications for the return and volume terms are sufficient to approximately recover the empirical map \(\hat m\). To this end, we compare the results of the reduced parametric model with the neural-network profiles discussed in Section~\ref{sec:aggregated}, and more specifically with the corresponding empirical profiles computed directly from the data and reported in Appendix~\ref{sec:data_nonlinear}.

\subsection{Calibration of the SHAP-Inspired Parametric Model}
 
We estimate the parameters separately for the two equations by nonlinear least squares, minimizing the mean squared prediction error on the training sample. To ensure numerical stability and preserve the qualitative structure of the model, the positivity constraints are imposed by reparametrization. In particular, the scale and steepness parameters are constrained to be positive, while the condition
\[
B_{r}^r>A_{r}^r,
\qquad
B_{r}^v>A_{r}^v
\]
can be imposed through
\[
B_{r}^r=A_{r}^r+\operatorname{softplus}(\delta_{r}^r),
\qquad
B_{r}^v=A_{r}^v+\operatorname{softplus}(\delta_{r}^v),
\]
where $\operatorname{softplus}(x)=\log(1+e^x)$ is a smooth, strictly
positive function, ensuring $B_r^r>A_r^r$ and $B_r^v>A_r^v$.
This condition guarantees that the correction terms can change sign at finite thresholds in \(|r_{t-1}|\). For instance, in the return equation the threshold is
\[
|r|^\star
=
\frac{1}{\lambda_{r}^r}
\operatorname{arctanh}\!\left(\frac{A_{r}^r}{B_{r}^r}\right),
\]
provided \(B_{r}^r>A_{r}^r\). Hence, below this threshold the correction reinforces the signed-volume regime, whereas above it the correction acts in the opposite direction.

The model is calibrated on the same preprocessed data used for the VAR and neural-network analysis.

\begin{figure}[t]
    \centering

    \begin{subfigure}{0.48\textwidth}
        \centering
        \includegraphics[width=\textwidth]{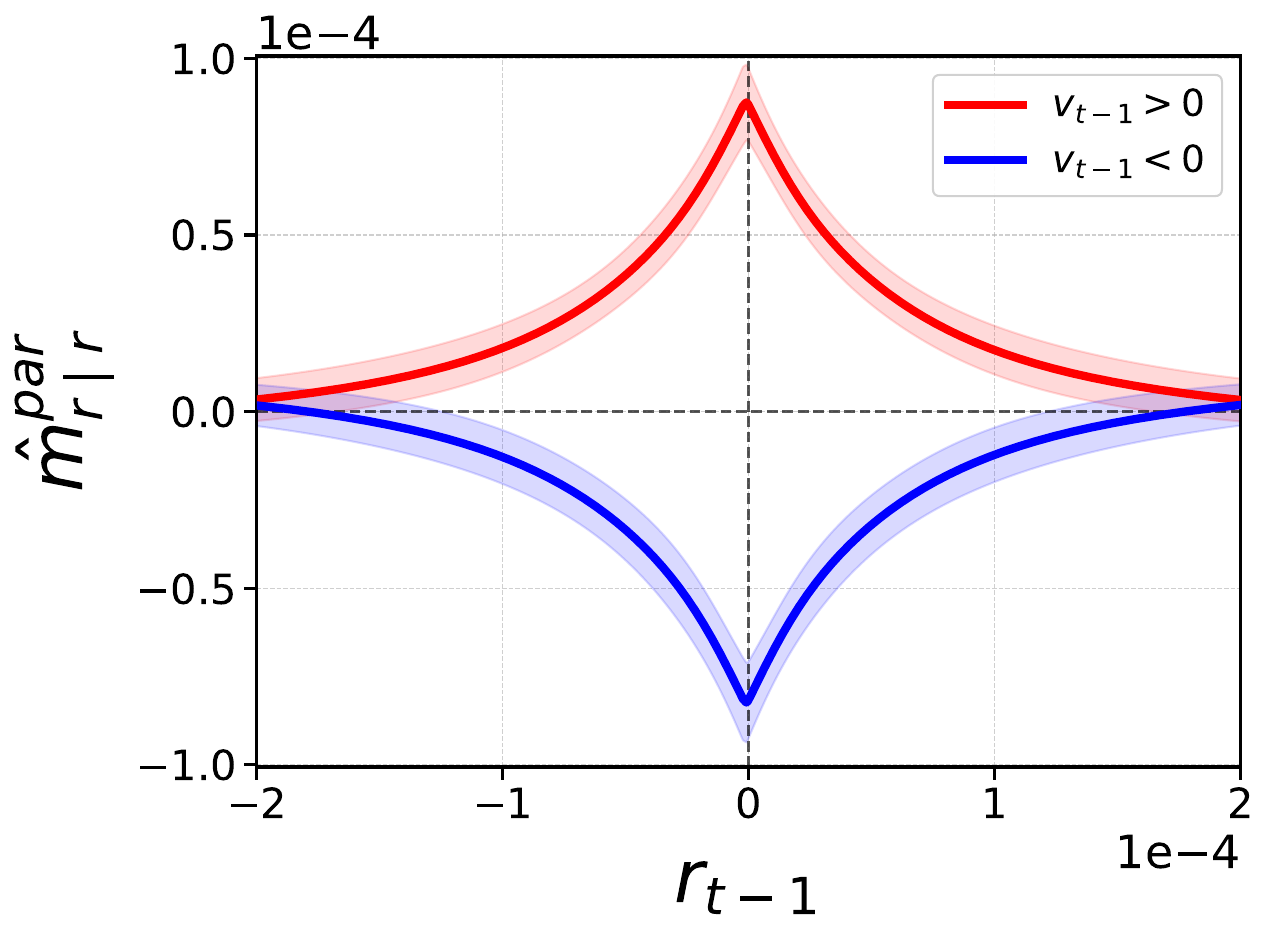}
        \caption{\(\hat r_t\) as a function of \(r_{t-1}\).}
        \label{fig:parametric_LT_r_lag_r}
    \end{subfigure}
    \hfill
    \begin{subfigure}{0.48\textwidth}
        \centering
        \includegraphics[width=\textwidth]{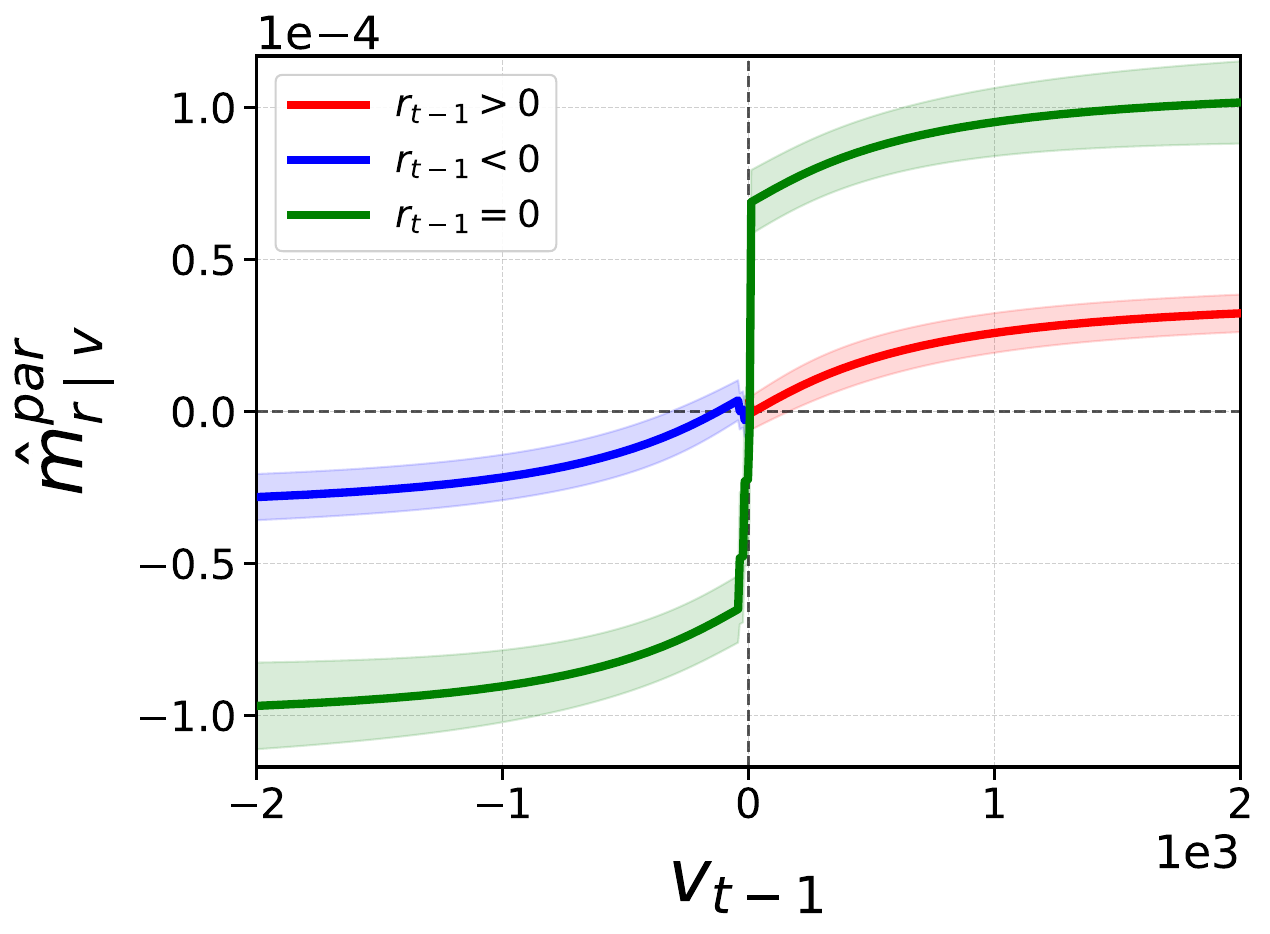}
        \caption{\(\hat r_t\) as a function of \(v_{t-1}\).}
        \label{fig:parametric_LT_r_lag_v}
    \end{subfigure}

    \vspace{0.35cm}

    \begin{subfigure}{0.48\textwidth}
        \centering
        \includegraphics[width=\textwidth]{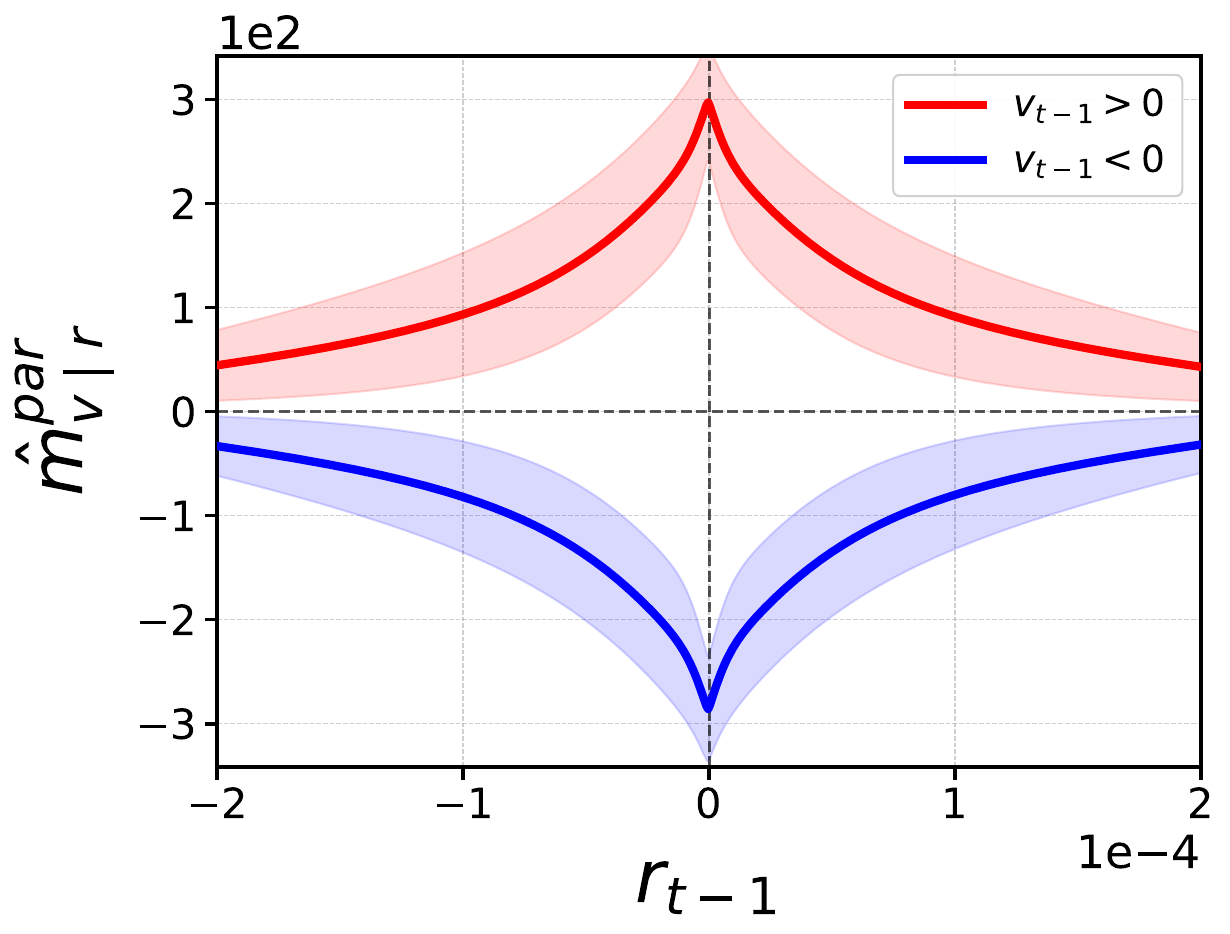}
        \caption{\(\hat v_t\) as a function of \(r_{t-1}\).}
        \label{fig:parametric_LT_v_lag_r}
    \end{subfigure}
    \hfill
    \begin{subfigure}{0.48\textwidth}
        \centering
        \includegraphics[width=\textwidth]{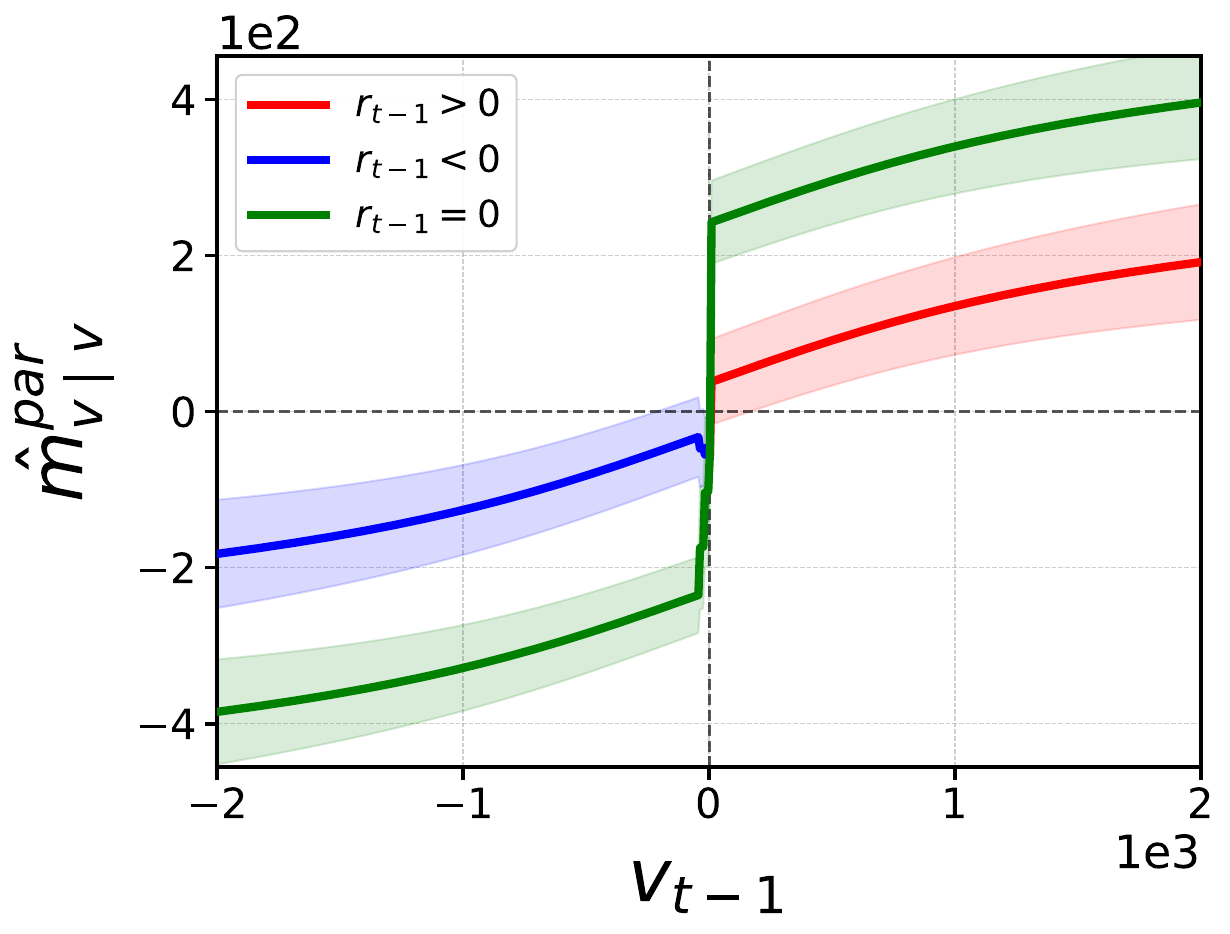}
        \caption{\(\hat v_t\) as a function of \(v_{t-1}\).}
        \label{fig:parametric_LT_v_lag_v}
    \end{subfigure}

    \caption{Fitted one-lag parametric response functions for large-tick stocks.
    The top row reports the response of \(\hat r_t\) to \(r_{t-1}\) and
    \(v_{t-1}\), while the bottom row reports the response of \(\hat v_t\)
    to \(r_{t-1}\) and \(v_{t-1}\).}
    \label{fig:parametric_LT_responses}
\end{figure}
\begin{figure}[t]
    \centering

    \begin{subfigure}{0.48\textwidth}
        \centering
        \includegraphics[width=\textwidth]{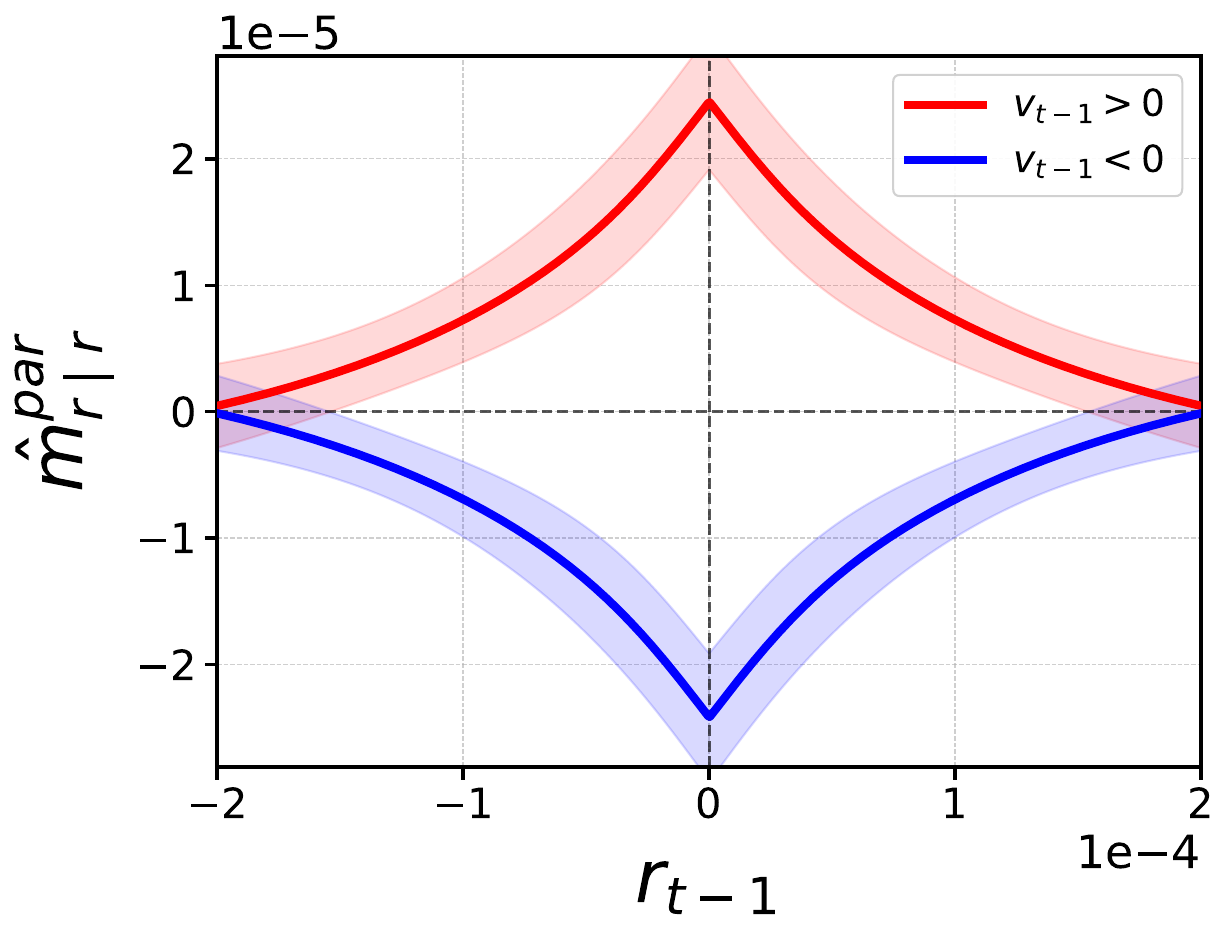}
        \caption{\(\hat r_t\) as a function of \(r_{t-1}\).}
        \label{fig:parametric_ST_r_lag_r}
    \end{subfigure}
    \hfill
    \begin{subfigure}{0.48\textwidth}
        \centering
        \includegraphics[width=\textwidth]{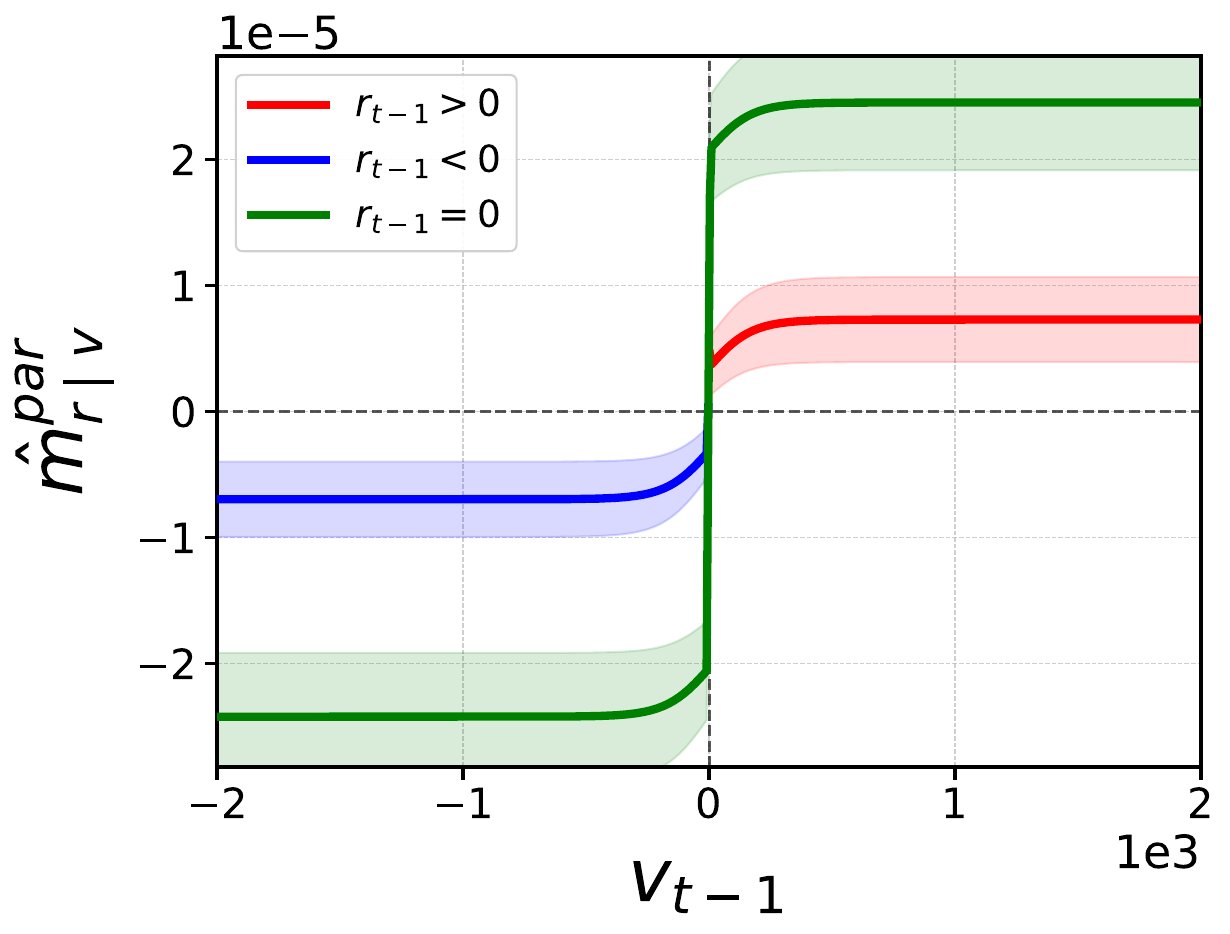}
        \caption{\(\hat r_t\) as a function of \(v_{t-1}\).}
        \label{fig:parametric_ST_r_lag_v}
    \end{subfigure}

    \vspace{0.35cm}

    \begin{subfigure}{0.48\textwidth}
        \centering
        \includegraphics[width=\textwidth]{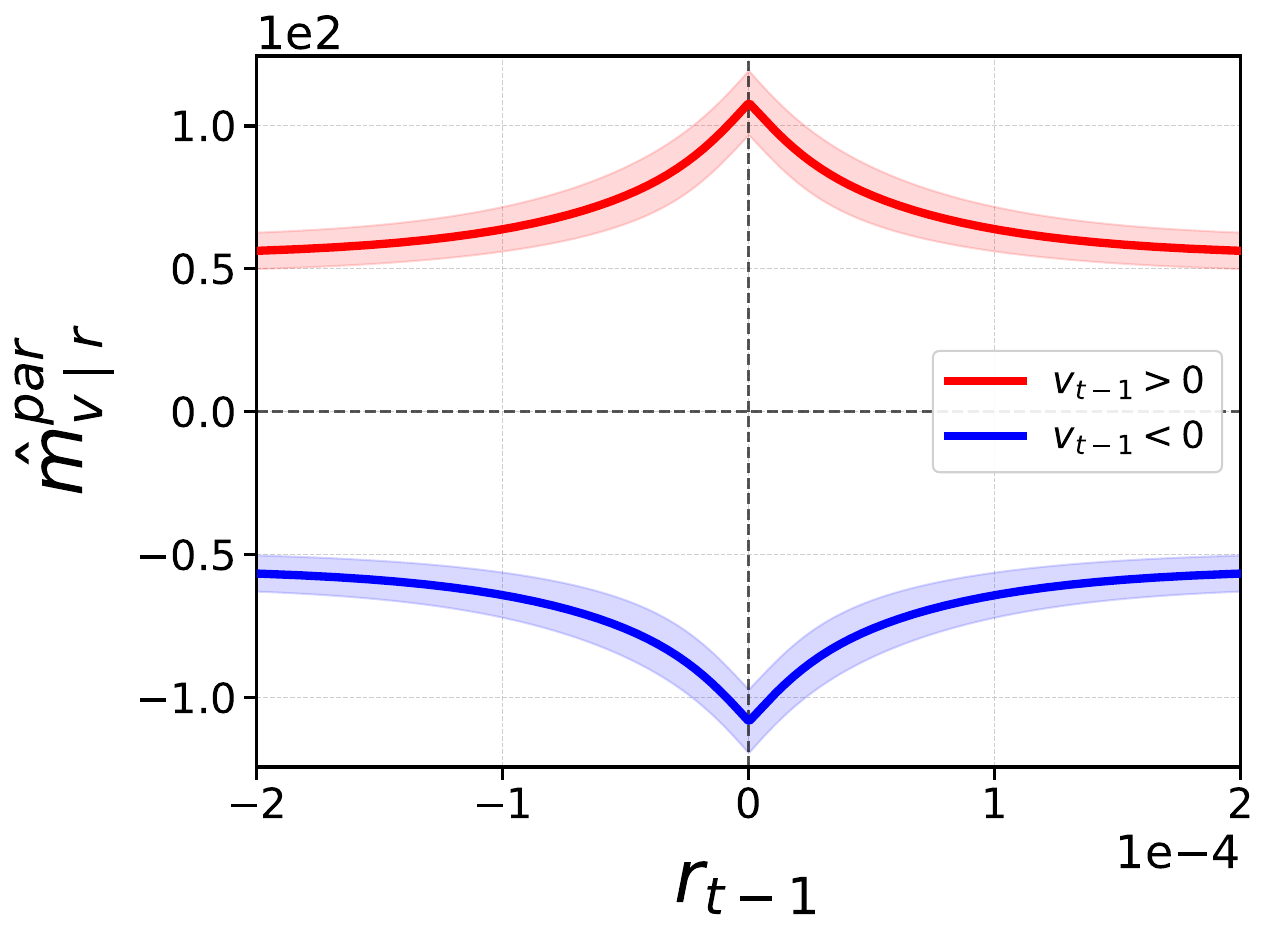}
        \caption{\(\hat v_t\) as a function of \(r_{t-1}\).}
        \label{fig:parametric_ST_v_lag_r}
    \end{subfigure}
    \hfill
    \begin{subfigure}{0.48\textwidth}
        \centering
        \includegraphics[width=\textwidth]{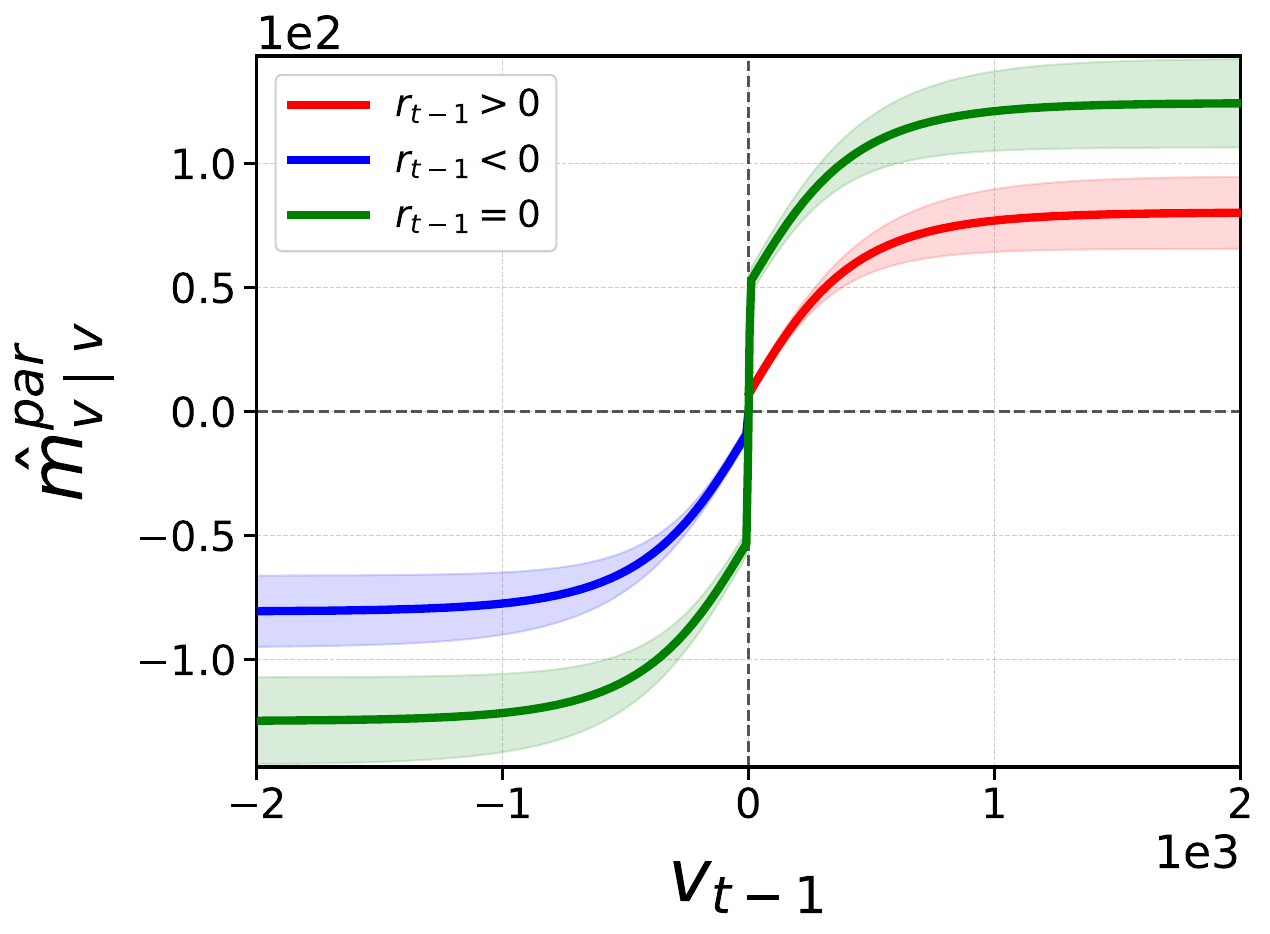}
        \caption{\(\hat v_t\) as a function of \(v_{t-1}\).}
        \label{fig:parametric_ST_v_lag_v}
    \end{subfigure}

    \caption{Fitted one-lag parametric response functions for small-tick stocks.
    The top row reports the response of \(\hat r_t\) to \(r_{t-1}\) and
    \(v_{t-1}\), while the bottom row reports the response of \(\hat v_t\)
    to \(r_{t-1}\) and \(v_{t-1}\).}
    \label{fig:parametric_ST_responses}
\end{figure}
The fitted response functions are reported in
Figures~\ref{fig:parametric_LT_responses}--\ref{fig:parametric_ST_responses}.
Overall, the one-lag parametric model is able to reproduce, at least
qualitatively, the main dependencies identified through the neural-network
analysis. More importantly, the same qualitative structures are consistent with
the aggregate dependencies reconstructed directly from the data and reported in
Appendix~\ref{sec:data_nonlinear}, Figures~\ref{fig:DATA_LT}--\ref{fig:DATA_ST}.

The agreement is not exact, especially when the parametric surfaces are compared
with the neural-network predictive surfaces. Some differences are visible also
with respect to the empirical averages, and are mainly related to asymmetries in
the response functions. These discrepancies may have different origins. They may
reflect residual learning errors or smoothing effects in the neural-network
prediction, finite-sample fluctuations in the empirical conditional averages, or
higher-order corrections that break the symmetric structure imposed in the
baseline reduced model. In this first specification, we do not include such
asymmetric corrections, in order to preserve a parsimonious and interpretable
form; they represent a natural direction for future extensions.

The important point is that, starting from the isolated Shapley contributions of
the individual regressors, the reduced model is able to recover aggregate
patterns that are comparable with those observed in the data. Thus, despite the
expected differences due to its reduced complexity, the parametric specification
captures the dominant qualitative structure of the return--volume dependence in
both large-tick and small-tick stocks.

Table~\ref{tab:var_mlp_reduced_r2_out} reports the out-of-sample
\(R^2\) obtained by the one-lag nonlinear parametric model for returns and
signed volumes, together with the corresponding results of the VAR and DNN
models estimated with \(p=20\) lags. The comparison is particularly informative
because the reduced model uses only the most recent return--order-flow pair,
\((r_{t-1},v_{t-1})\), whereas the VAR and the DNN exploit the full lagged
state up to twenty lags.

The main finding is that the Shapley-motivated one-lag nonlinear specification
already captures most of the predictive structure in the data. For returns, the
reduced model systematically improves on the linear VAR benchmark across all
stocks. The gains are especially pronounced for large-tick stocks: for example,
the out-of-sample \(R^2\) increases from \(0.0494\) to \(0.0934\) for BAC, from
\(0.0764\) to \(0.1156\) for CSCO, and from \(0.0356\) to \(0.0845\) for PFE.
These improvements indicate that the relevant predictability is not only a
lagged linear effect, but is largely associated with the nonlinear interaction
between the most recent return and signed volume.

The comparison with the DNN is even more revealing. Despite its much lower
dimensionality, the reduced model often matches or outperforms the neural
network. This is the case for most large-tick stocks in the return equation and
for several small-tick stocks as well, such as AMZN, GILD, and NVDA. The DNN
retains an advantage in some small-tick cases, most notably AAPL and TSLA for
returns, suggesting that additional nonlinear or longer-memory effects may still
be relevant for these stocks. Nevertheless, the fact that a transparent
two-variable parametric specification is competitive with, and frequently
superior to, a twenty-lag neural network is one of the central empirical results
of the paper.

A similar pattern appears for signed volumes. The reduced model improves on the
VAR benchmark for every stock and is particularly effective for large-tick
stocks, where it also dominates the DNN in all cases. For small-tick stocks the
DNN remains stronger in some cases, notably GILD and NVDA, while the two models perform identically for TSLA. Overall, Table~\ref{tab:var_mlp_reduced_r2_out} shows that most of the
out-of-sample predictability can be recovered from a parsimonious nonlinear
description of the latest return--order-flow state. This provides direct
empirical support for the reduced-form structure identified by the Shapley
analysis and motivates the extension, considered next, in which the same
nonlinear building blocks are enriched with multiple lagged inputs.
\begin{table}[ht]
\centering
\small
\setlength{\tabcolsep}{3pt}
\begin{tabular}{clccc|ccc}
\toprule
 & Ticker
& \multicolumn{3}{c}{\(R^2_{\mathrm{out}}(r)\)}
& \multicolumn{3}{c}{\(R^2_{\mathrm{out}}(v)\)} \\
\cmidrule(lr){3-5}\cmidrule(lr){6-8}
& & VAR & DNN & Reduced & VAR & DNN & Reduced \\
\midrule
\multirow{5}{*}{\rotatebox[origin=c]{90}{large tick}}
& BAC   & 0.0494 & 0.0591 & 0.0934 & 0.0299 & 0.0215 & 0.0623 \\
& INTC  & 0.0814 & 0.1217 & 0.1065 & 0.0198 & 0.0288 & 0.0644 \\
& CSCO  & 0.0764 & 0.0982 & 0.1156 & 0.0328 & 0.0273 & 0.0735 \\
& CMCSA & 0.0604 & 0.0859 & 0.1020 & 0.0350 & 0.0486 & 0.0820 \\
& PFE   & 0.0356 & 0.0367 & 0.0845 & 0.0183 & 0.0091 & 0.0478 \\
\midrule
\multirow{5}{*}{\rotatebox[origin=c]{90}{small tick}}
& AMZN  & 0.0246 & 0.0341 & 0.0528 & 0.0685 & 0.0705 & 0.0892 \\
& AAPL  & 0.0643 & 0.0951 & 0.0858 & 0.1147 & 0.1372 & 0.1421 \\
& GILD  & 0.0499 & 0.0429 & 0.0723 & 0.1055 & 0.1173 & 0.1165 \\
& TSLA  & 0.0433 & 0.0906 & 0.0826 & 0.1115 & 0.1475 & 0.1475 \\
& NVDA  & 0.0166 & 0.0454 & 0.0498 & 0.0872 & 0.1068 & 0.0967 \\
\bottomrule
\end{tabular}
\caption{Out-of-sample \(R^2\) for the VAR, DNN, and reduced parametric models, separately for returns \(r\) and signed volumes ($v$). The VAR and DNN models are estimated at lag \(p=20\), while the reduced parametric model is the one-lag specification obtained from the Shapley-based reduced-form analysis. Stocks are grouped according to the large-tick/small-tick classification used in Table~\ref{tab:spread_ticks}.}
\label{tab:var_mlp_reduced_r2_out}
\end{table}

\subsection{Multi-lag extension} 

Although the one-lag specification is the most direct translation of the SHAP analysis, it is natural to ask whether the same reduced-form structure
can be extended to longer histories. A fully flexible multi-lag version would
assign independent nonlinear parameters to each lag. For example, the return
equation would contain terms of the form
\[
A_{rv,\ell}\tanh(\beta_{rv,\ell}v_{t-\ell})
+
\operatorname{sign}(v_{t-\ell})
\left[
A_{rr,\ell}
-
B_{rr,\ell}\tanh(\lambda_{rr,\ell}|r_{t-\ell}|)
\right],
\qquad \ell=1,\dots,p.
\]
This specification is the most general extension of the one-lag model, but it
also introduces a large number of nonlinear parameters. As a consequence, it
risks fitting lag-specific noise rather than persistent predictive structure,
especially in light of the SHAP lag-importance analysis, which shows that most
of the predictive contribution is concentrated at the first lags and then
decays smoothly.

We therefore impose a structured multi-lag restriction motivated by the
empirical patterns documented above. The key assumption is that the qualitative
nonlinear shape identified at the first lag remains the same across lags, while
the strength of the corresponding contribution decreases with lag distance.
Thus, instead of estimating an independent nonlinear function for each lag, we
share the nonlinear shape parameters across lags and let lag-specific
intensities follow a geometric decay. This preserves the economic structure
suggested by the SHAP analysis, while allowing the model to incorporate
longer-memory effects without introducing a separate set of nonlinear
parameters for every lag.
 
The return equation becomes
\begin{equation}
\begin{split}
r_t
=
\alpha_r
&+
A_{rv}
\sum_{\ell=1}^{p}
\rho_{rv}^{\ell-1}
\tanh(\beta_{rv}v_{t-\ell})\\
&+
\sum_{\ell=1}^{p}
\rho_{rr}^{\ell-1}
\operatorname{sign}(v_{t-\ell})
\left[
A_{rr}
-
B_{rr}\tanh(\lambda_{rr}|r_{t-\ell}|)
\right]
+
\varepsilon_t^r,
\end{split}
\label{eq:parametric_return_decay}
\end{equation}
with \(0<\rho_{rv}<1\) and \(0<\rho_{rr}<1\). Similarly, the signed-volume equation is
\begin{equation}
\begin{split}
v_t
=
\alpha_v
&+
A_{vv}
\sum_{\ell=1}^{p}
\rho_{vv}^{\ell-1}
\tanh(\beta_{vv}v_{t-\ell})\\
&+
\sum_{\ell=1}^{p}
\rho_{vr}^{\ell-1}
\operatorname{sign}(v_{t-\ell})
\left[
A_{vr}
-
B_{vr}\tanh(\lambda_{vr}|r_{t-\ell}|)
\right]
+
\varepsilon_t^v,
\end{split}
\label{eq:parametric_volume_decay}
\end{equation}
with \(0<\rho_{vv}<1\) and \(0<\rho_{vr}<1\).

This specification separates the nonlinear shape of the dependence from its temporal decay. The parameters \(A\), \(\beta\), \(B\), and \(\lambda\) determine the common nonlinear response, while the \(\rho\)'s determine how quickly the corresponding contribution fades with the lag. In this way, the model remains parsimonious even for large \(p\): the number of nonlinear parameters does not grow linearly with the lag length. At the same time, it incorporates longer memory in a way that is consistent with the SHAP evidence of decreasing lag importance.

Table~\ref{tab:var_mlp_reduced_multilag_r2_out} reports the out-of-sample
\(R^2\) obtained by the VAR, the deep neural network, the one-lag reduced
parametric model, and the multi-lag parametric specification with exponential
decay and \(p=20\). The table shows that the reduced parametric models provide a
clear improvement over the linear VAR benchmark. This holds for both returns and
signed volumes and across both large-tick and small-tick stocks.

For returns, the one-lag reduced model already captures most of the predictive
gain. The multi-lag extension produces only moderate additional improvements:
the \(R^2\) increases slightly for most stocks, while remaining essentially
unchanged for some small-tick names such as TSLA and NVDA. This pattern is
consistent with the Shapley analysis, which shows that the dominant nonlinear
dependence is concentrated at the first lag. Additional lags therefore contain
some residual predictive information, but their contribution to return
forecasting is limited. Importantly, the reduced parametric specifications are
often comparable to the deep neural network and outperform it for several
stocks, especially among large-tick names such as BAC, CSCO, CMCSA, and PFE.

The gains from the multi-lag extension are more pronounced for signed volumes.
While the one-lag reduced model already improves substantially over the VAR
benchmark, the multi-lag specification further increases the out-of-sample
\(R^2\), particularly for small-tick stocks. This suggests that signed-volume dynamics contain a more persistent component
than returns, so that longer histories provide useful additional information
beyond the first lag. Consistently, for the signed-volume equation, the multi-lag
parametric model achieves higher out-of-sample \(R^2\) than the deep
neural-network benchmark for all stocks in the table.

Overall, these results show that the explainability-guided parametric
specification retains the main nonlinear mechanisms learned by the neural
network while offering a more structured and interpretable representation. The
one-lag version already captures most of the nonlinear predictability of
returns, whereas the multi-lag extension is especially useful for signed-volume
forecasting, where persistence across lags plays a more important role.

\begin{table}[ht]
\centering
\small
\setlength{\tabcolsep}{2.5pt}
\begin{tabular}{clcccc|cccc}
\toprule
 & Ticker
& \multicolumn{4}{c}{\(R^2_{\mathrm{out}}(r)\)}
& \multicolumn{4}{c}{\(R^2_{\mathrm{out}}(v)\)} \\
\cmidrule(lr){3-6}\cmidrule(lr){7-10}
& & VAR & DNN & Reduced & Multi-lag
  & VAR & DNN & Reduced & Multi-lag \\
\midrule
\multirow{5}{*}{\rotatebox[origin=c]{90}{large tick}}
& BAC   & 0.0494 & 0.0591 & 0.0934 & 0.0945 & 0.0299 & 0.0215 & 0.0623 & 0.0623 \\
& INTC  & 0.0814 & 0.1217 & 0.1065 & 0.1073 & 0.0198 & 0.0288 & 0.0644 & 0.0651 \\
& CSCO  & 0.0764 & 0.0982 & 0.1156 & 0.1168 & 0.0328 & 0.0273 & 0.0735 & 0.0745 \\
& CMCSA & 0.0604 & 0.0859 & 0.1020 & 0.1031 & 0.0350 & 0.0486 & 0.0820 & 0.0828 \\
& PFE   & 0.0356 & 0.0367 & 0.0845 & 0.0855 & 0.0183 & 0.0091 & 0.0478 & 0.0479 \\
\midrule
\multirow{5}{*}{\rotatebox[origin=c]{90}{small tick}}
& AMZN  & 0.0246 & 0.0341 & 0.0528 & 0.0529 & 0.0685 & 0.0705 & 0.0892 & 0.1047 \\
& AAPL  & 0.0643 & 0.0951 & 0.0858 & 0.0866 & 0.1147 & 0.1372 & 0.1421 & 0.1552 \\
& GILD  & 0.0499 & 0.0429 & 0.0723 & 0.0731 & 0.1055 & 0.1173 & 0.1165 & 0.1248 \\
& TSLA  & 0.0433 & 0.0906 & 0.0826 & 0.0826 & 0.1115 & 0.1475 & 0.1475 & 0.1589 \\
& NVDA  & 0.0166 & 0.0454 & 0.0498 & 0.0498 & 0.0872 & 0.1068 & 0.0967 & 0.1141 \\
\bottomrule
\end{tabular}
\caption{Out-of-sample \(R^2\) for the VAR, DNN, one-lag reduced parametric
model, and multi-lag parametric model, separately for returns \(r\) and signed
volumes \(v\). The VAR and DNN models are estimated with \(p=20\) lags. The
reduced parametric model is the one-lag specification obtained from the
Shapley-based reduced-form analysis, while the multi-lag parametric model uses
\(p=20\) lags with a shared nonlinear shape and exponential lag decay. Stocks
are grouped according to the large-tick/small-tick classification used in
Table~\ref{tab:spread_ticks}.}
\label{tab:var_mlp_reduced_multilag_r2_out}
\end{table}

Additional evidence on the role of the lag length is reported in
Appendix~\ref{app:LagDependenceMulti}. There we study the evolution of the
in-sample and out-of-sample \(R^2\) of the multi-lag parametric model as a
function of the maximum lag \(
\ell\). The results show that predictive performance
increases only mildly after the first few lags and then rapidly saturates,
confirming that the dominant nonlinear dependence is short-lived. This supports
the use of the exponential-decay parametrization, which incorporates residual
longer-lag information without introducing a separate set of nonlinear
parameters for each lag.

\section{Conclusions}\label{sec:conclusions}

In this paper, we studied the joint high-frequency dynamics of returns and
signed volumes through the combination of nonlinear forecasting models,
explainability techniques, and interpretable parametric modeling. The starting
point of the analysis was the comparison between a linear VAR benchmark and a
deep feed-forward neural network trained on five large-tick and five small-tick
stocks. The neural network generally improves upon the VAR benchmark, especially
for return prediction, suggesting that relevant nonlinear dependencies are
present in the data and are not fully captured by a linear specification.

Before interpreting the nonlinear model, we analyzed the dependence of
forecasting performance on the lag order. For the VAR benchmark, the
out-of-sample performance increases at short lags and then rapidly saturates.
The neural network exhibits a less regular dependence on the lag length, with
fluctuations in out-of-sample performance, but the overall evidence again points
to a short-memory structure. We therefore use \(p=20\) as a conservative lag
choice: it is sufficiently large to include the relevant short-horizon
dependencies, while avoiding an excessive expansion of the input space.

We then used SHAP-based explainability to understand whether the nonlinear model
learns economically meaningful relationships between the regressors. The
Shapley lag-importance analysis shows that, across both tick-size regimes, the
dominant contributions are concentrated at the most recent lags. This motivates
the construction of projected predictive surfaces conditional on the first
return--volume pair, \((r_{t-1},v_{t-1})\). These surfaces allow us to study the
aggregate dependence of the neural-network predictions on the most relevant
lagged variables.

A first important result is that the model-implied predictive surfaces are
qualitatively consistent with the corresponding empirical conditional averages
computed directly from the data. This provides a crucial validation step: the
nonlinear structures learned by the neural network are not merely artifacts of
the architecture or of the training procedure, but reflect regularities that are
already present in the data. Once this consistency is established, the Shapley
decomposition can be used to go beyond aggregate conditional averages and
separate the predictive response into the isolated contributions of individual
regressors.

The isolated Shapley components reveal a clear and interpretable structure. The
contribution of lagged signed volume is strongly nonlinear, sign-preserving and
saturating. Positive lagged signed volume contributes positively to future
returns and signed volumes, while negative lagged signed volume contributes
negatively. This term does not display a substantial spurious dependence on the
lagged return, indicating that the direct volume contribution is well described
as a sign-like saturating function of \(v_{t-1}\).

The return-related contribution is more complex and state-dependent. It depends
on the magnitude of the lagged return, but it is also strongly organized by the
sign of the previous signed volume. In particular, when the previous trade does
not move the price, \(r_{t-1}=0\), the model tends to predict continuation: both
the next return and the next signed volume are shifted in the same direction as
\(v_{t-1}\). Conversely, when the previous trade is associated with a non-zero
price change, the model tends to generate an opposite pressure: the next return
and signed volume are shifted against the sign of \(v_{t-1}\). This mechanism
can be interpreted as a transition from order-flow persistence when price
adjustment has not yet occurred to attenuation or reversal when the previous
order-flow pressure has already produced a price move. The same qualitative
structure is observed for both small-tick and large-tick stocks, although with
expected differences in intensity, smoothness, and asymmetry.

We also analyzed the contemporaneous residual dependence between return and
signed-volume innovations. Since the reduced-form models use lagged information
only, the instantaneous impact of unexpected order flow remains encoded in the
residuals. By estimating a nonparametric structural projection
\(u_t^r=\phi(u_t^v)+\epsilon_t^r\), we find that the residual impact function is
again increasing, sign-preserving, and saturating, for both large-tick and
small-tick stocks. This confirms that, after removing lagged nonlinear
predictability, a sizeable instantaneous order-flow channel remains in the
return residuals.

Building on these observations, we introduced a SHAP-inspired nonlinear
parametric model. The model contains the main dependencies extracted from the
explainability analysis: a saturating sign-like term in lagged signed volume and
a state-dependent correction driven by the magnitude of lagged returns and by
the sign of previous order flow. The one-lag version of the model provides a
compact Markovian representation of the dominant nonlinear mechanisms identified
by the neural network. Despite its simplicity, it reconstructs the main
qualitative features of the model-implied and empirical predictive surfaces.

A central empirical result of the paper is that this reduced parametric model
outperforms the linear VAR benchmark and achieves predictive performance
comparable to that of the deep neural network, and in several cases even higher.
This shows that much of the forecasting advantage of the neural network does not
come from an opaque high-dimensional structure, but from a small set of stable
and interpretable nonlinear mechanisms. In this sense, the neural network acts
as a discovery device, while the explainability analysis translates its learned
dependencies into a parsimonious econometric specification.

Finally, we extended the reduced model to multiple lags by assuming that the
same nonlinear shapes apply across lagged observations, with intensities that
decay exponentially with lag distance. This multi-lag specification preserves
the interpretability of the one-lag model while incorporating residual
longer-memory information. Its performance improves relative to the one-lag version, with the clearest gains appearing in the signed-volume equation, but the gain saturates rapidly as the
maximum lag increases. This is consistent with both the VAR lag analysis and the
SHAP lag-importance profiles: longer histories contain some additional
predictive information, but the dominant nonlinear contribution is concentrated
at short horizons. Moreover, the structured parametric lag decay provides a more
stable dependence on the lag length than the unconstrained neural-network
benchmark.

Overall, the paper proposes a constructive use of explainability in financial
modeling. Starting from a flexible nonlinear model calibrated on data, we use
SHAP values and aggregated conditional response surfaces analysis to extract isolated, economically
meaningful dependencies among the regressors. These dependencies are then
translated into interpretable parametric models that retain most of the
predictive power of the neural network while offering greater transparency and
stability. The approach has two complementary benefits. First, it provides
evidence that the neural network is learning meaningful financial mechanisms
rather than spurious statistical patterns. Second, it turns these mechanisms
into explicit parametric forms that can be calibrated, interpreted, compared
with standard benchmarks, and potentially used in simulation or structural
market-microstructure models.

Several directions remain open. First, the parametric model could be extended to
include richer state variables from the limit order book, such as queue
imbalance, depth, spread regimes, or measures of liquidity at the best quotes. Second, the framework could be tested on
longer time periods, additional assets, and different market regimes to assess
the stability of the extracted nonlinear mechanisms. These extensions would
further clarify how machine-learning explainability can be used to build
structural models of price formation in modern electronic markets.

\section*{Acknowledgements}
This publication has received funding from the  European Union – NextGenerationEU – PNRR – Project: “SoBigData.it – Strengthening the Italian RI for Social Mining and Big Data Analytics” – Prot. IR0000013.

The research was carried out also with the contribution of research groups linked to the Horizon 2020 Program under the scheme “INFRAIA-01-2018-2019 – Integrating Activities for Advanced Communities”, Grant Agreement n.871042, ``SoBigData++: European Integrated Infrastructure for Social Mining and Big Data Analytics”, to the HORIZON Europe Program under the scheme ``INFRA-2021-DEV-02- SoBigData RI Preparatory Phase Project" Grant Agreement n. 101079043 — SoBigData RI PPP.

\newpage
\bibliographystyle{unsrt}
\bibliography{Biblio}

@article{bouchaud2003fluctuations,
  title={Fluctuations and response in financial markets: the subtle nature of random price changes},
  author={Bouchaud, Jean-Philippe and Gefen, Yuval and Potters, Marc and Wyart, Matthieu},
  journal={Quantitative Finance},
  volume={4},
  number={2},
  pages={176--190},
  year={2004}
}

@article{toth2011anomalous,
  title={Anomalous price impact and the critical nature of liquidity in financial markets},
  author={T\'oth, Bence and Lemperiere, Yohann and Deremble, Cyril and de Lataillade, Johan and Kockelkoren, Julien and Bouchaud, Jean-Philippe},
  journal={Physical Review X},
  volume={1},
  number={2},
  pages={021006},
  year={2011}
}

@article{lillo2004long,
  title={Long memory in the efficient market},
  author={Lillo, Fabrizio and Farmer, J. Doyne},
  journal={Studies in Nonlinear Dynamics \& Econometrics},
  volume={8},
  number={3},
  year={2004}
}

@article{almgren2001optimal,
  title={Optimal execution of portfolio transactions},
  author={Almgren, Robert and Chriss, Neil},
  journal={Journal of Risk},
  volume={3},
  pages={5--40},
  year={2001}
}

@article{bucci2019are,
  title={Are trading invariants really invariant? Trading costs matter},
  author={Bucci, Andrea and Mastromatteo, Iacopo and Benzaquen, Michael and Bouchaud, Jean-Philippe},
  journal={Quantitative Finance},
  year={2019}
}

@article{sirignano2019deep,
  title={Deep learning for limit order books},
  author={Sirignano, Justin and Cont, Rama},
  journal={Quantitative Finance},
  volume={19},
  number={4},
  pages={549--570},
  year={2019}
}

@article{zhang2019deeplob,
  title={DeepLOB: Deep convolutional neural networks for limit order books},
  author={Zhang, Zihao and Zohren, Stefan and Roberts, Stephen},
  journal={IEEE Transactions on Signal Processing},
  volume={67},
  number={11},
  pages={3001--3012},
  year={2019}
}

@article{lundberg2017unified,
  title={A unified approach to interpreting model predictions},
  author={Lundberg, Scott M and Lee, Su-In},
  journal={Advances in Neural Information Processing Systems},
  volume={30},
  year={2017}
}

@article{hasbrouck1991measuring,
  title={Measuring the information content of stock trades},
  author={Hasbrouck, Joel},
  journal={Journal of Finance},
  volume={46},
  number={1},
  pages={179--207},
  year={1991}
}

@article{Naviglio03042026,
author = {M. Naviglio and G. Bormetti and F. Campigli and G. Rodikov and F. Lillo},
title = {Why is the estimation of metaorder impact with public market data so challenging?},
journal = {Quantitative Finance},
volume = {26},
number = {4},
pages = {505--523},
year = {2026},
publisher = {Routledge},
doi = {10.1080/14697688.2026.2628221},


URL = { 
    
        https://doi.org/10.1080/14697688.2026.2628221
    
    

},
eprint = { 
    
        https://doi.org/10.1080/14697688.2026.2628221
    
    

}

}

@misc{kingma2014adam,
  author        = {Kingma, Diederik P. and Ba, Jimmy},
  title         = {Adam: A Method for Stochastic Optimization},
  year          = {2014},
  eprint        = {1412.6980},
  archivePrefix = {arXiv},
  primaryClass  = {cs.LG}
}

@article{cybenko1989approximation,
  author  = {Cybenko, George},
  title   = {Approximation by Superpositions of a Sigmoidal Function},
  journal = {Mathematics of Control, Signals and Systems},
  volume  = {2},
  number  = {4},
  pages   = {303--314},
  year    = {1989},
  doi     = {10.1007/BF02551274}
}

@article{hornik1989multilayer,
  author  = {Hornik, Kurt and Stinchcombe, Maxwell and White, Halbert},
  title   = {Multilayer Feedforward Networks are Universal Approximators},
  journal = {Neural Networks},
  volume  = {2},
  number  = {5},
  pages   = {359--366},
  year    = {1989},
  doi     = {10.1016/0893-6080(89)90020-8}
}

@book{goodfellow2016deep,
  author    = {Goodfellow, Ian and Bengio, Yoshua and Courville, Aaron},
  title     = {Deep Learning},
  publisher = {MIT Press},
  year      = {2016},
  url       = {http://www.deeplearningbook.org}
}

@article{lecun1998gradient,
  author  = {LeCun, Yann and Bottou, L{\'e}on and Bengio, Yoshua and Haffner, Patrick},
  title   = {Gradient-Based Learning Applied to Document Recognition},
  journal = {Proceedings of the IEEE},
  volume  = {86},
  number  = {11},
  pages   = {2278--2324},
  year    = {1998},
  doi     = {10.1109/5.726791}
}

@article{hochreiter1997long,
  author  = {Hochreiter, Sepp and Schmidhuber, J{\"u}rgen},
  title   = {Long Short-Term Memory},
  journal = {Neural Computation},
  volume  = {9},
  number  = {8},
  pages   = {1735--1780},
  year    = {1997},
  doi     = {10.1162/neco.1997.9.8.1735}
}

@article{friedman2001greedy,
  author  = {Friedman, Jerome H.},
  title   = {Greedy Function Approximation: A Gradient Boosting Machine},
  journal = {The Annals of Statistics},
  volume  = {29},
  number  = {5},
  pages   = {1189--1232},
  year    = {2001},
  doi     = {10.1214/aos/1013203451}
}

@article{eisler2012price,
  author        = {Eisler, Zolt{\'a}n and Bouchaud, Jean-Philippe and Kockelkoren, Julien},
  title         = {The Price Impact of Order Book Events: Market Orders, Limit Orders and Cancellations},
  journal       = {Quantitative Finance},
  volume        = {12},
  number        = {9},
  pages         = {1395--1419},
  year          = {2012},
  publisher     = {Taylor \& Francis},
  doi           = {10.1080/14697688.2010.528444},
  eprint        = {0904.0900},
  archivePrefix = {arXiv},
  primaryClass  = {q-fin.TR}
}

@article{bouchaud2004fluctuations,
  author  = {Bouchaud, Jean-Philippe and Gefen, Yuval and Potters, Marc and Wyart, Matthieu},
  title   = {Fluctuations and Response in Financial Markets: The Subtle Nature of `Random' Price Changes},
  journal = {Quantitative Finance},
  volume  = {4},
  number  = {2},
  pages   = {176--190},
  year    = {2004},
  doi     = {10.1088/1469-7688/4/2/007}
}

@article{toth2015equity,
  author  = {T{\'o}th, Bence and Palit, Imon and Lillo, Fabrizio and Farmer, J. Doyne},
  title   = {Why is Equity Order Flow so Persistent?},
  journal = {Journal of Economic Dynamics and Control},
  volume  = {51},
  pages   = {218--239},
  year    = {2015},
  doi     = {10.1016/j.jedc.2014.11.007}
}

@article{lillo2003master,
  author  = {Lillo, Fabrizio and Farmer, J. Doyne and Mantegna, Rosario N.},
  title   = {Master Curve for Price-Impact Function},
  journal = {Nature},
  volume  = {421},
  number  = {6919},
  pages   = {129--130},
  year    = {2003},
  doi     = {10.1038/421129a}
}

@article{bouchaud2009price,
  author  = {Bouchaud, Jean-Philippe and Farmer, J. Doyne and Lillo, Fabrizio},
  title   = {How Markets Slowly Digest Changes in Supply and Demand},
  journal = {Handbook of Financial Markets: Dynamics and Evolution},
  pages   = {57--160},
  year    = {2009},
  publisher = {Elsevier},
  doi     = {10.1016/B978-012374258-2.50006-3}
}

@article{cont2014price,
  author  = {Cont, Rama and Kukanov, Arseniy and Stoikov, Sasha},
  title   = {The Price Impact of Order Book Events},
  journal = {Journal of Financial Econometrics},
  volume  = {12},
  number  = {1},
  pages   = {47--88},
  year    = {2014},
  doi     = {10.1093/jjfinec/nbt003}
}

@inproceedings{shrikumar2017learning,
  author    = {Shrikumar, Avanti and Greenside, Peyton and Kundaje, Anshul},
  title     = {Learning Important Features Through Propagating Activation Differences},
  booktitle = {Proceedings of the 34th International Conference on Machine Learning},
  pages     = {3145--3153},
  year      = {2017},
  volume    = {70},
  series    = {Proceedings of Machine Learning Research},
  publisher = {PMLR}
}

@incollection{kilian2017nonlinear,
  author    = {Kilian, Lutz and L{\"u}tkepohl, Helmut},
  title     = {Nonlinear Structural {VAR} Models},
  booktitle = {Structural Vector Autoregressive Analysis},
  publisher = {Cambridge University Press},
  address   = {Cambridge},
  pages     = {609--658},
  year      = {2017},
  chapter   = {18},
  doi       = {10.1017/9781108164818.019}
}

@article{cordoni2024identification,
  author  = {Cordoni, Francesco and Doremus, Nicolas and Moneta, Alessio},
  title   = {Identification of Vector Autoregressive Models with Nonlinear Contemporaneous Structure},
  journal = {Journal of Economic Dynamics and Control},
  volume  = {161},
  pages   = {104846},
  year    = {2024},
  doi     = {10.1016/j.jedc.2024.104846}
}

\appendix

\section{Residual Contemporaneous Dependence and Structural Identification}
\label{app:residual_structural_identification}

In this Appendix we provide additional details on the residual structural
interpretation used in the main text. The key point is that the reduced-form
models considered in the paper use lagged information only. Therefore,
contemporaneous effects are not included as ordinary regressors, but appear as
dependence between reduced-form residuals.

The reduced-form model can be written as
\begin{equation}
\begin{split}
r_t &= f_r(\mathcal{X}_t) + u_t^r,\\
v_t &= f_v(\mathcal{X}_t) + u_t^v,
\end{split}
\label{eq:app_reduced_form_residuals}
\end{equation}
where $\mathcal{X}_t$ denotes the lagged information set, while $u_t^r$ and
$u_t^v$ are the reduced-form innovations. The Hasbrouck-type
structural interpretation is obtained by imposing a recursive contemporaneous
ordering. In trade time, the signed volume is observed at the event and can be
regarded as causally prior to the associated quote revision. This motivates
\begin{equation}
\begin{split}
v_t &= f_v(\mathcal{X}_t) + \epsilon_t^v,\\
r_t &= f_r(\mathcal{X}_t) + \psi(\epsilon_t^v) + \epsilon_t^r.
\end{split}
\label{eq:app_structural_triangular_general}
\end{equation}
In the linear Hasbrouck model, \(\psi(\epsilon_t^v)=b_0\epsilon_t^v\). Hence,
\begin{equation}
u_t^v=\epsilon_t^v,
\qquad
u_t^r=b_0u_t^v+\epsilon_t^r.
\label{eq:app_linear_residual_decomposition}
\end{equation}
Thus, the contemporaneous impact coefficient is recovered from the residual
dependence between \(u_t^r\) and \(u_t^v\), rather than by inserting \(v_t\)
directly into the reduced-form forecasting equation. This is the usual logic of
structural VAR identification: reduced-form residuals are correlated because
they are mixtures of underlying structural shocks, and additional restrictions
are required to recover those shocks~\cite{hasbrouck1991measuring,kilian2017nonlinear}.

Directly fitting the return equation by inserting \(v_t\) as an ordinary
contemporaneous regressor is generally not the cleanest approach when the goal
is to separate lagged predictability from instantaneous impact. In the
reduced-form formulation, the model estimates
\[
 f_r(\mathcal{X}_t)\simeq \mathbb{E}[r_t\mid \mathcal{X}_t],
\qquad
 f_v(\mathcal{X}_t)\simeq \mathbb{E}[v_t\mid \mathcal{X}_t],
\]
and the residuals are the corresponding innovations,
\[
u_t^r
=
r_t-\mathbb{E}[r_t\mid \mathcal{X}_t],
\qquad
u_t^v
=
v_t-\mathbb{E}[v_t\mid \mathcal{X}_t].
\]
Thus, \(u_t^v\) represents the unexpected component of order flow, namely the
part of \(v_t\) that is not predictable from past returns and volumes.

By contrast, if one fits
\[
r_t = G(\mathcal{X}_t,v_t)+\eta_t,
\]
then the model receives as input
\[
v_t
=
\mathbb{E}[v_t\mid \mathcal{X}_t]
+
u_t^v.
\]
Hence, the return equation receives both the predictable component of order
flow, already determined by lagged information, and the genuinely
contemporaneous innovation \(u_t^v\). These two effects are mixed inside the
same function \(G\). This can lead to an incorrect interpretation of the learned
dependence: what appears to be an instantaneous effect of \(v_t\) may partly
come from its predictable lagged component, while the true contemporaneous
innovation is not isolated.

In a linear model this mixing is less problematic, because the predictable
component can be reabsorbed into the reduced-form lagged predictor. Suppose,
for example, that the structural model is
\begin{equation}
\begin{split}
v_t &= f_v(\mathcal{X}_t)+\epsilon_t^v,\\
r_t &= g_r(\mathcal{X}_t) + b_0 v_t + \epsilon_t^r.
\end{split}
\label{eq:app_linear_direct_contemporaneous}
\end{equation}
Substituting the first equation into the second gives
\begin{equation}
r_t
=
g_r(\mathcal{X}_t)
+
b_0 f_v(\mathcal{X}_t)
+
b_0\epsilon_t^v
+
\epsilon_t^r.
\label{eq:app_linear_substitution}
\end{equation}
The predictable component \(f_v(\mathcal{X}_t)\) is absorbed into the lagged
reduced-form predictor, while the innovation
\(\epsilon_t^v=u_t^v\) remains in the residual component. Therefore, in the
linear case, the direct contemporaneous formulation and the reduced-form
residual formulation are tightly connected.

This equivalence is much less transparent in nonlinear models. If one directly
fits
\begin{equation}
r_t
=
G\!\left(\mathcal{X}_t,v_t\right)
+
\eta_t,
\label{eq:app_direct_nonlinear_contemporaneous}
\end{equation}
then, since \(v_t=f_v(\mathcal{X}_t)+u_t^v\), the nonlinear function \(G\) acts
jointly on the predictable component of order flow and on its contemporaneous
innovation:
\[
G\!\left(\mathcal{X}_t, f_v(\mathcal{X}_t)+u_t^v\right).
\]
In general, this expression cannot be decomposed uniquely into a purely lagged
term plus a purely contemporaneous innovation term. As a result, direct
conditioning on \(v_t\) may mix lagged predictable order-flow effects with
instantaneous impact. This is precisely the effect that the residual
decomposition avoids.

For this reason, especially in nonlinear models, we estimate first the
reduced-form predictive model using only lagged information,
\begin{equation}
x_t=f_\theta(\mathcal{X}_t)+u_t,
\qquad
x_t=(r_t,v_t),
\label{eq:app_nonlinear_reduced_form}
\end{equation}
and then analyze the residual dependence. Even when \(f_\theta\) is a single
vector-valued nonlinear map, the residual vector can be written componentwise as
\[
u_t=
\begin{pmatrix}
u_t^r\\
u_t^v
\end{pmatrix}
=
\begin{pmatrix}
r_t- f_r(\mathcal{X}_t)\\
v_t- f_v(\mathcal{X}_t)
\end{pmatrix}.
\]
A residual dependence between \(u_t^r\) and \(u_t^v\) indicates that, after
removing the lagged predictable dynamics, an instantaneous relation remains
between unexpected order flow and unexpected price changes.

In the most general nonlinear structural representation, one could write the
reduced-form residuals as nonlinear functions of structural shocks,
\begin{equation}
u_t
=
H_t(\epsilon_t^r,\epsilon_t^v).
\label{eq:app_general_nonlinear_residual_map}
\end{equation}
Recovering the shocks then requires identifying restrictions on \(H_t\). This
is precisely the point emphasized by nonlinear structural VAR models: when
contemporaneous relations are nonlinear, the mapping from reduced-form
innovations to structural shocks is no longer characterized by a constant impact
matrix, and identification requires additional assumptions~\cite{kilian2017nonlinear,cordoni2024identification}.

In our market microstructure setting, we impose the same recursive economic
ordering used in the Hasbrouck framework: order-flow innovations are allowed to
affect return innovations contemporaneously, but not vice versa within the same
trade-time event. This leads to
\begin{equation}
u_t^v=\epsilon_t^v,
\qquad
u_t^r=\phi(u_t^v)+\epsilon_t^r.
\label{eq:app_nonlinear_residual_structural}
\end{equation}
The function \(\phi\) captures the contemporaneous impact of the unexpected
component of signed volume on the unexpected component of returns.

This decomposition also clarifies the distinction between reduced-form
forecasting and structural conditioning. If one is interested in pure
forecasting before the event at time \(t\), the relevant prediction is
\[
\widehat r_t^{\,\mathrm{lagged}}
=
 f_r(\mathcal{X}_t),
\]
which uses only information available up to \(t-1\). This is the object used
when comparing the predictive performance of the VAR, the nonlinear parametric
model, and the neural network. By contrast, if the signed volume \(v_t\) is
observed before the associated quote revision, one can form
\[
\widehat r_t^{\,\mathrm{struct}}
=
 f_r(\mathcal{X}_t)
+
\hat\phi\!\left(
v_t- f_v(\mathcal{X}_t)
\right),
\]
which is a prediction conditional on the contemporaneous realization of order
flow, rather than a purely lagged forecast.

\section{Lag-order dependence of forecasting performance}
\label{app:lag_selection}

\begin{figure}[!htb]
\centering
\begin{subfigure}{0.44\textwidth}
  \centering
  \includegraphics[width=\linewidth]{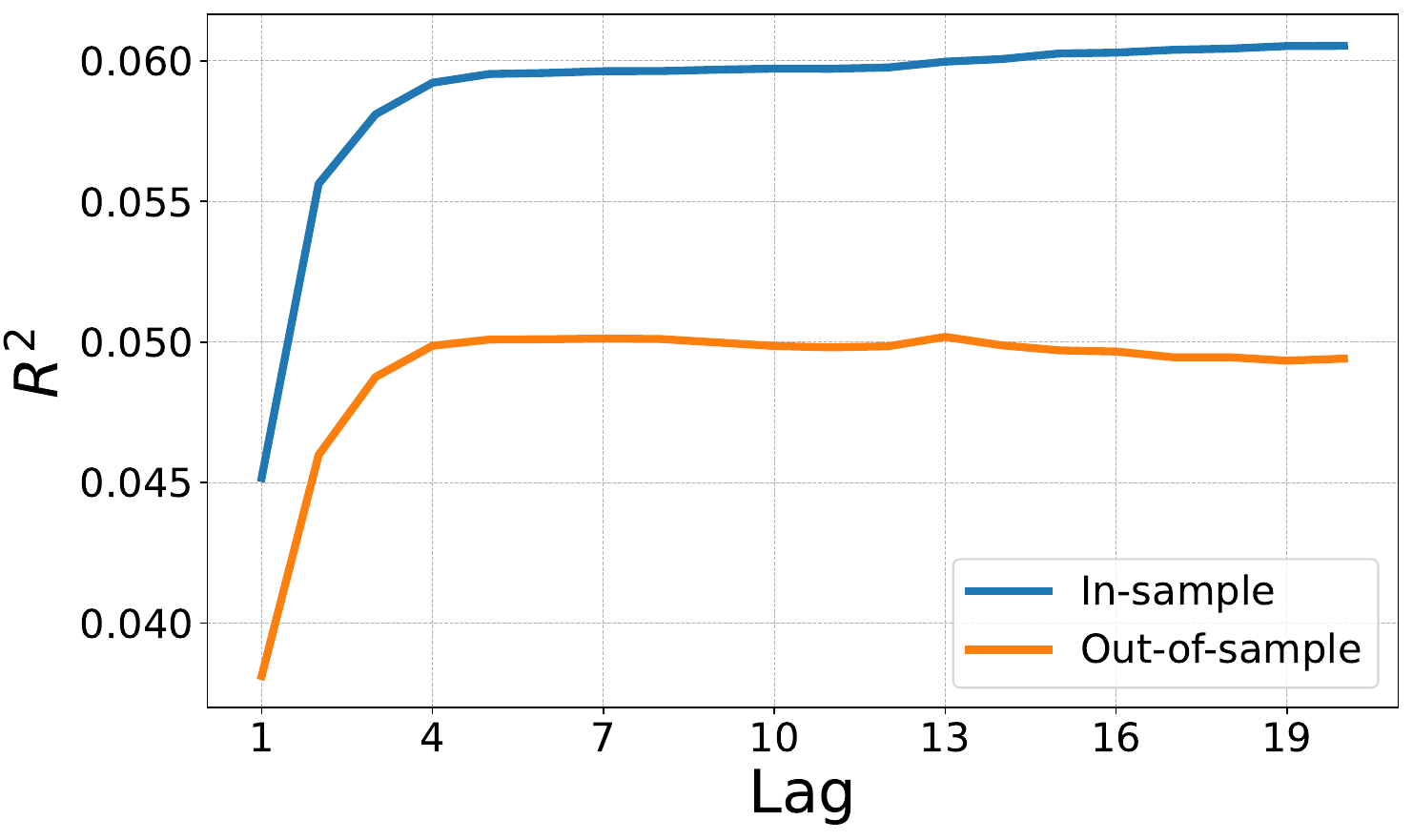}
\end{subfigure}
\begin{subfigure}{0.44\textwidth}
  \centering
  \includegraphics[width=\linewidth]{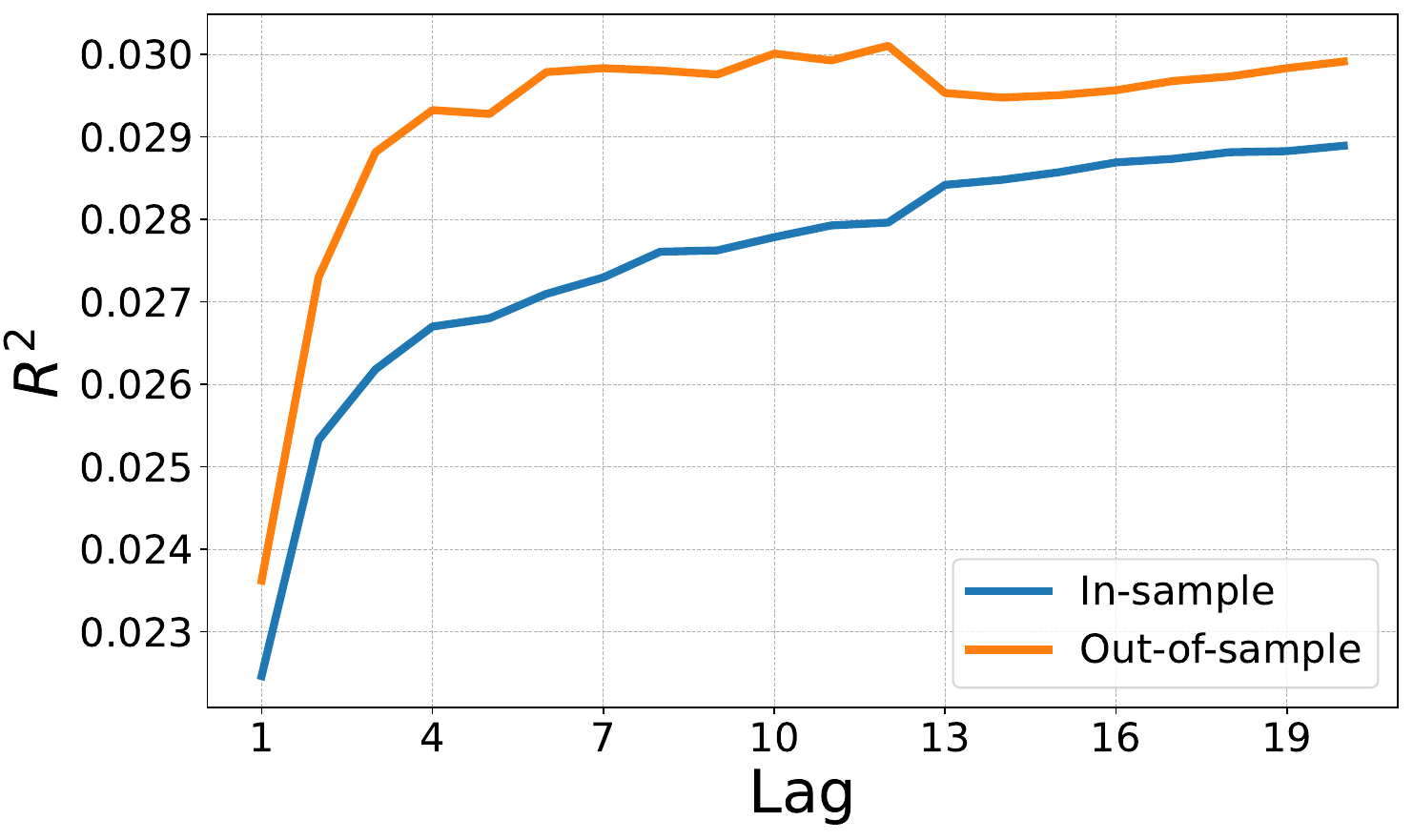}
\end{subfigure}
\caption{Predictive performance of VAR model of Eq.~\eqref{hasbrouckmodel} as a function of the lag order $p$ for BAC. The panels report the in-sample and out-of-sample ($R^2$) as functions of the lag order. BAC is shown as a representative example, since the same qualitative behavior is observed across the other stocks in the dataset.}
\label{fig:VAR_results_BAC}
\end{figure}

In this appendix we provide a more detailed discussion of the dependence of the
forecasting performance on the lag order \(p\). The purpose of this analysis is
not to identify an asset-specific optimal lag length, but rather to verify whether
increasing the memory of the model leads to systematic predictive gains. This
serves as a robustness check for the choice \(p=20\) adopted in the main text.

For each value of \(p\), the models are trained using the lagged state vector
\begin{equation}
    X_t^{(p)}
    =
    \left(
    r_{t-1},v_{t-1},
    r_{t-2},v_{t-2},
    \ldots,
    r_{t-p},v_{t-p}
    \right),
\end{equation}
where \(r_t\) denotes returns and \(v_t\) denotes signed volumes. We then evaluate
the in-sample and out-of-sample \(R^2\) for the prediction of the target variables.
This allows us to compare how the amount of past information included in the
model affects predictive performance.

Figure~\ref{fig:VAR_results_BAC} reports the lag-dependence for the VAR
benchmark in the case of the BAC stock. The behavior is rather regular. At small lag orders, increasing \(p\)
substantially improves both in-sample and out-of-sample performance, since the
model is progressively allowed to exploit short-term linear dependencies in the
joint dynamics of returns and signed volumes. After the first few lags, however,
the out-of-sample \(R^2\) reaches a relatively stable level and further increases
in \(p\) do not lead to systematic improvements. In some cases, the out-of-sample
performance even slightly decreases at larger lag orders.

This behavior is natural for a linear VAR model. Increasing the lag order amounts
to adding additional linear regressors to the specification. Once the dominant
short-lag dependencies have been included, the marginal predictive content of
further lagged variables becomes small. At the same time, the number of parameters
to estimate increases with \(p\). Longer lag structures may therefore introduce
additional estimation noise without providing a compensating increase in useful
predictive information. The resulting pattern is an initial improvement followed
by a saturation of the out-of-sample performance. A similar qualitative behavior is observed for the other stocks in the sample. The BAC results are therefore reported as a representative illustration of the general lag-dependence pattern, rather than as an asset-specific finding.

\begin{figure}[!htb]
\centering
\begin{subfigure}{0.44\textwidth}
  \centering
  \includegraphics[width=\linewidth]{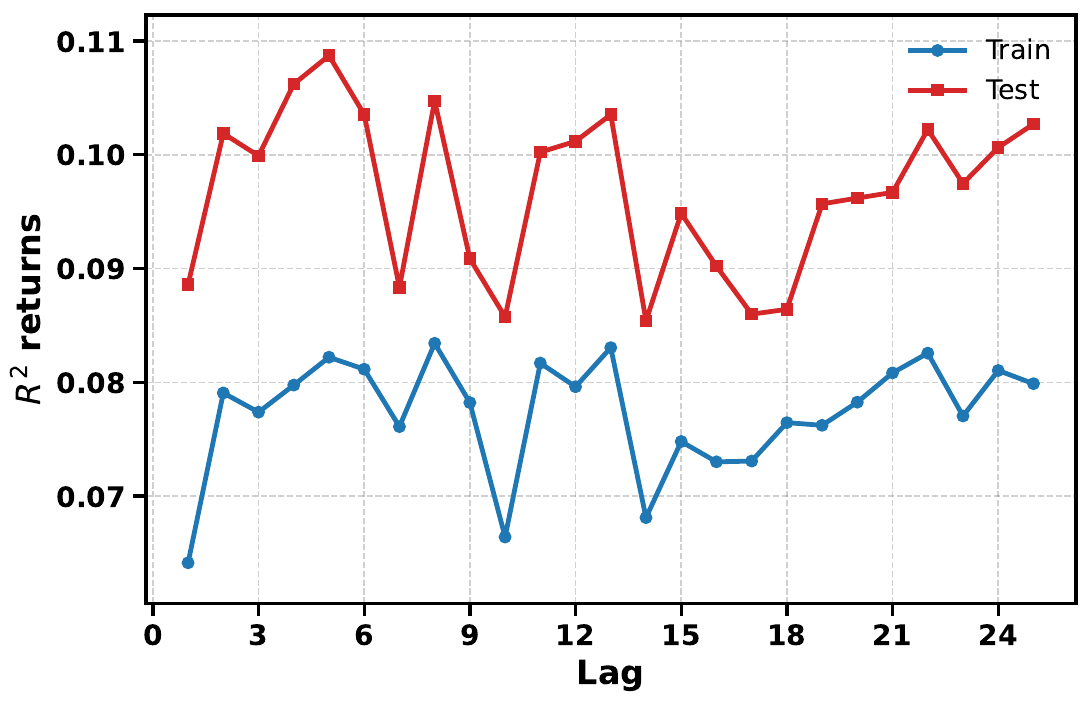}
\end{subfigure}
\begin{subfigure}{0.44\textwidth}
  \centering
  \includegraphics[width=\linewidth]{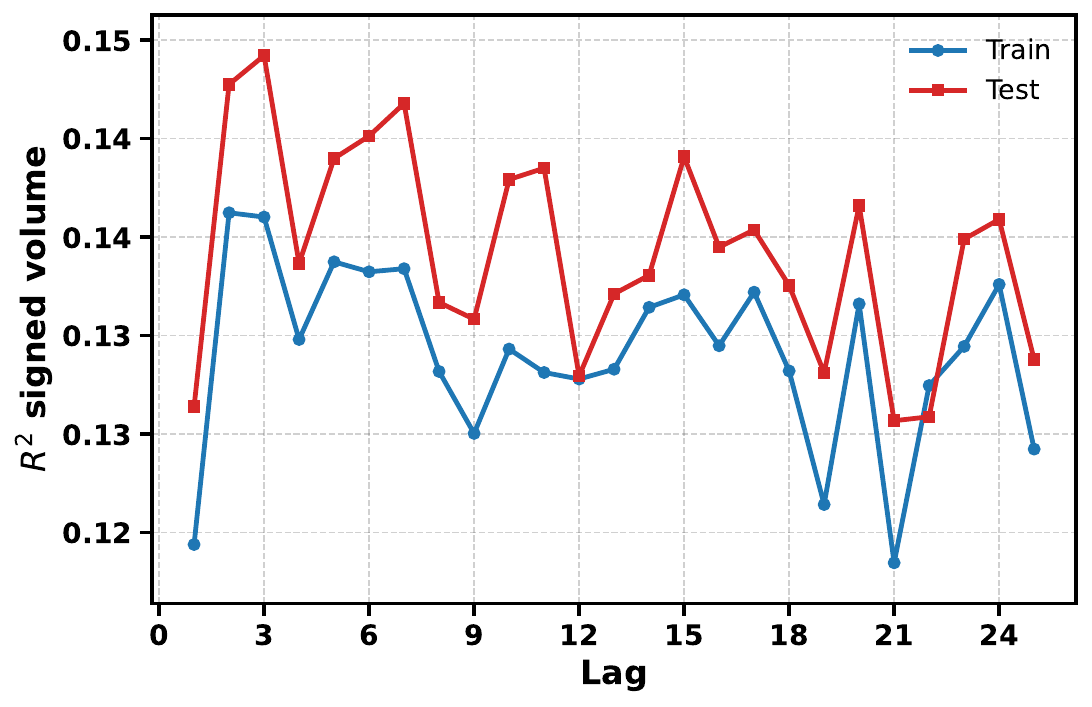}
\end{subfigure}
\caption{Lag-dependent predictive performance for AAPL using the neural network model. The left panel reports the in-sample and out-of-sample ($R^2$) for returns as a function of the lag order, while the right panel shows the corresponding results for signed volumes. AAPL is shown as a representative example, as similar qualitative patterns are observed across all stocks.}
\label{fig:NN_AAPL_lag}
\end{figure}

The neural-network model displays a related, but less regular, behavior, as shown
in Figure~\ref{fig:NN_AAPL_lag}. In this case, increasing the lag order does
not produce a smooth monotonic improvement of the test \(R^2\). After the first
lags, the performance fluctuates around a comparable range of values, with local
increases and decreases depending on the specific lag length. These oscillations
are more pronounced than in the VAR case and should not be interpreted as evidence
that particular lag orders are structurally privileged.
\begin{figure}[!htb]
    \centering
    \includegraphics[width=0.45\textwidth]{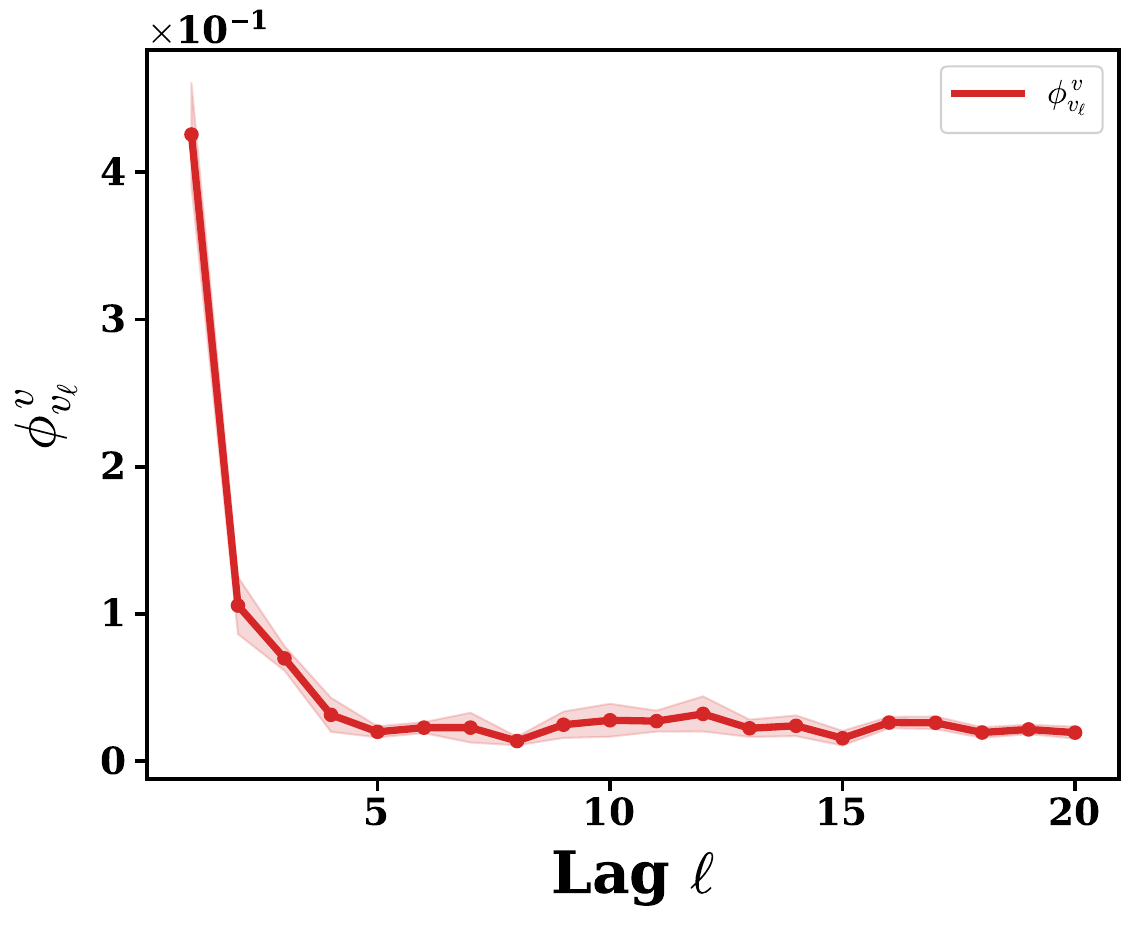}
    \includegraphics[width=0.45\textwidth]{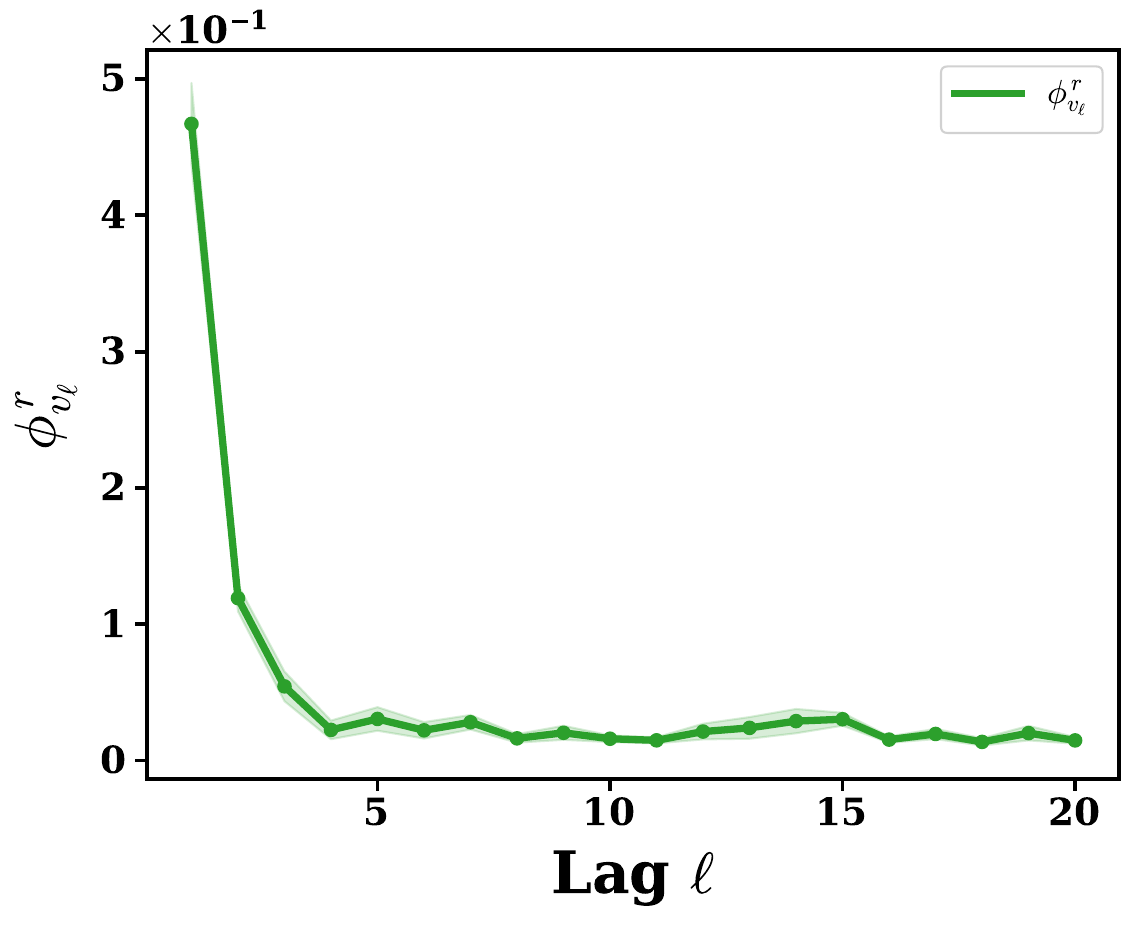}
    
    \includegraphics[width=0.45\textwidth]{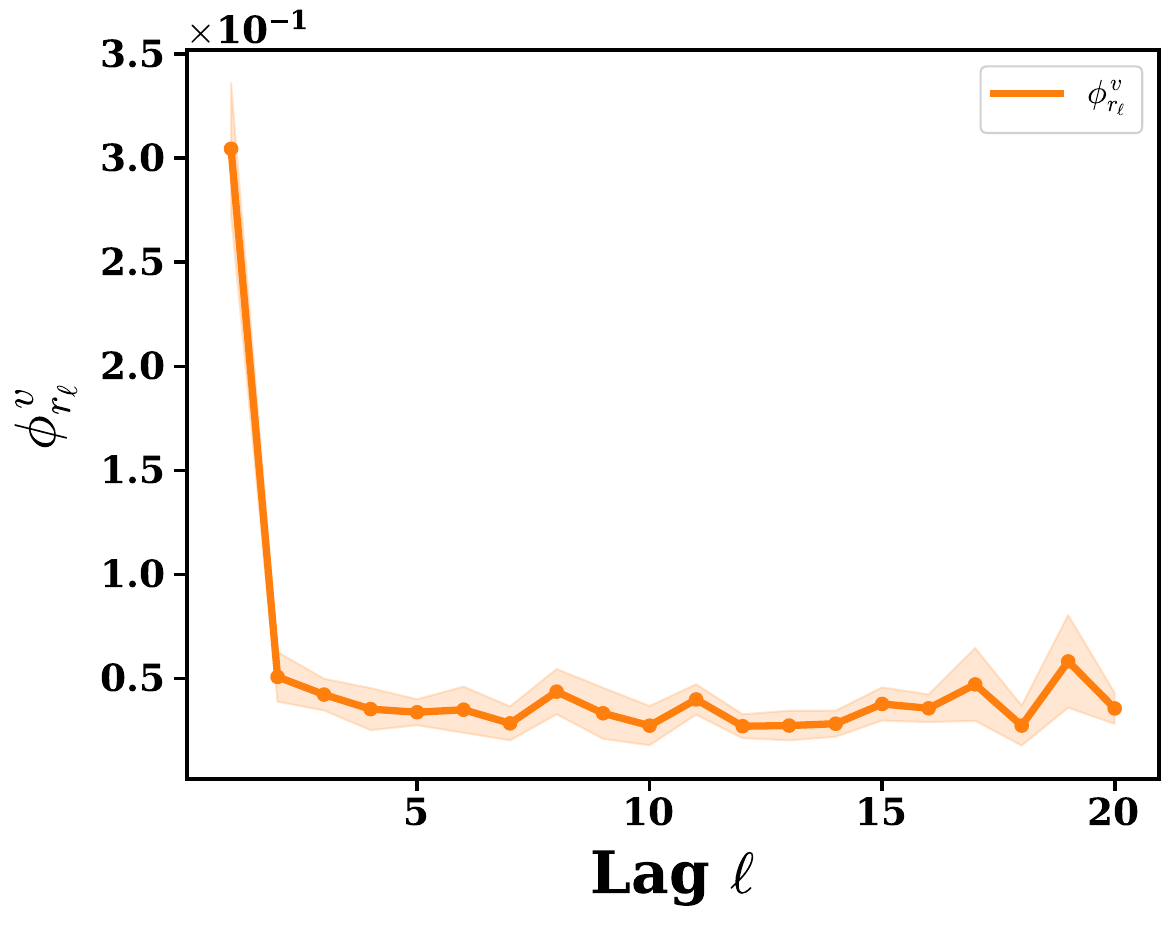}
    \includegraphics[width=0.45\textwidth]{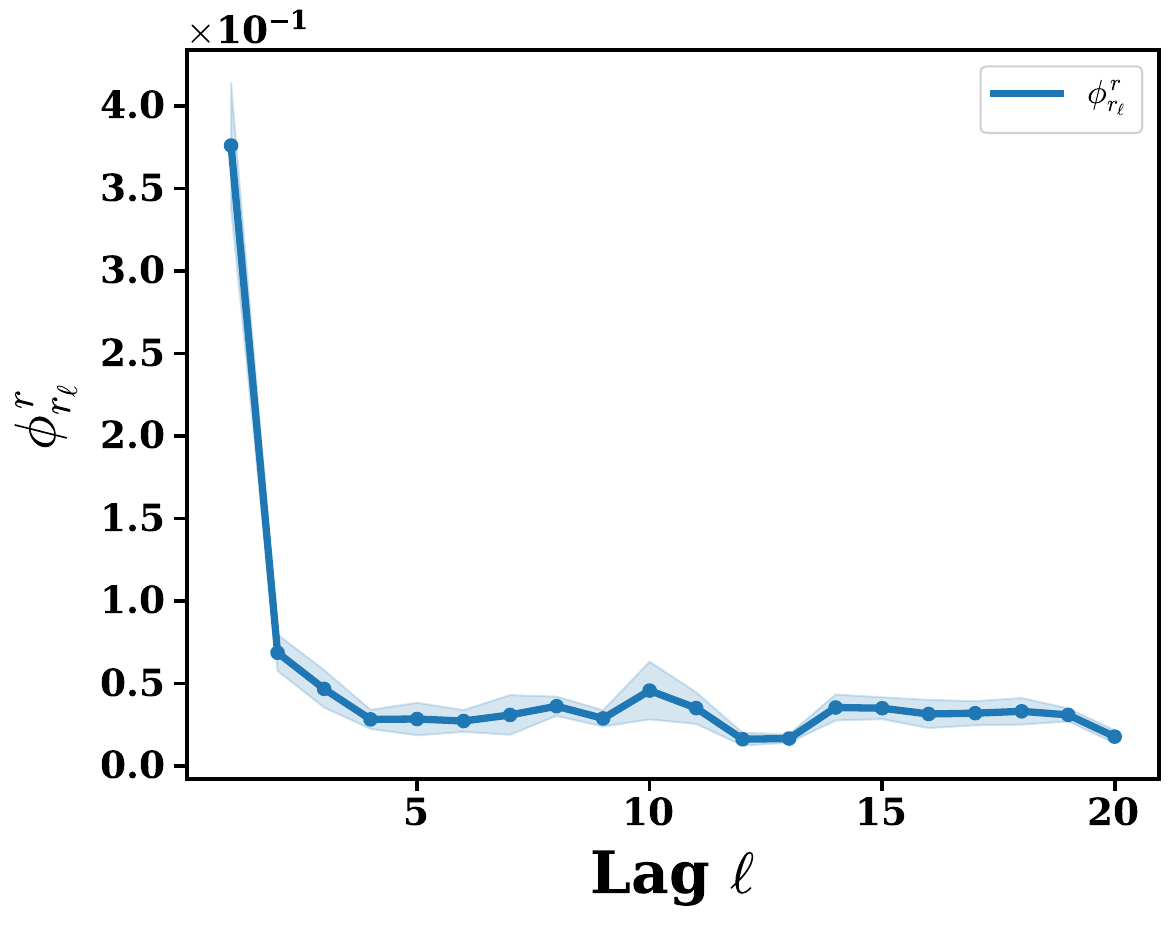}
    \caption{Normalized lag importance for large-tick stocks (LT). Each curve
    represents the average fraction of total Shapley magnitude explained by lag
    \(\ell\).}
    \label{fig:lt_shap_lags}
\end{figure}

\begin{figure}[!htb]
    \centering
    \includegraphics[width=0.45\textwidth]{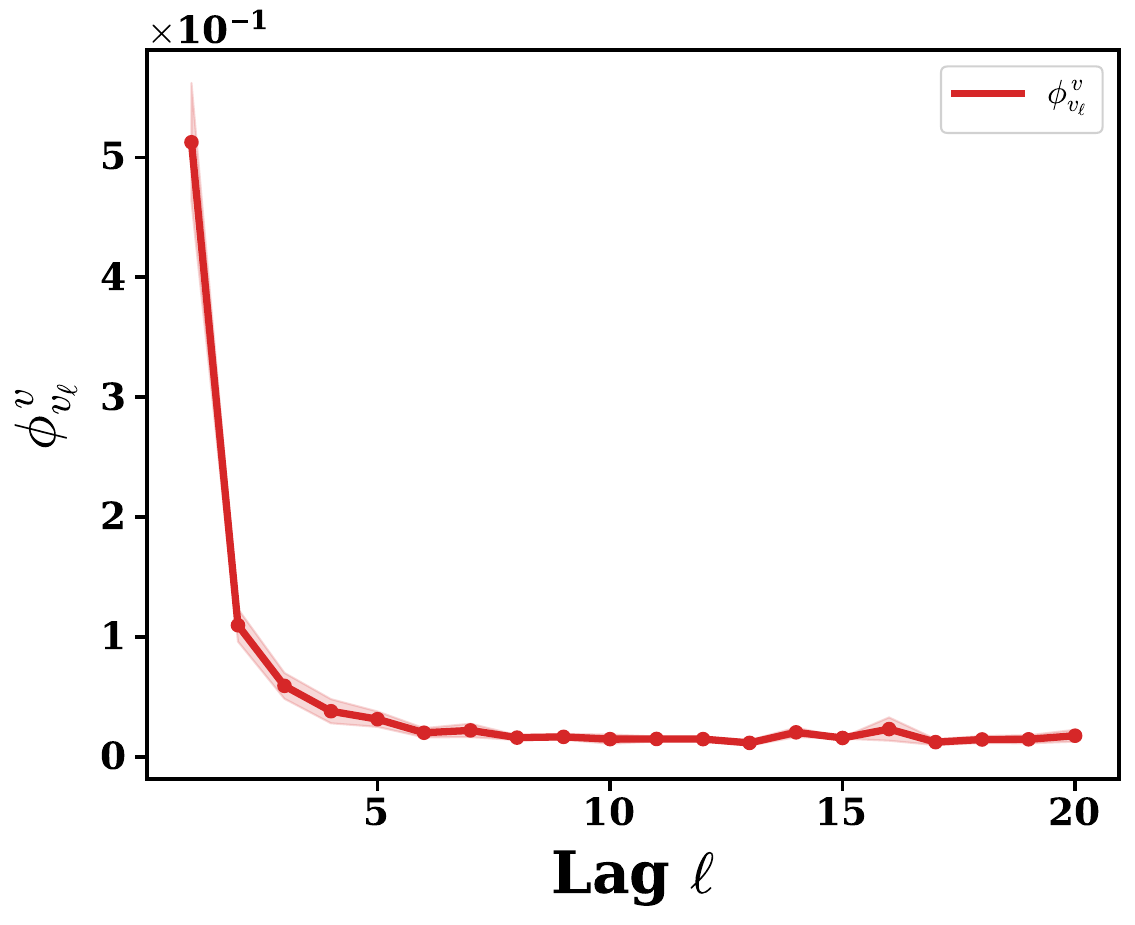}
    \includegraphics[width=0.45\textwidth]{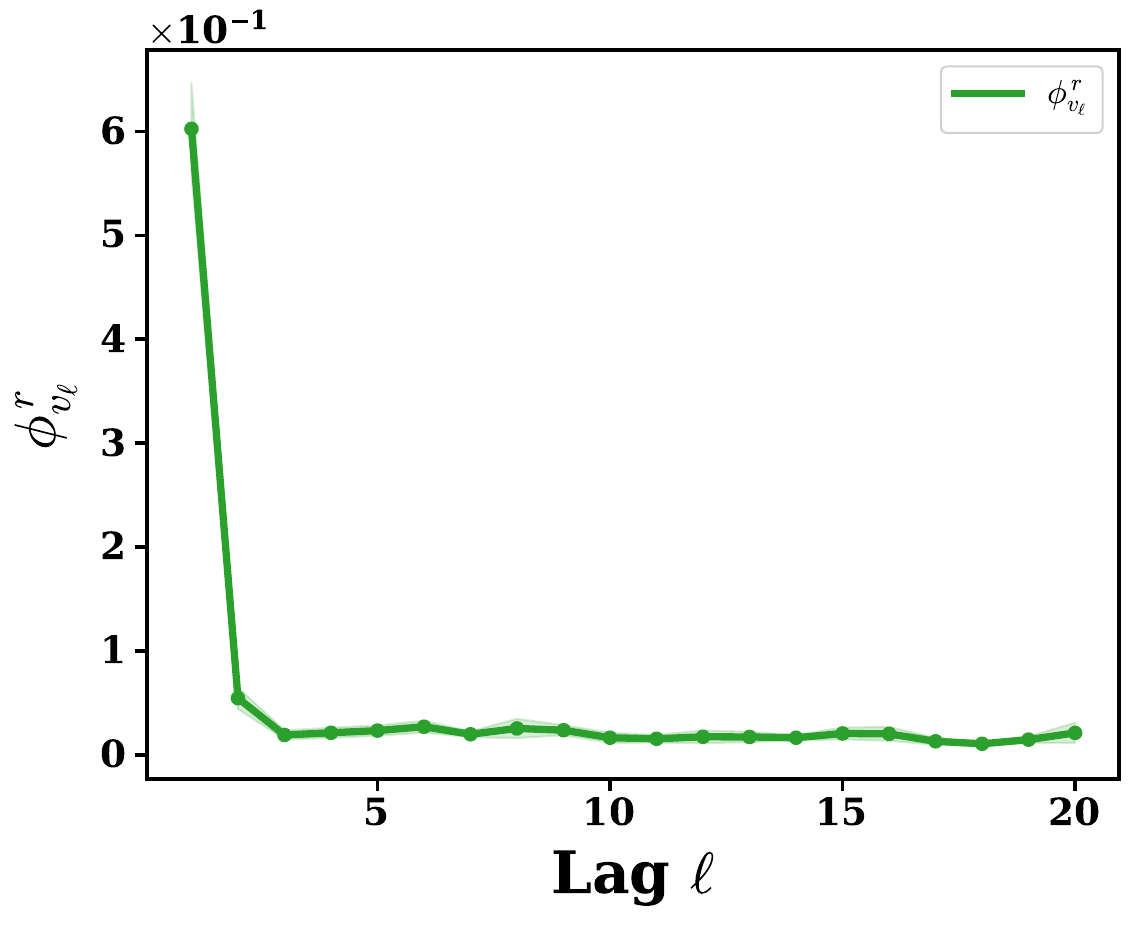}
    
    \includegraphics[width=0.45\textwidth]{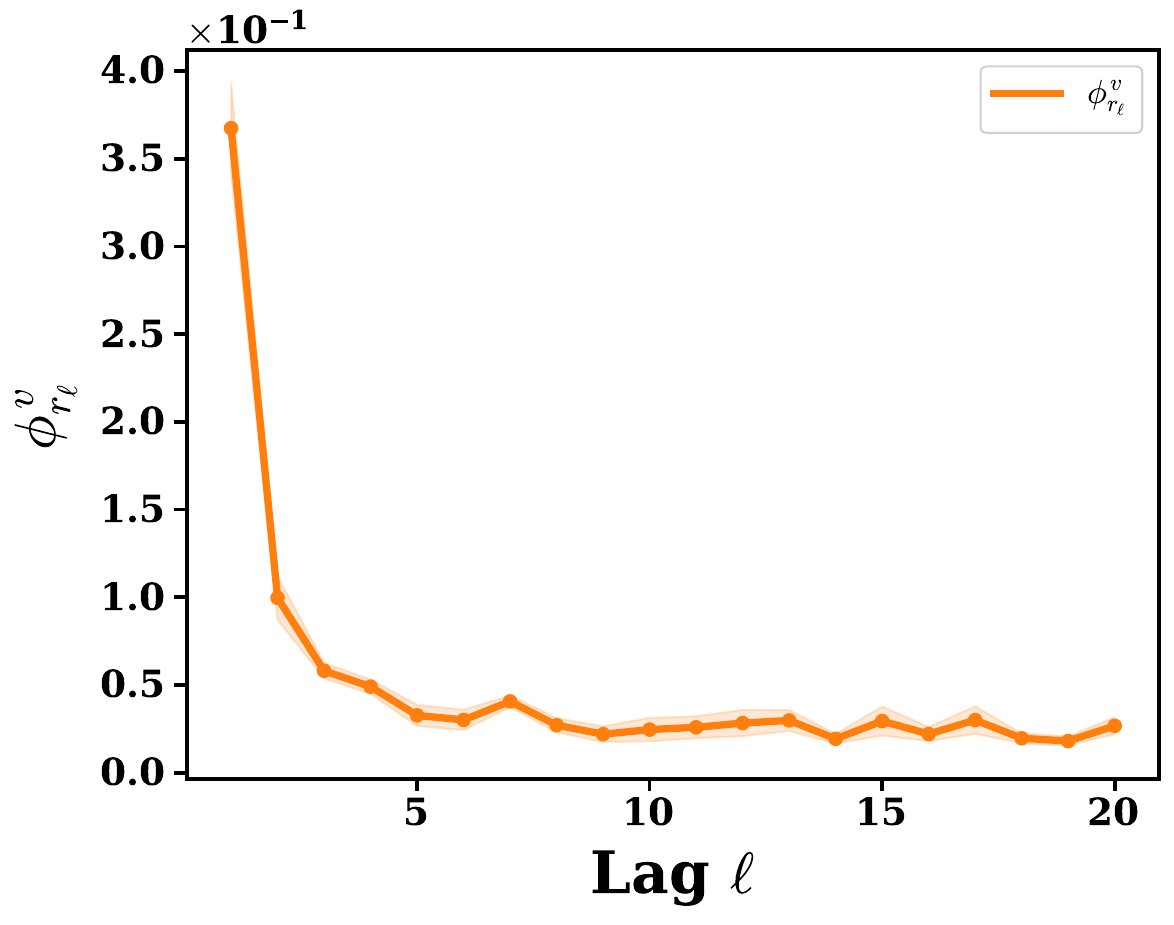}
    \includegraphics[width=0.45\textwidth]{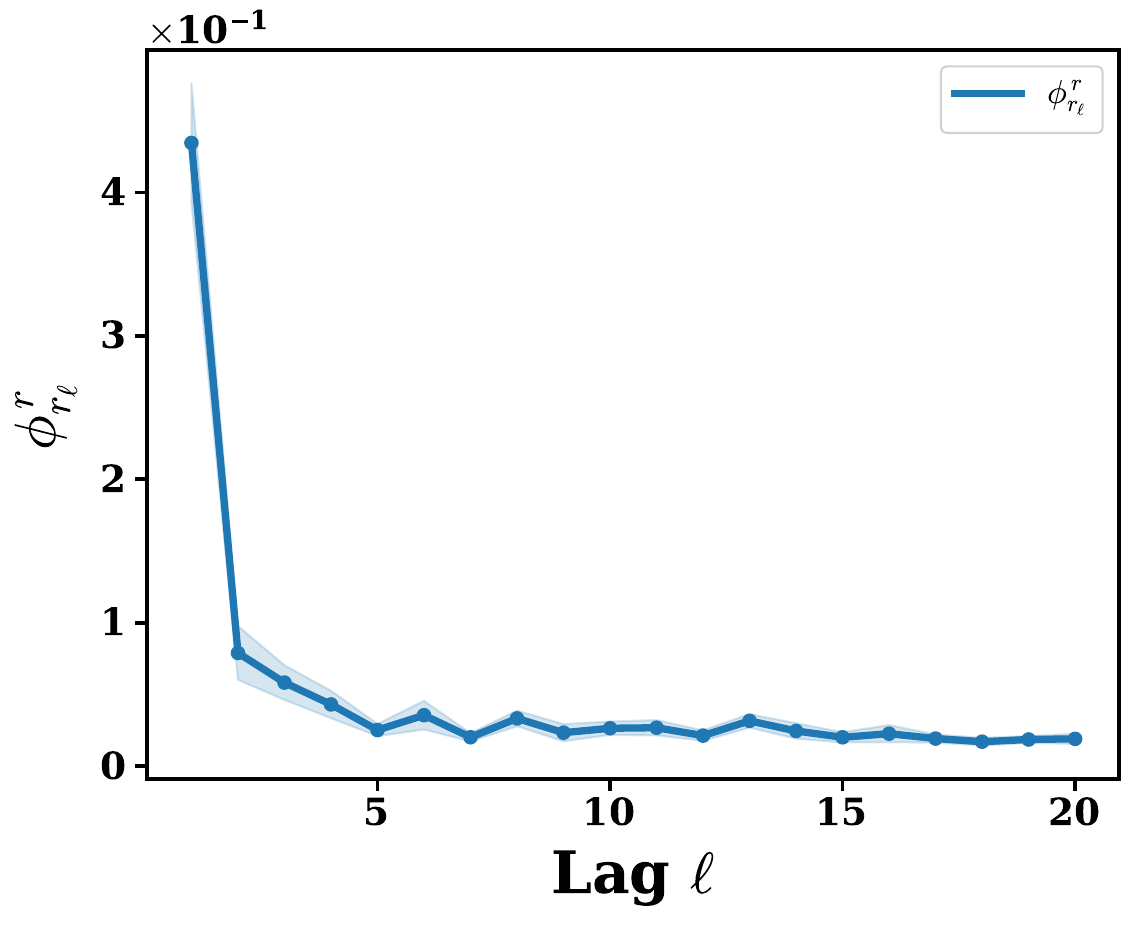}
    \caption{Normalized lag importance for small-tick stocks (ST). Each curve
    represents the average fraction of total Shapley magnitude explained by lag
    \(\ell\).}
    \label{fig:st_shap_lags}
\end{figure}
The reason is that changing \(p\) in the neural-network setting does more than
simply adding a few additional coefficients. For each lag order, the model is
trained on a different input representation \(X_t^{(p)}\). Therefore, the network
does not only receive a longer history, but also explores a different nonlinear
mapping from the lagged state variables to the target variable. Additional distant
lags may contain weak residual information, especially for signed volumes, but
they can also act as noisy inputs. Because the model is flexible, such weak signals
can interact nonlinearly with the most recent variables and with finite-sample
fluctuations. This can generate a noisy dependence of the test performance on
\(p\), even when there is no genuine systematic gain from using longer histories.

This distinction also explains why the VAR curves are typically smoother than the
neural-network curves. In the linear case, the effect of increasing \(p\) is more
controlled: the model class changes by adding linear terms. In the nonlinear case,
instead, the effective model explored by training can vary more substantially
across lag orders, because the network can combine the additional inputs through
nonlinear interactions. Hence, a larger lag order does not necessarily imply a
better forecast, even if the training set contains more explanatory variables.

Overall, the lag-dependence analysis supports two conclusions. First, the relevant
predictive memory of the system is concentrated at relatively short horizons:
after the initial lags, increasing \(p\) does not provide robust out-of-sample
improvements. Second, the noisy oscillations observed for the neural network are
consistent with the interaction between weak long-lag signals, nonlinear model
flexibility and finite-sample estimation effects, rather than with the existence
of sharply optimal lag lengths.

On the basis of these results, we use \(p=20\) in the main analysis. This choice is
conservative: it is large enough to include the short-memory effects that drive the
initial improvement in predictive performance, but not so large as to excessively
increase the input dimension and the number of parameters. Moreover, the absence of
systematic gains at larger lag orders indicates that the main empirical conclusions
are not driven by a fine tuning of the memory length.

\section{Shapley Lag-Importance Profiles}
\label{app:shap_lag_importance}

To quantify how the explanatory contribution of the neural network is
distributed across lags, we construct an aggregated measure of lag importance
based on mean absolute Shapley values.

For each ticker \(i\) and each prediction block, for example
\(r\rightarrow r_t\), \(v\rightarrow r_t\), \(r\rightarrow v_t\), or
\(v\rightarrow v_t\), we compute
\begin{equation}
I_i(\ell)
=
\mathbb{E}\big[|\phi_{j_\ell}^{f}(x_t)|\big],
\end{equation}
where \(j_\ell\) denotes the input feature corresponding to lag \(\ell\),
either \(r_{t-\ell}\) or \(v_{t-\ell}\), and the expectation is taken over all
observations for that ticker.

Since raw Shapley magnitudes are not directly comparable across tickers and
outputs, we normalize them within each ticker and block:
\begin{equation}
\tilde I_i(\ell)
=
\frac{I_i(\ell)}
{\sum_{k=1}^{p} I_i(k)}.
\label{eq:app_normalized_shap_importance}
\end{equation}
By construction,
\[
\sum_{\ell=1}^{p}\tilde I_i(\ell)=1,
\]
so that \(\tilde I_i(\ell)\) can be interpreted as the fraction of the total
Shapley magnitude attributed to lag \(\ell\). We then average these profiles
across tickers, separately for large-tick and small-tick stocks:
\begin{equation}
\bar I(\ell)
=
\frac{1}{N}
\sum_{i=1}^{N}
\tilde I_i(\ell).
\end{equation}

Figures~\ref{fig:lt_shap_lags} and~\ref{fig:st_shap_lags} report the resulting
profiles for LT and ST stocks. The first lag dominates across all prediction
blocks and both tick-size regimes. This suggests that most of the nonlinear
structure learned by the neural network is concentrated in the most recent
return--volume pair \((r_{t-1},v_{t-1})\).

At the same time, the decay is not identical across channels. Lagged signed
volumes generally display a more persistent profile than lagged returns,
especially in the \(v\rightarrow v_t\) block, consistently with the persistence
of order flow. Lagged returns instead tend to decay more rapidly, indicating
that their predictive contribution is mainly concentrated at very short
horizons.

The comparison between LT and ST stocks also shows some differences. In LT
stocks, volume lags retain a relatively more gradual contribution in the
\(v\rightarrow r_t\) and \(v\rightarrow v_t\) blocks. In ST stocks, the effect
of lagged volumes on returns is more concentrated at the first lag, while
lagged returns contribute somewhat more persistently to future signed volumes.
This suggests that the interaction between past price changes and subsequent
trading activity is richer in small-tick assets.

Overall, these results justify focusing the main predictive-surface analysis on
the first lag, while also motivating the multi-lag parametric extension with
decaying lag weights introduced in the main text.

\section{Empirical surfaces and validation}
\label{sec:data_nonlinear}

The binning procedure depends on the variable placed on the horizontal axis. When the response is plotted as a function of \(v_{t-1}\), the support of signed volumes is partitioned into uniform bins within each ticker. Since these bins are defined over ticker-specific ranges, they are not directly comparable across assets. When the response is plotted as a function of \(r_{t-1}\), uniform binning is less appropriate, as returns are concentrated around a finite number of clusters due to price discreteness. In this case, a Gaussian kernel density estimator is used to identify local maxima of the empirical distribution, and bins are defined through the intervals between consecutive peaks. This results in a data-driven clustering of returns. Each ticker-specific curve is then interpolated linearly on a common grid, separately for each conditioning regime. At each grid point, the cross-sectional average of the model prediction is computed across tickers. To avoid distortions from sparsely populated regions, only bins with at least 150 average observations are retained in the final plots, while the histograms are kept complete. The shaded bands represent cross-sectional uncertainty and are constructed as \(\pm\) one standard error of the mean across tickers. These bands capture the dispersion of the estimated relationship across assets and should not be interpreted as confidence intervals at the observation level.

The results of the previous analysis suggest that the predictive relationships captured by the model are strongly nonlinear and state-dependent. In this section, we want to validate these results by tracing them back to empirical regularities directly observed in the data. 

Of course, when relying only on the observed data, we can recover only the empirical analogues of the aggregated conditional response surfaces defined in Eqs.~\eqref{eq:model_surface_v_binned} and~\eqref{eq:model_surface_r_binned}. What is lost is the additive attribution of the prediction across the different input variables, which is provided by the Shapley decomposition. In this sense, a well-calibrated model that captures genuine regularities in the data can be used not only for prediction, but also as an interpretative device: it allows us to decompose the observed predictive structure and to assess the relative contribution of returns, signed volumes, and the remaining lagged variables.

The empirical results are constructed starting from the time series of returns $r_t$ and signed volumes $v_t$, considered separately for large-tick (LT) and small-tick (ST) stocks.

We then study bivariate relationships between a response variable at time $t$ and a lagged explanatory variable at time $t-1$, considering the four configurations:
\[
(r_t \text{ vs } r_{t-1}), \quad
(r_t \text{ vs } v_{t-1}), \quad
(v_t \text{ vs } r_{t-1}), \quad
(v_t \text{ vs } v_{t-1}).
\]

In addition to unconditional relationships, we also consider conditional ones, splitting the sample according to the sign of the second lagged variable. In particular:
\begin{itemize}
    \item when conditioning on $r_{t-1}$, we consider the regimes $r_{t-1}<0$, $r_{t-1}=0$, and $r_{t-1}>0$;
    \item when conditioning on $v_{t-1}$, we consider the regimes $v_{t-1}<0$ and $v_{t-1}>0$.
\end{itemize}

For each ticker, the explanatory variable is discretized into bins and, within each bin, we compute the empirical mean of the response variable. The resulting curves represent bin-wise averages of the data. To obtain bucket-level results (LT and ST), the ticker-specific curves are interpolated onto a common grid and averaged cross-sectionally, using the same procedure adopted in the previous sections. The final plots display the average curve, a confidence band based on the standard error, and histograms representing the average number of observations per bin.

\begin{figure}[!htb]
\centering
\begin{subfigure}{0.45\textwidth}
  \includegraphics[width=\linewidth]{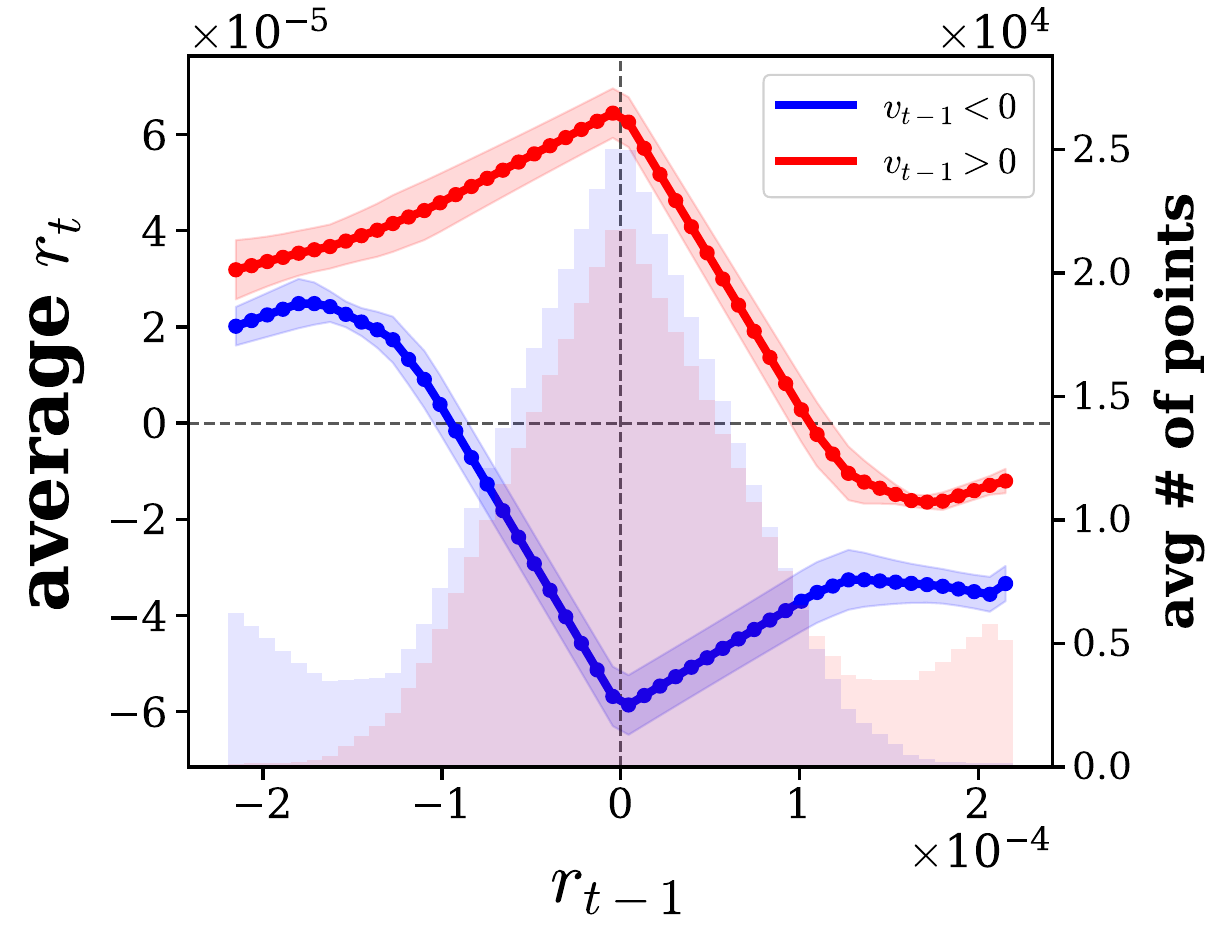}
\end{subfigure}
\begin{subfigure}{0.45\textwidth}
  \includegraphics[width=\linewidth]{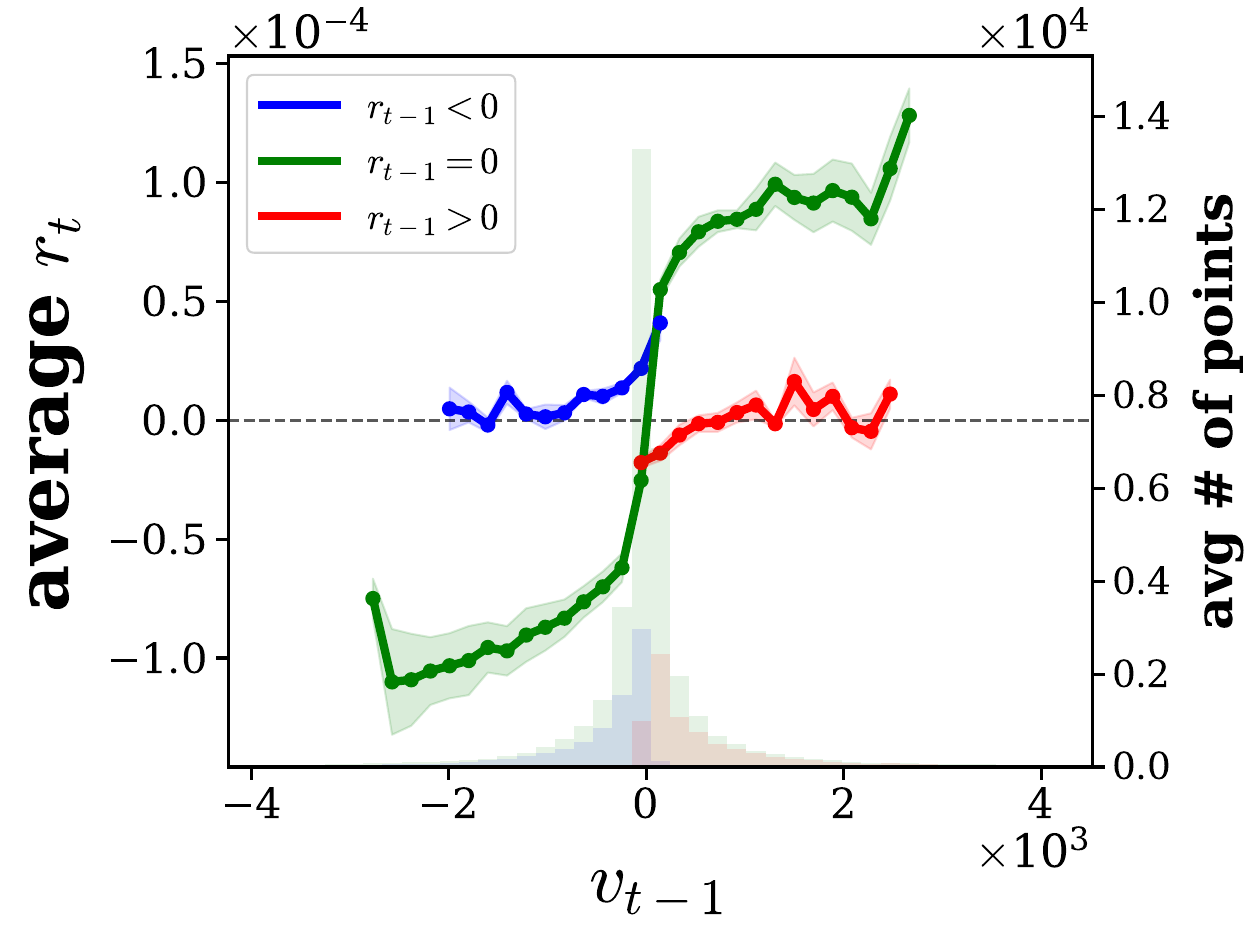}
\end{subfigure}

\vspace{0.3cm}

\begin{subfigure}{0.45\textwidth}
  \includegraphics[width=\linewidth]{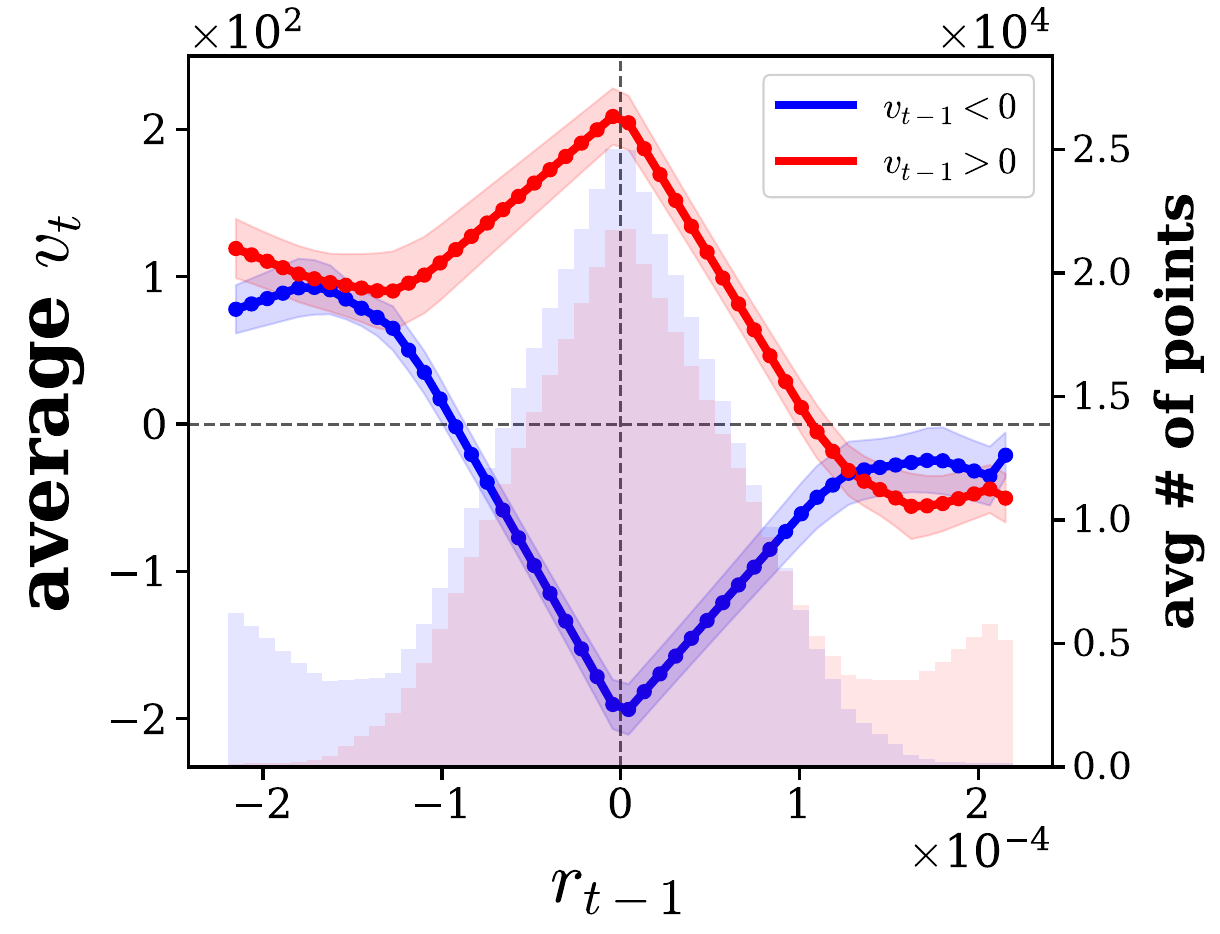}
\end{subfigure}
\begin{subfigure}{0.45\textwidth}
  \includegraphics[width=\linewidth]{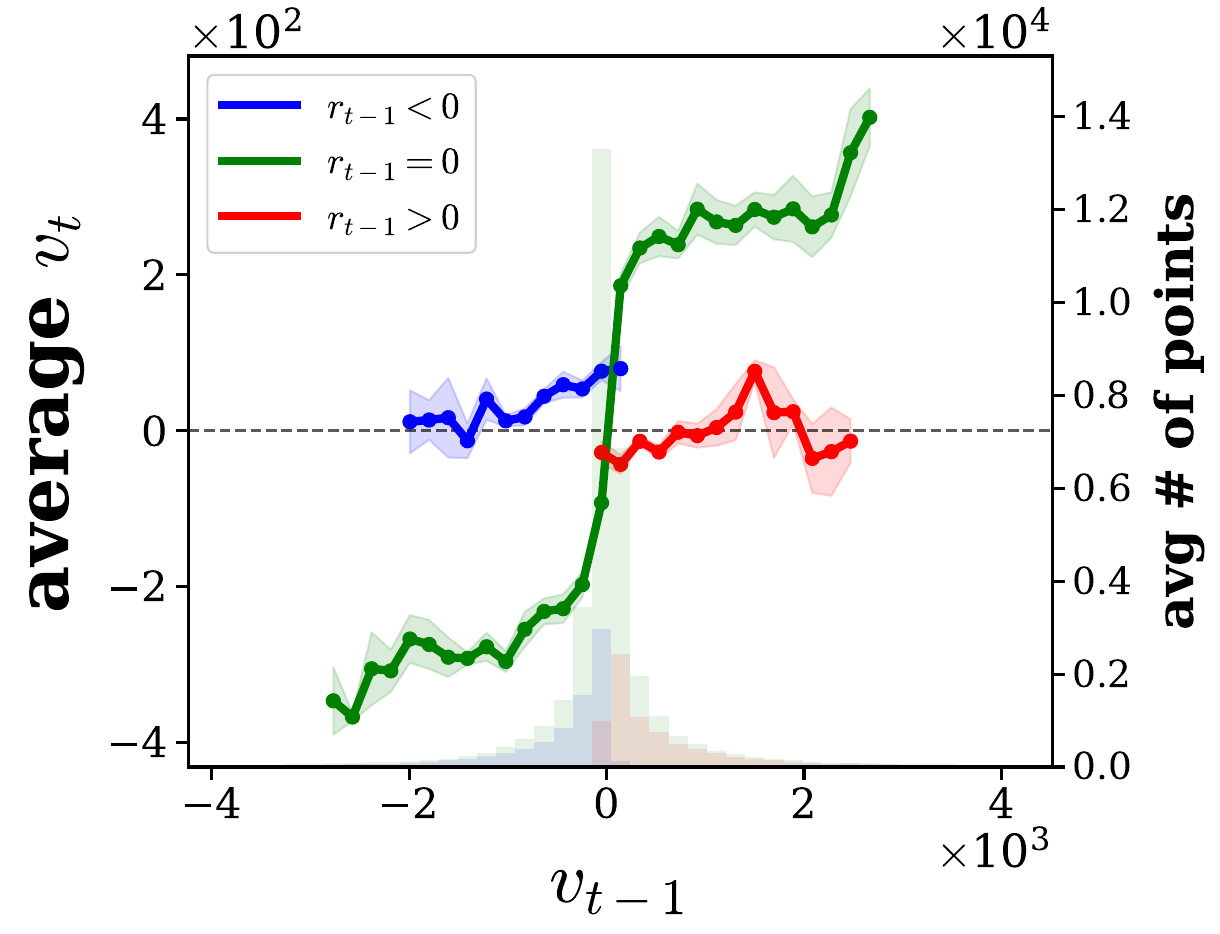}
\end{subfigure}

\caption{Empirical nonlinear relationships for large-tick (LT) stocks.}
\label{fig:DATA_LT}
\end{figure}

\begin{figure}[!htb]
\centering
\begin{subfigure}{0.45\textwidth}
  \includegraphics[width=\linewidth]{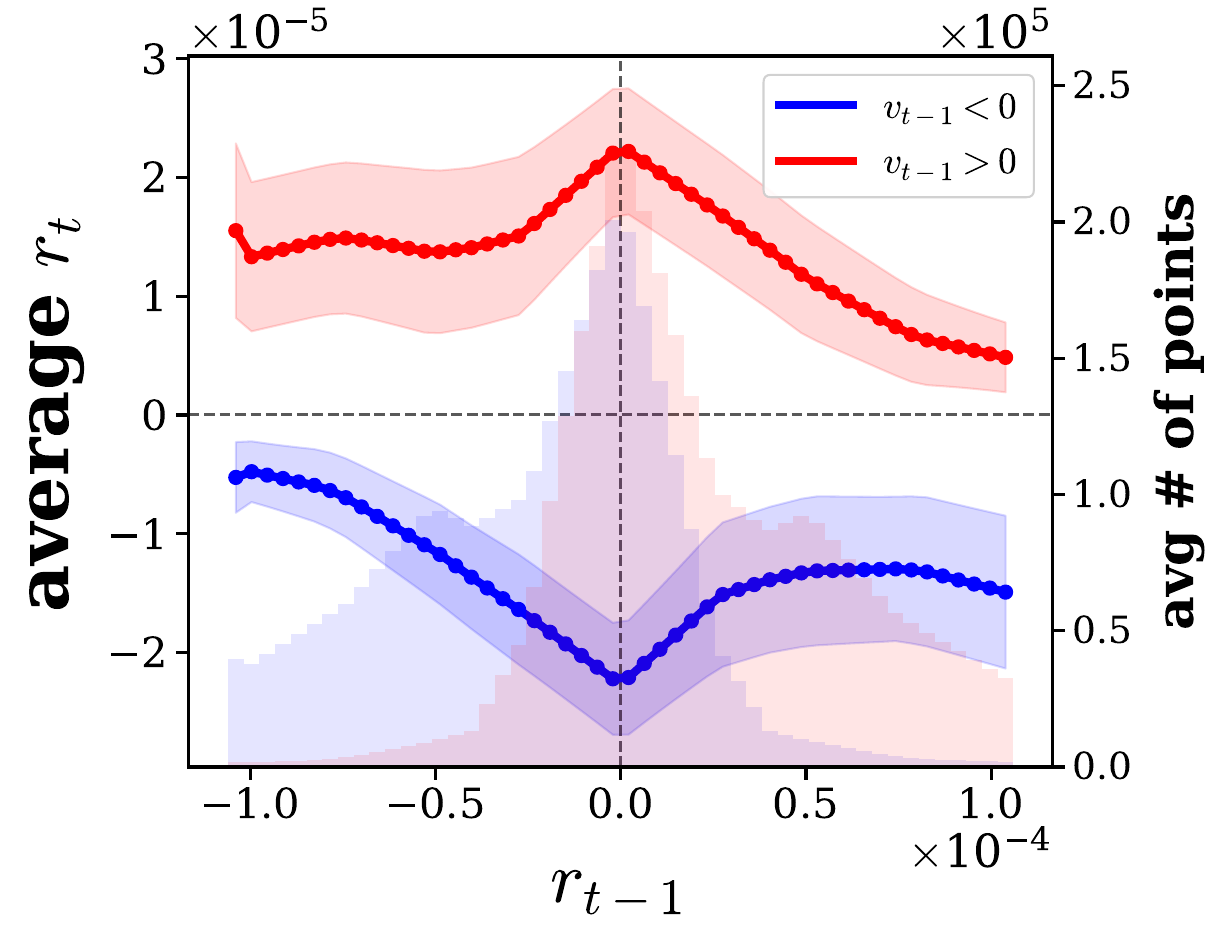}
\end{subfigure}
\begin{subfigure}{0.45\textwidth}
  \includegraphics[width=\linewidth]{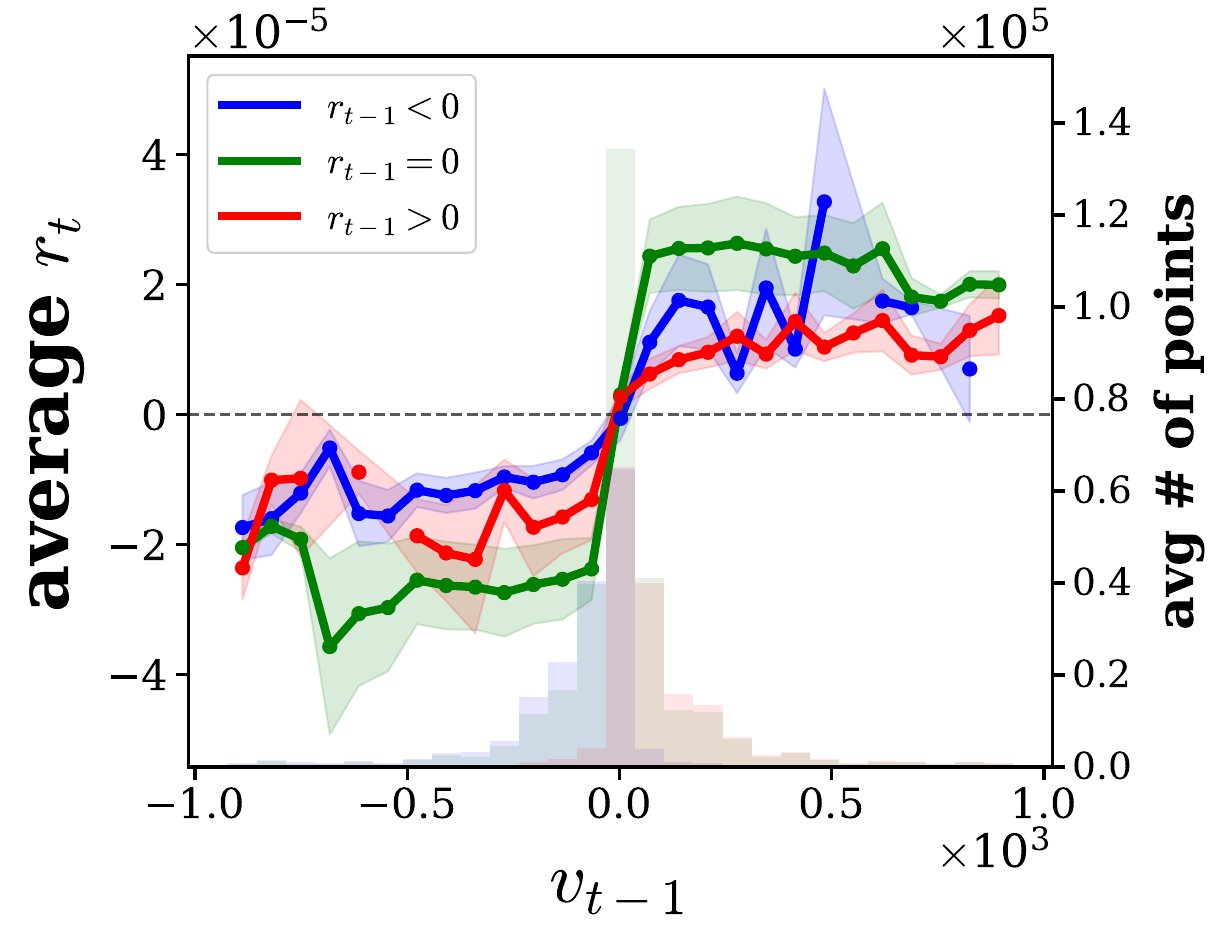}
\end{subfigure}

\vspace{0.3cm}

\begin{subfigure}{0.45\textwidth}
  \includegraphics[width=\linewidth]{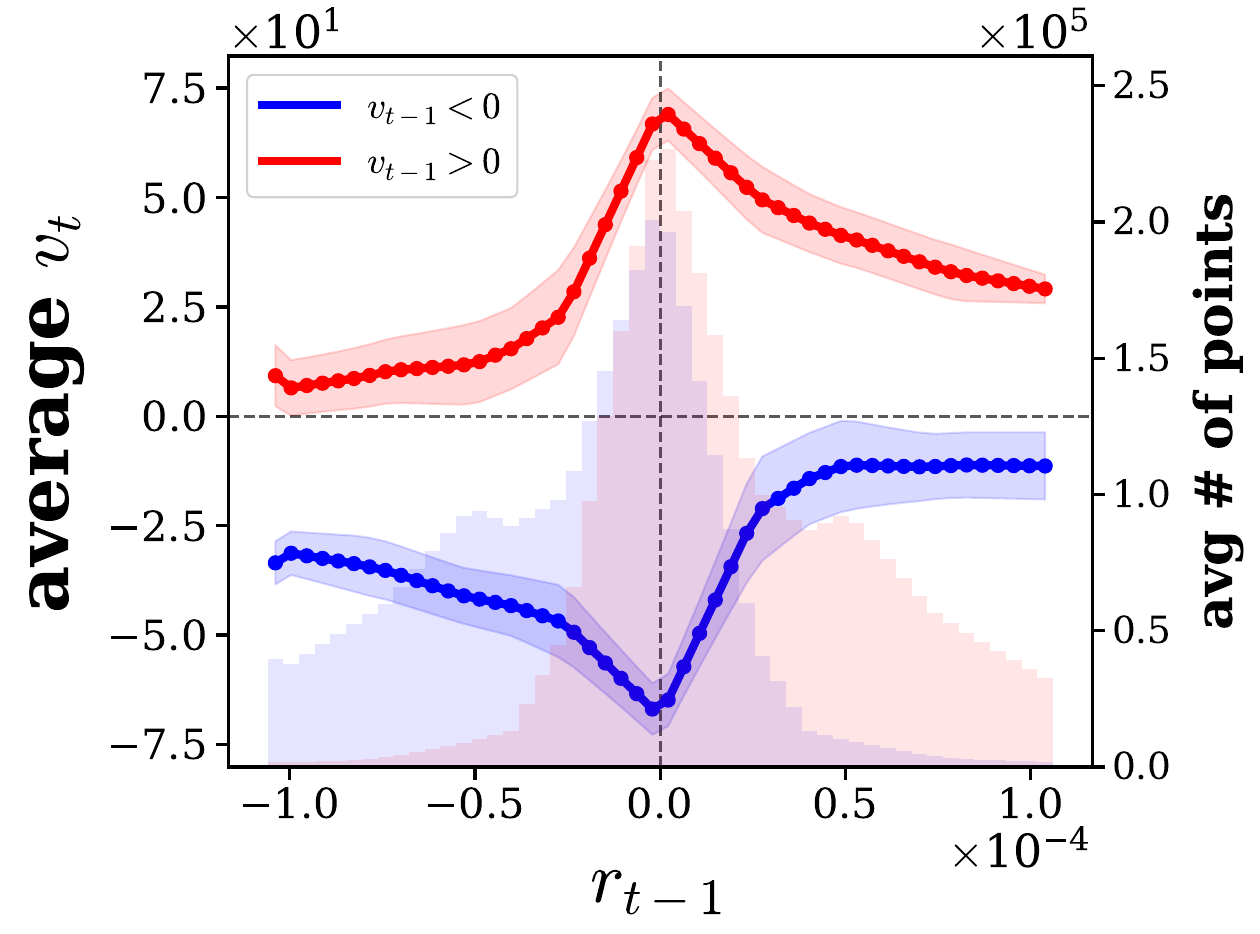}
\end{subfigure}
\begin{subfigure}{0.45\textwidth}
  \includegraphics[width=\linewidth]{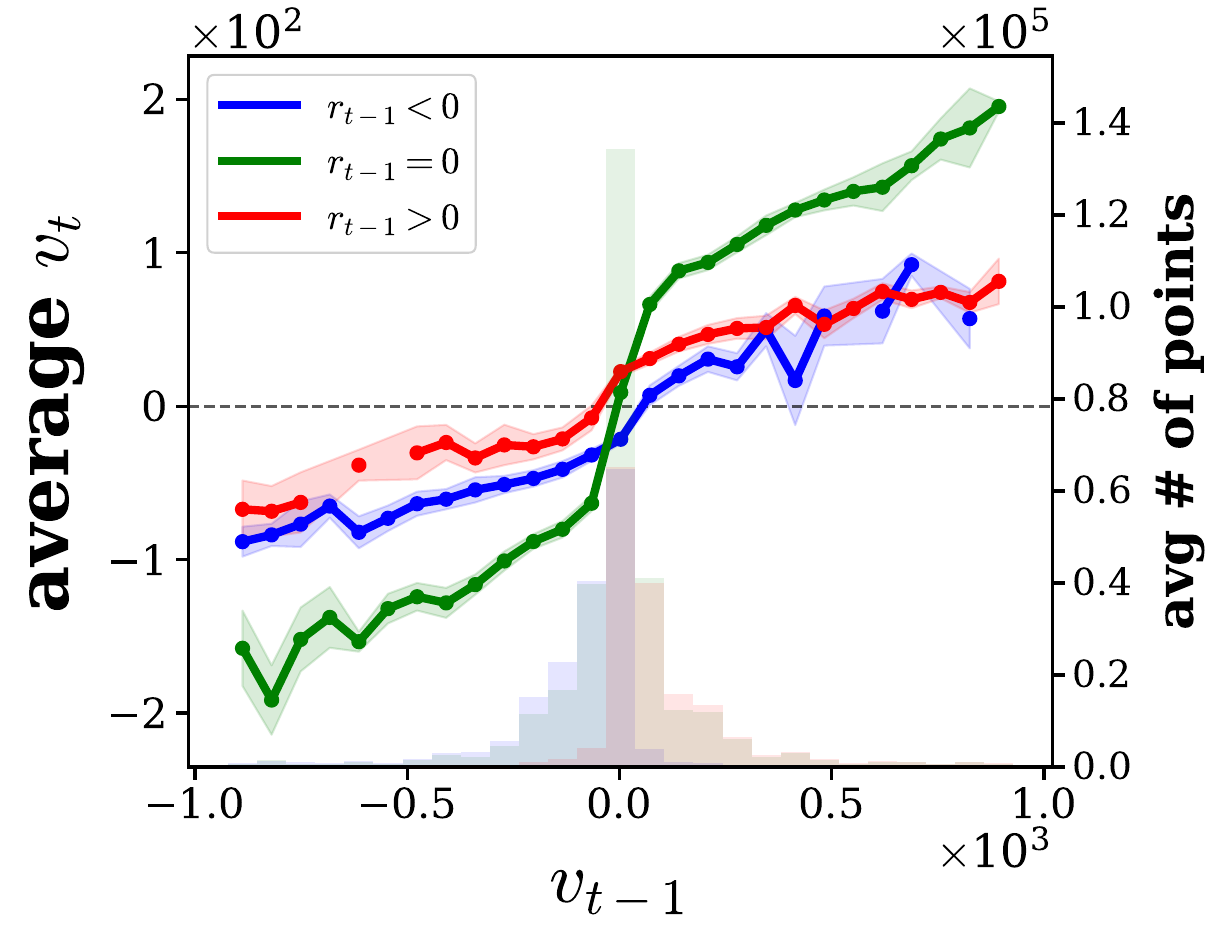}
\end{subfigure}

\caption{Empirical nonlinear relationships for small-tick (ST) stocks.}
\label{fig:DATA_ST}
\end{figure}

The empirical relationships for large-tick (LT) and small-tick (ST) stocks are shown in Figures~\ref{fig:DATA_LT} and~\ref{fig:DATA_ST}. Overall, the agreement between the model-implied aggregated conditional response surfaces and the empirical ones reconstructed directly from the data provides an important sanity check. It indicates that the nonlinear model is capturing dependence patterns that are genuinely present in the data, rather than introducing spurious nonlinearities through its flexibility. This validation is important for the interpretation developed in the previous sections: once the model is shown to reproduce the empirical predictive structure, we are more confident that the Shapley decomposition offers a meaningful way to separate the contributions of individual lagged variables and to identify which components drive the observed relationships.

\section{VAR-implied predictive surfaces}
\label{app:var_predictive_surfaces}

In this appendix we repeat the analysis of the model-implied predictive surfaces
using the linear VAR benchmark instead of the neural-network model. The goal is to
provide a direct comparison between the nonlinear response functions learned by the
DNN and the corresponding structures implied by a purely linear specification.

Starting from the VAR model in Eq.~\eqref{hasbrouckmodel}, the one-step-ahead
predictions for returns and signed volumes are linear functions of the lagged
return--volume state. Therefore, when the prediction is viewed as a function of
\((r_{t-1},v_{t-1})\), the resulting predictive surface is affine in the current
lagged variables, up to the contribution of the remaining lags. In particular, the
dependence on \(v_{t-1}\) is controlled by the first-lag volume coefficients,
whereas the dependence on \(r_{t-1}\) is controlled by the first-lag return
coefficients. This makes the VAR surfaces a natural benchmark for assessing which
patterns can already be explained by linear lagged dependencies and which ones
require the additional flexibility of the neural network.

\begin{figure}[!htb]
\centering
\begin{subfigure}{0.45\textwidth}
  \includegraphics[width=\linewidth]{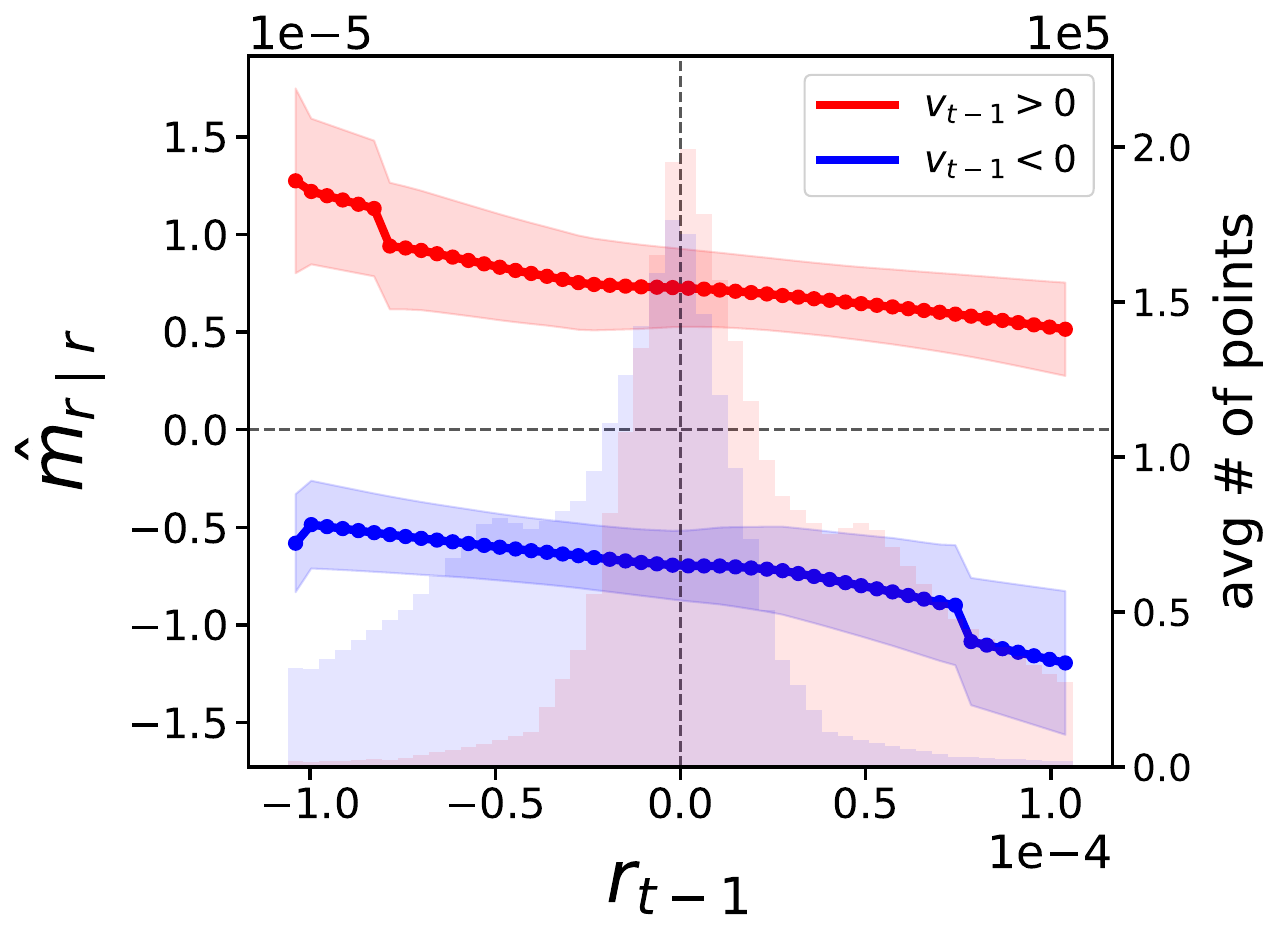}
\end{subfigure}
\begin{subfigure}{0.45\textwidth}
  \includegraphics[width=\linewidth]{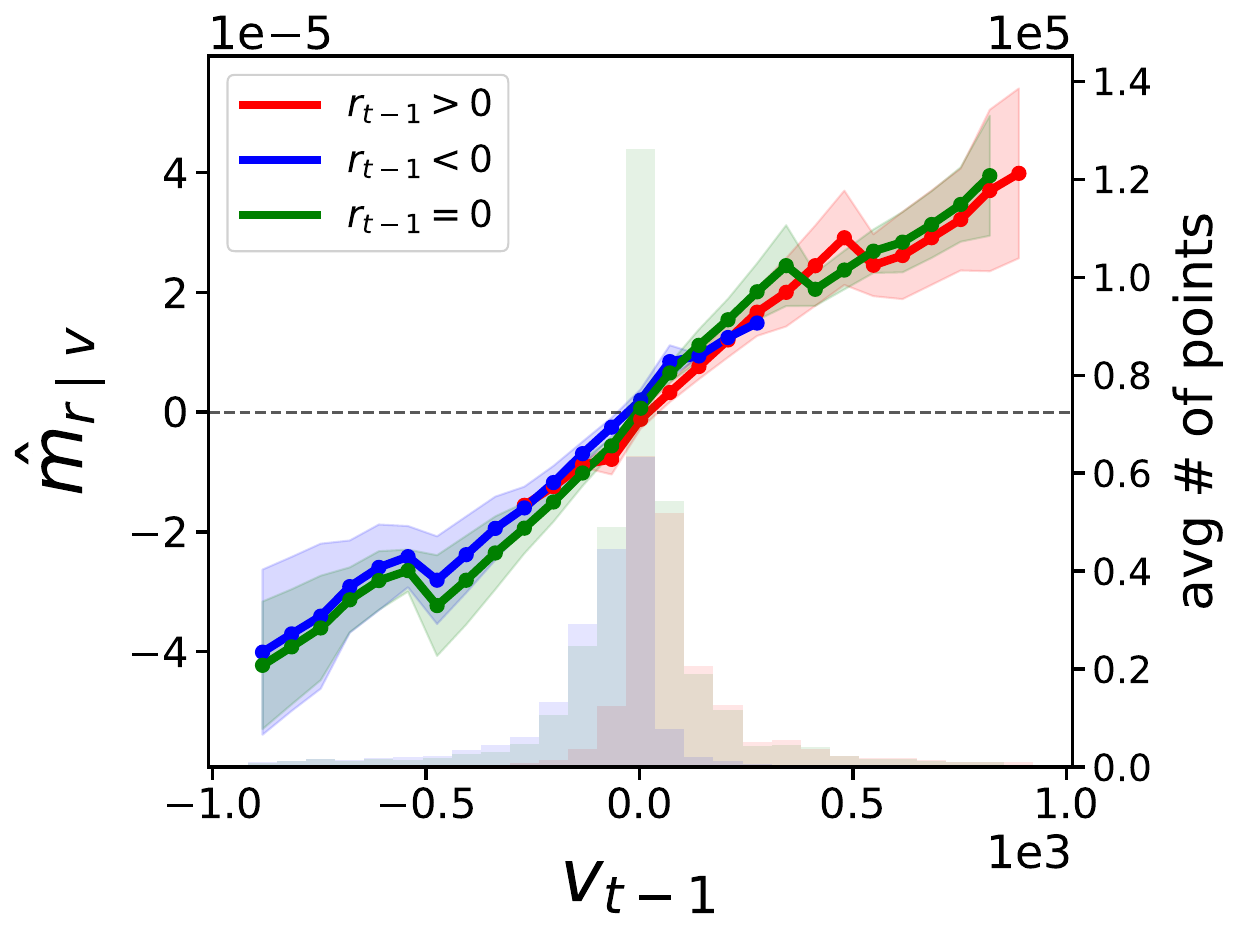}
\end{subfigure}

\vspace{0.3cm}

\begin{subfigure}{0.45\textwidth}
  \includegraphics[width=\linewidth]{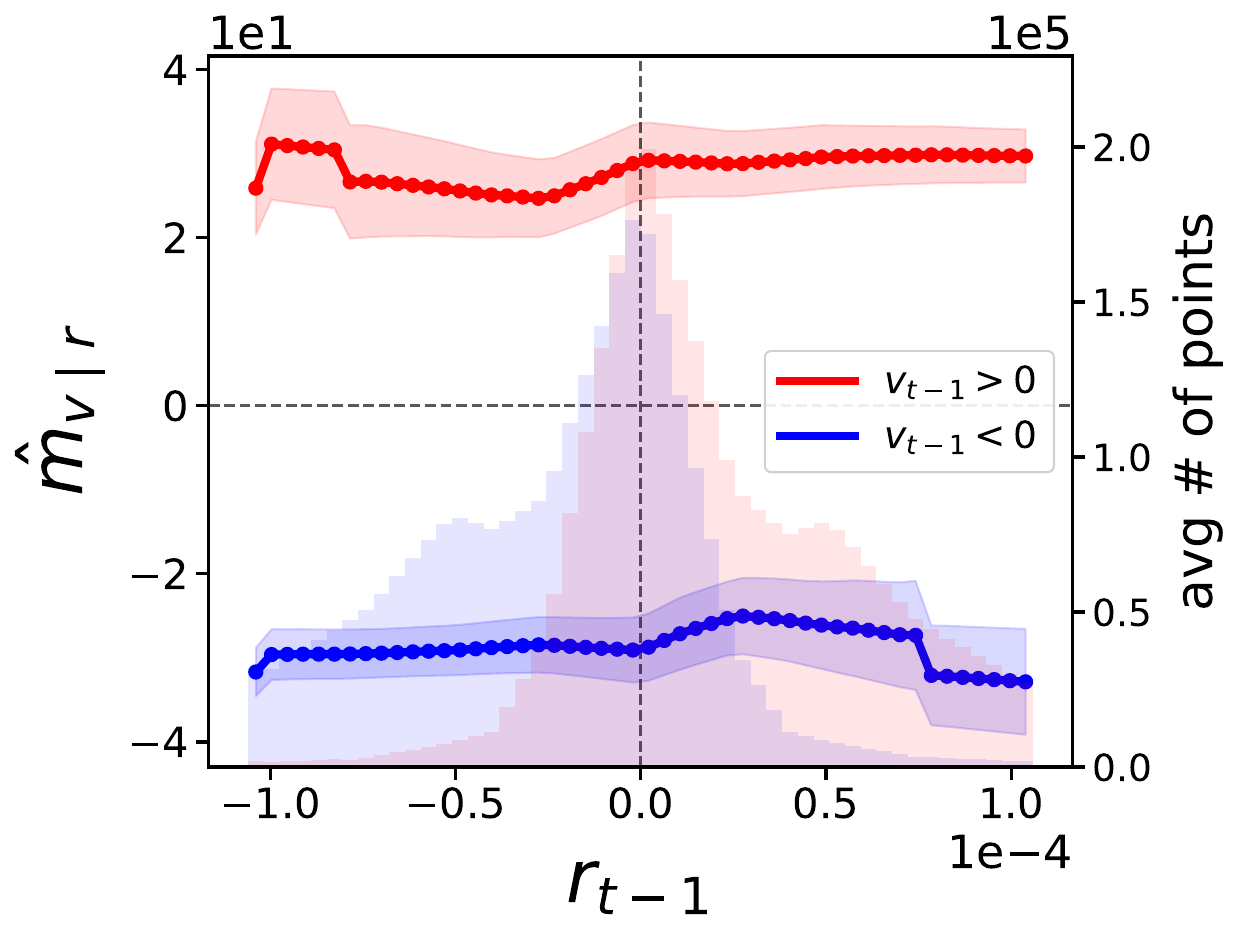}
\end{subfigure}
\begin{subfigure}{0.45\textwidth}
  \includegraphics[width=\linewidth]{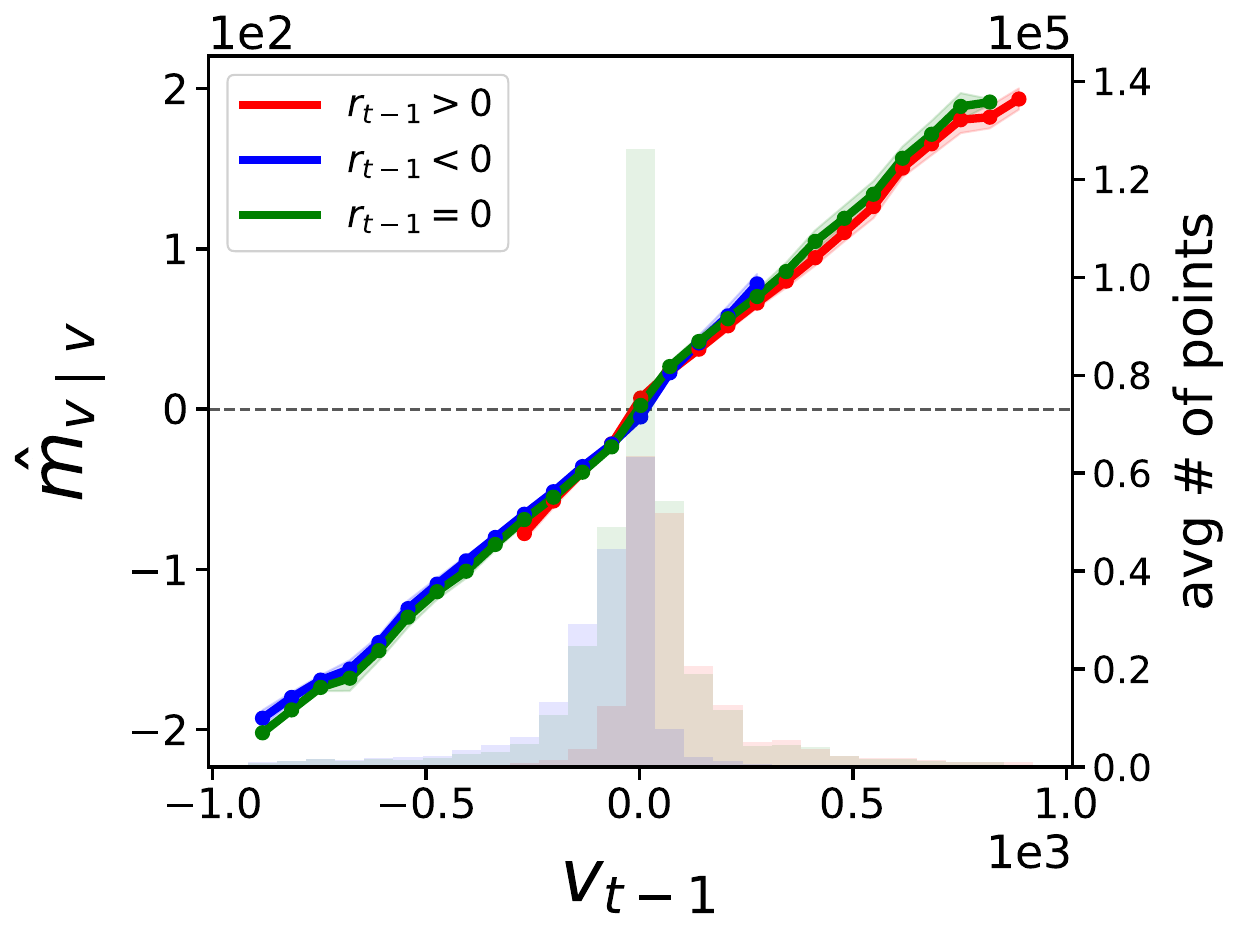}
\end{subfigure}

\caption{VAR-implied binned predictive surfaces for small-tick (ST) stocks. The panels are organized as in Figure~\ref{fig:PDP_ST}: predicted returns and signed volumes are represented as functions of either \(r_{t-1}\) or \(v_{t-1}\), conditional on the sign regime of the other lagged variable.}
\label{fig:VAR_PDP_ST}
\end{figure}

\begin{figure}[!htb]
\centering
\begin{subfigure}{0.45\textwidth}
  \includegraphics[width=\linewidth]{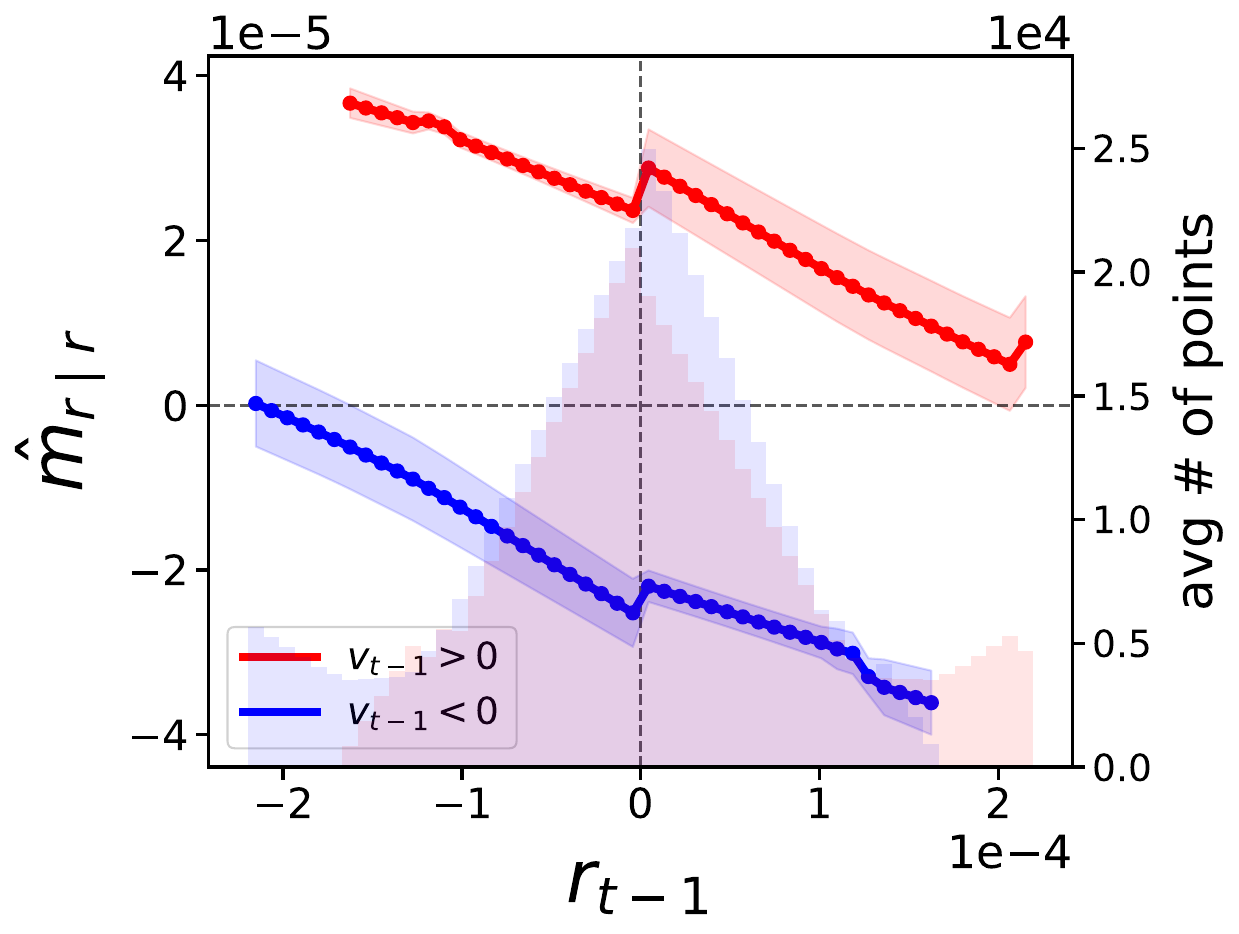}
\end{subfigure}
\begin{subfigure}{0.45\textwidth}
  \includegraphics[width=\linewidth]{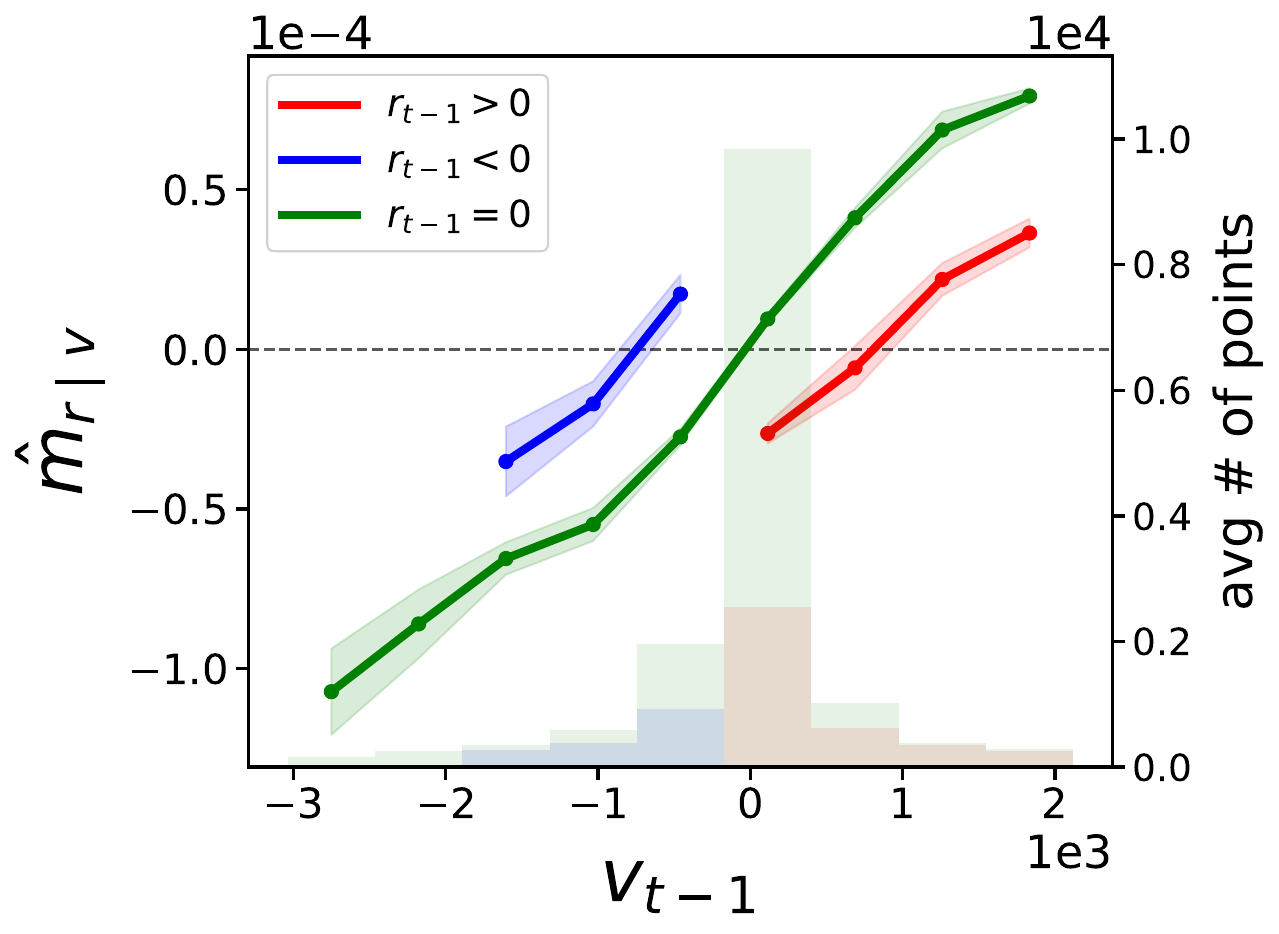}
\end{subfigure}

\vspace{0.3cm}

\begin{subfigure}{0.45\textwidth}
  \includegraphics[width=\linewidth]{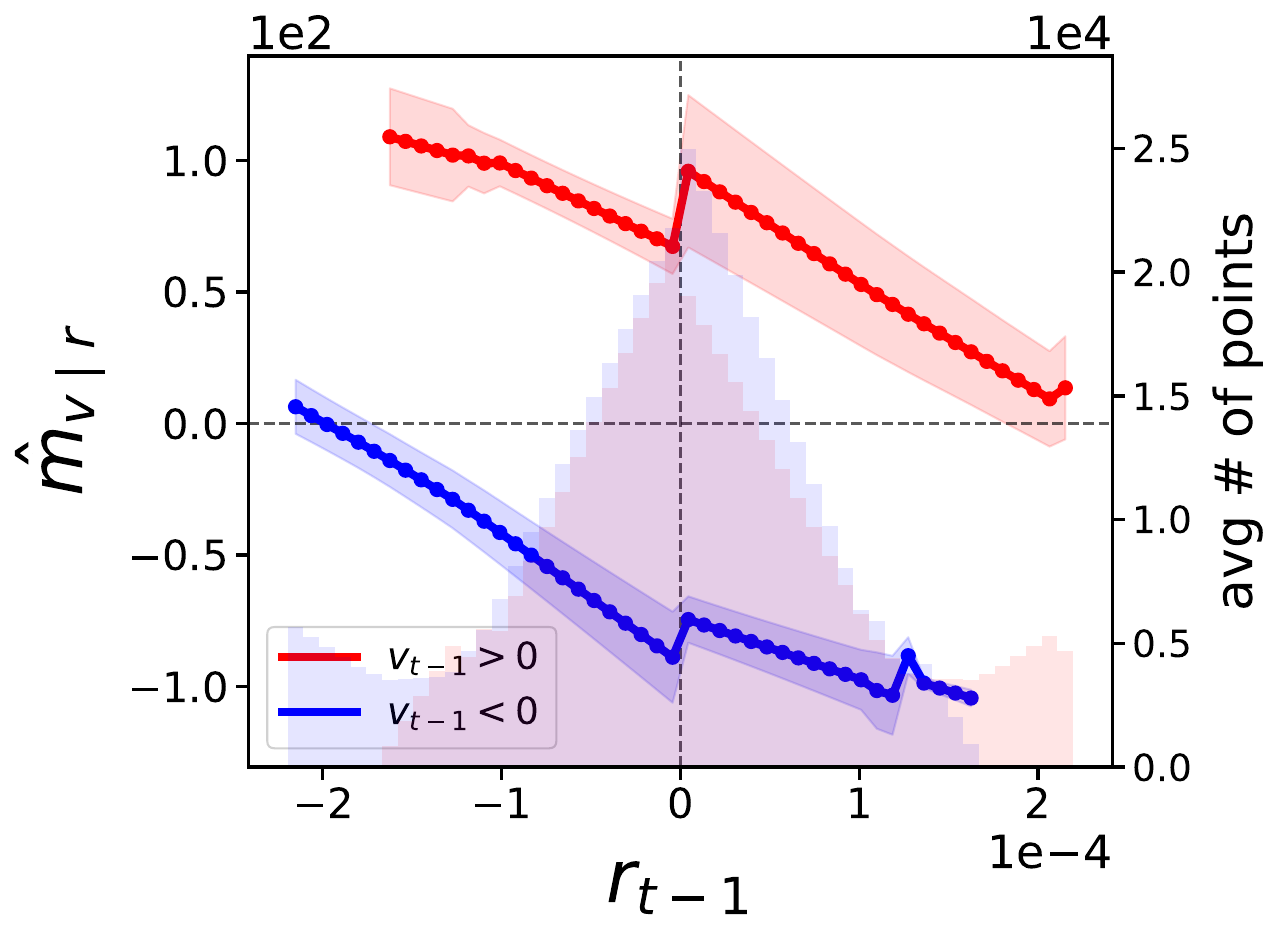}
\end{subfigure}
\begin{subfigure}{0.45\textwidth}
  \includegraphics[width=\linewidth]{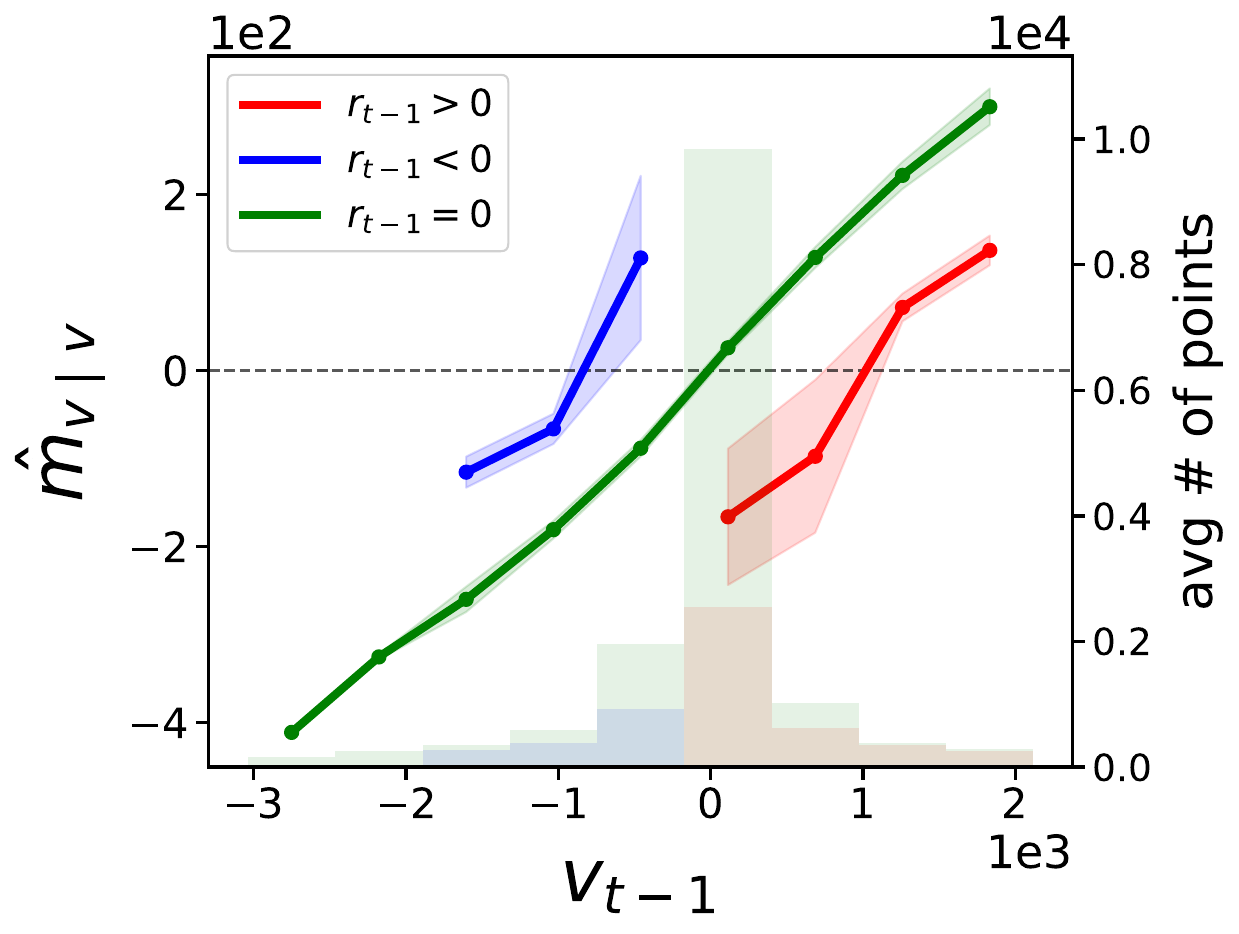}
\end{subfigure}

\caption{VAR-implied binned predictive surfaces for large-tick (LT) stocks. The panels are organized as in Figure~\ref{fig:PDP_LT}: predicted returns and signed volumes are represented as functions of either \(r_{t-1}\) or \(v_{t-1}\), conditional on the sign regime of the other lagged variable.}
\label{fig:VAR_PDP_LT}
\end{figure}

We construct the binned VAR surfaces using the same procedure adopted in the main
text for the DNN. Namely, we compute conditional averages of the VAR predictions
within bins of either \(v_{t-1}\) or \(r_{t-1}\), further conditioning on the sign
regime of the other lagged variable. The same binning and clustering procedure is
used, so that the comparison with the DNN surfaces is not affected by differences
in empirical support. The resulting surfaces are reported in
Figures~\ref{fig:VAR_PDP_ST} and~\ref{fig:VAR_PDP_LT}, respectively for small-tick
and large-tick stocks.

Figures~\ref{fig:VAR_PDP_ST} and~\ref{fig:VAR_PDP_LT} show that the VAR captures
the main first-order directional effects in the data. In particular, when the
response is represented as a function of \(v_{t-1}\), the predicted signed volume
is approximately monotone increasing: positive lagged signed volume predicts
positive future signed volume, while negative lagged signed volume predicts
negative future signed volume. This is the linear counterpart of order-flow
persistence. Similarly, the predicted return tends to increase with \(v_{t-1}\),
consistently with the idea that buy-initiated volume is associated with positive
short-horizon price changes and sell-initiated volume with negative ones.

However, because the VAR is linear, these relations are necessarily smoother and
more constrained than those obtained with the neural network. The sharp transitions
around \(v_{t-1}=0\), the flattening at large absolute values of signed volume, and
the more pronounced state dependence observed in the DNN surfaces cannot be fully
reproduced by the linear benchmark. In the VAR case, the dependence on
\(v_{t-1}\) is essentially governed by a single linear slope, and the differences
between the curves associated with \(r_{t-1}<0\), \(r_{t-1}=0\), and
\(r_{t-1}>0\) mainly appear as level shifts induced by the conditioning variable
and by correlations with the remaining regressors.

A similar interpretation applies to the panels where the horizontal axis is
\(r_{t-1}\). The VAR produces approximately linear responses in lagged returns,
with the curves separated according to the sign of \(v_{t-1}\). This confirms that
lagged signed volume remains an important state variable also in the linear model.
Nevertheless, the role of \(r_{t-1}\) is constrained by the linear structure: it can
only affect the prediction proportionally. As a consequence, the VAR cannot
reproduce the more flexible attenuation patterns observed in the neural-network
surfaces, where large absolute lagged returns tend to weaken the continuation of
the order-flow signal.

The comparison across tick-size regimes is also informative. For small-tick stocks,
shown in Figure~\ref{fig:VAR_PDP_ST}, the VAR recovers the average direction of the
main relationships, but it provides a substantially smoother representation than
the DNN. This is especially visible in the response to lagged signed volume, where
the neural network displays sharper nonlinearities and stronger state dependence.
For large-tick stocks, shown in Figure~\ref{fig:VAR_PDP_LT}, the difference between
the VAR and the DNN is less pronounced. This is consistent with the idea that
large-tick dynamics are more strongly constrained by price discreteness, so that a
linear benchmark already captures a relevant part of the directional dependence
between recent order-flow pressure and price changes.

Overall, the VAR predictive surfaces confirm that part of the return--volume
dynamics can be described by linear lagged dependencies. In particular, the linear
model captures the sign of price impact and the persistence of signed order flow.
At the same time, the comparison with the neural-network surfaces shows that the
DNN is not merely reproducing the VAR relation with additional noise. Rather, it
learns a richer response characterized by sharper transitions, saturation effects,
and state-dependent modulation of the predictive signal. This provides a useful
benchmark for the explainability analysis in the main text: deviations from the
VAR surfaces can be interpreted as evidence of nonlinear dependencies learned by
the neural network beyond the linear specification.

\section{Large tick Shapley Components}
\label{app:LT_Shaps}

\begin{figure}[!htb]
\centering
\begin{subfigure}{0.45\textwidth}
  \includegraphics[width=\linewidth]{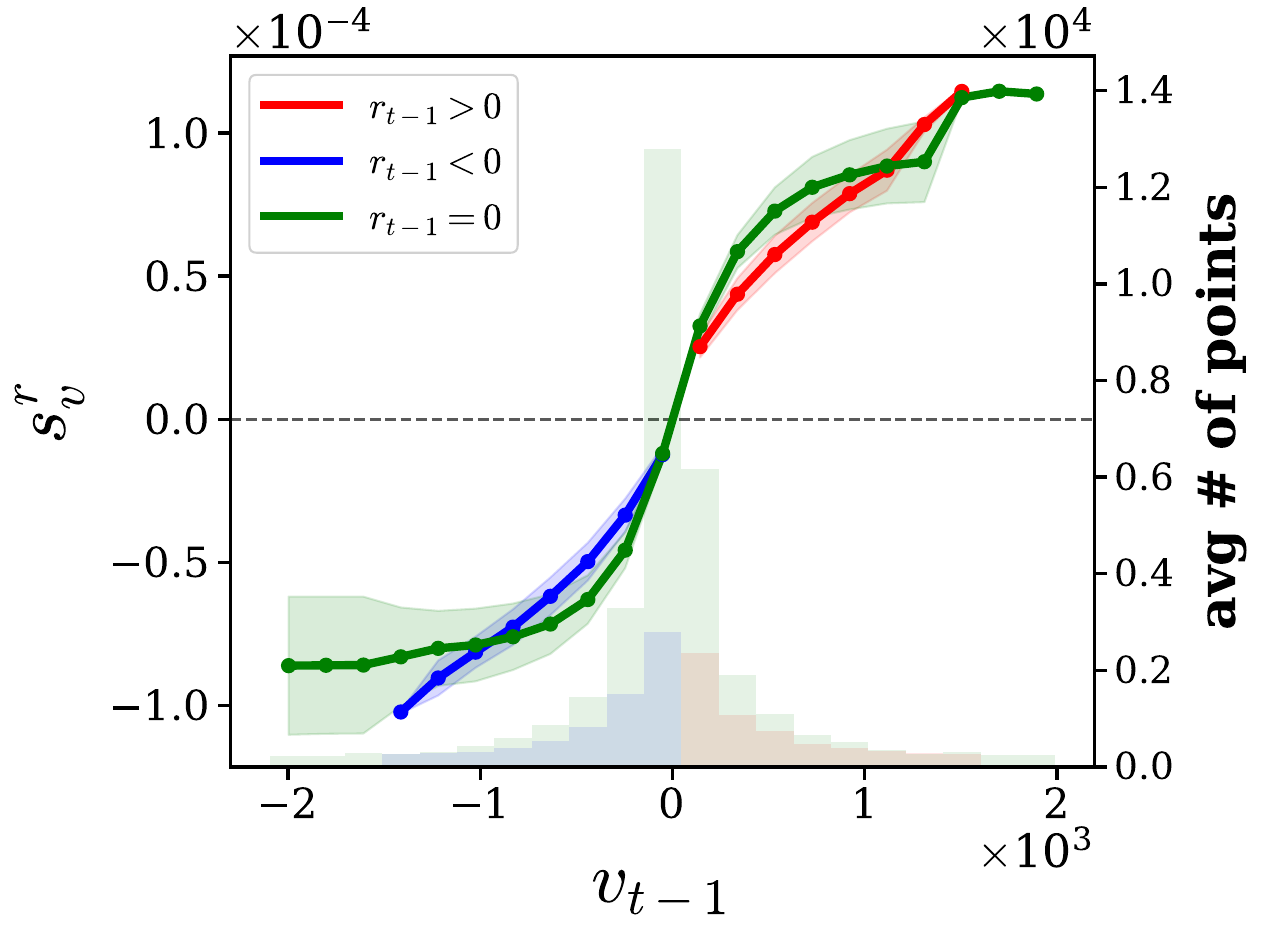}
\end{subfigure}
\begin{subfigure}{0.45\textwidth}
  \includegraphics[width=\linewidth]{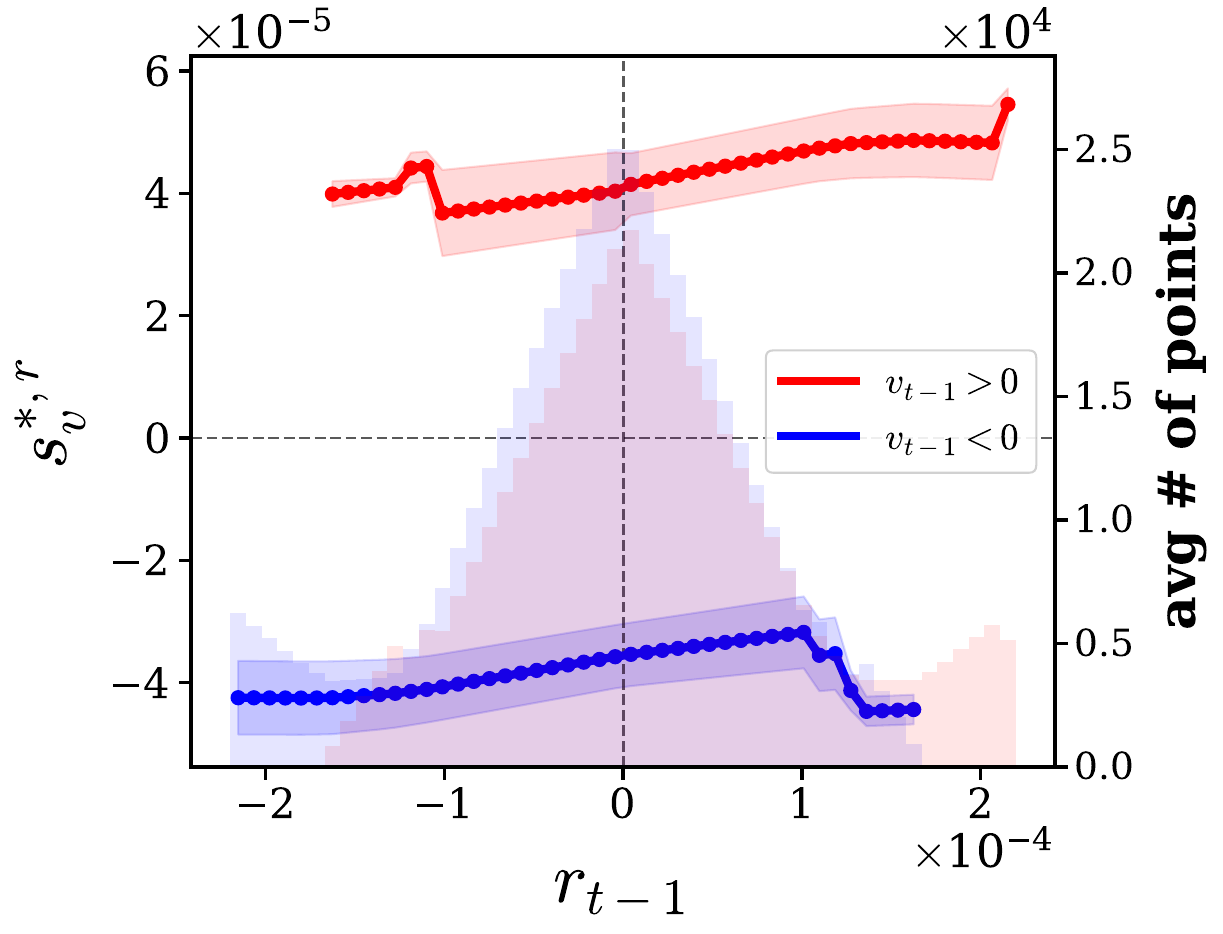}
\end{subfigure}

\vspace{0.3cm}

\begin{subfigure}{0.45\textwidth}
  \includegraphics[width=\linewidth]{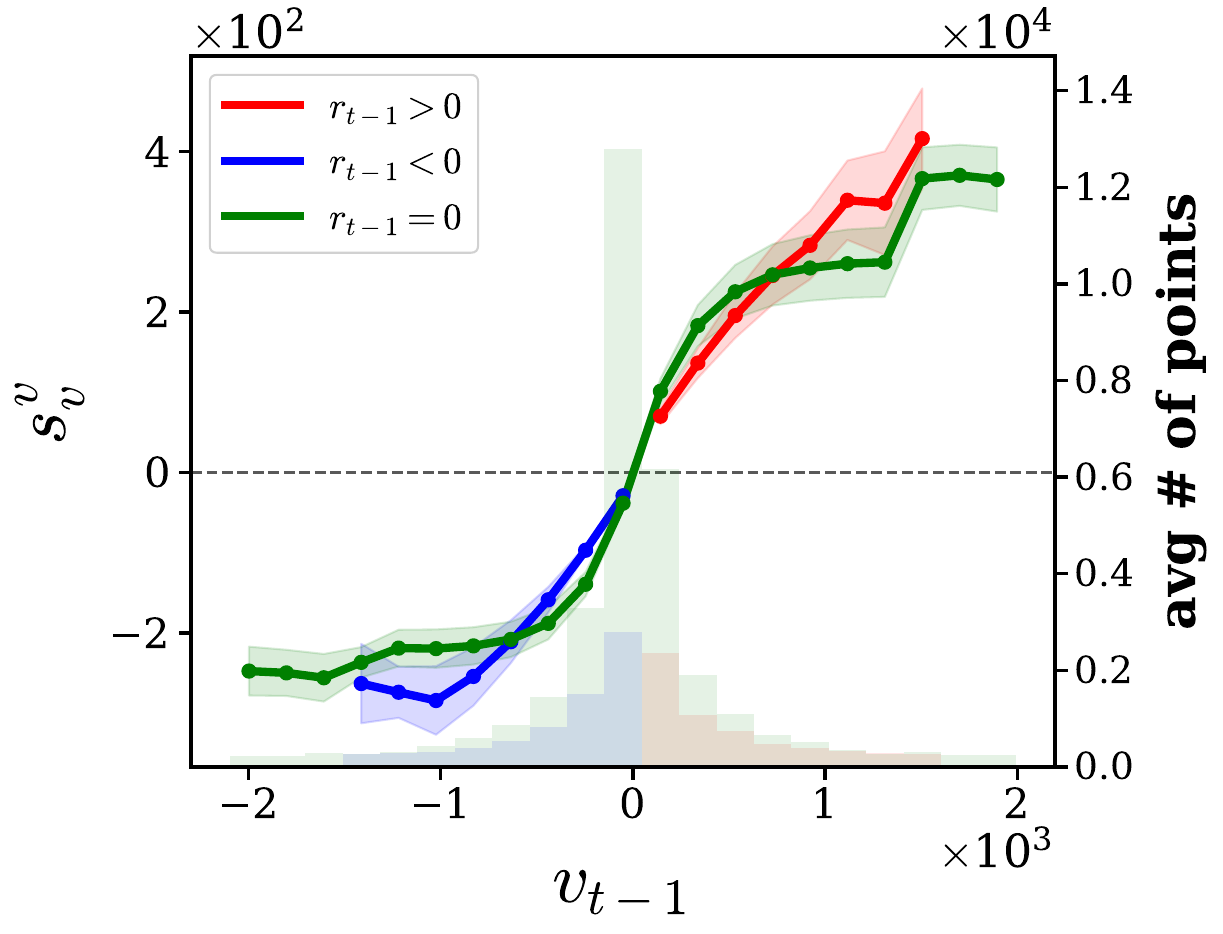}
\end{subfigure}
\begin{subfigure}{0.45\textwidth}
  \includegraphics[width=\linewidth]{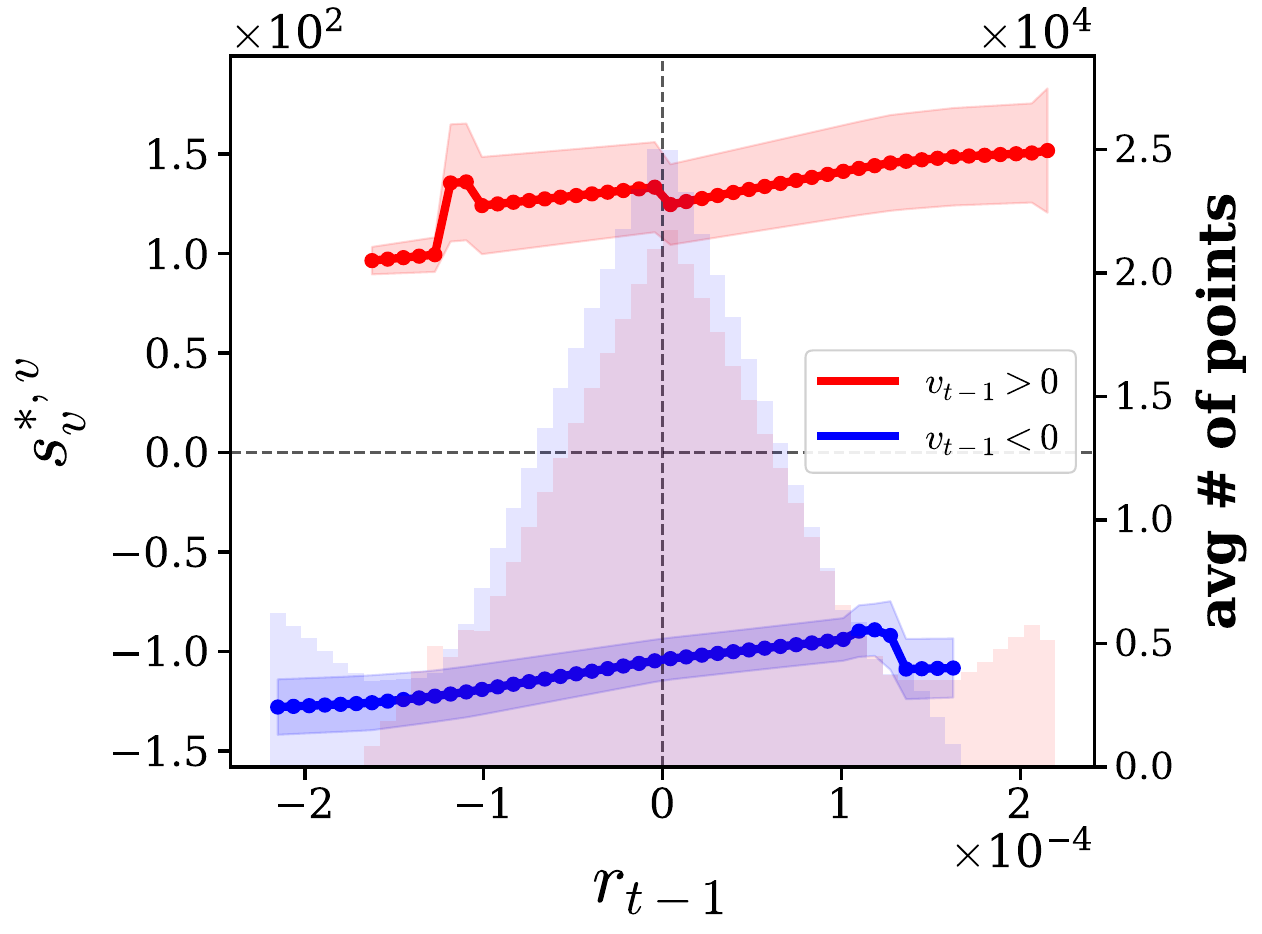}
\end{subfigure}

\caption{Average Shapley profiles for large tick (LT) stocks for the volume term. Top panels refer to the return prediction \(\hat r_t\), while bottom panels refer to the signed-volume prediction \(\hat v_t\). Left panels show the contribution of the conditioning variable \(v_{t-1}\) when the horizontal axis is \(v_{t-1}\). Right panels show the contribution of the conditioning variable \(v_{t-1}\) when the horizontal axis is \(r_{t-1}\). Shaded bands denote \(\pm\) one cross-sectional standard error, while histograms report the average empirical support.}
\label{fig:SHAP_star_LT}
\end{figure}

\begin{figure}[!htb]
\centering
\begin{subfigure}{0.45\textwidth}
  \includegraphics[width=\linewidth]{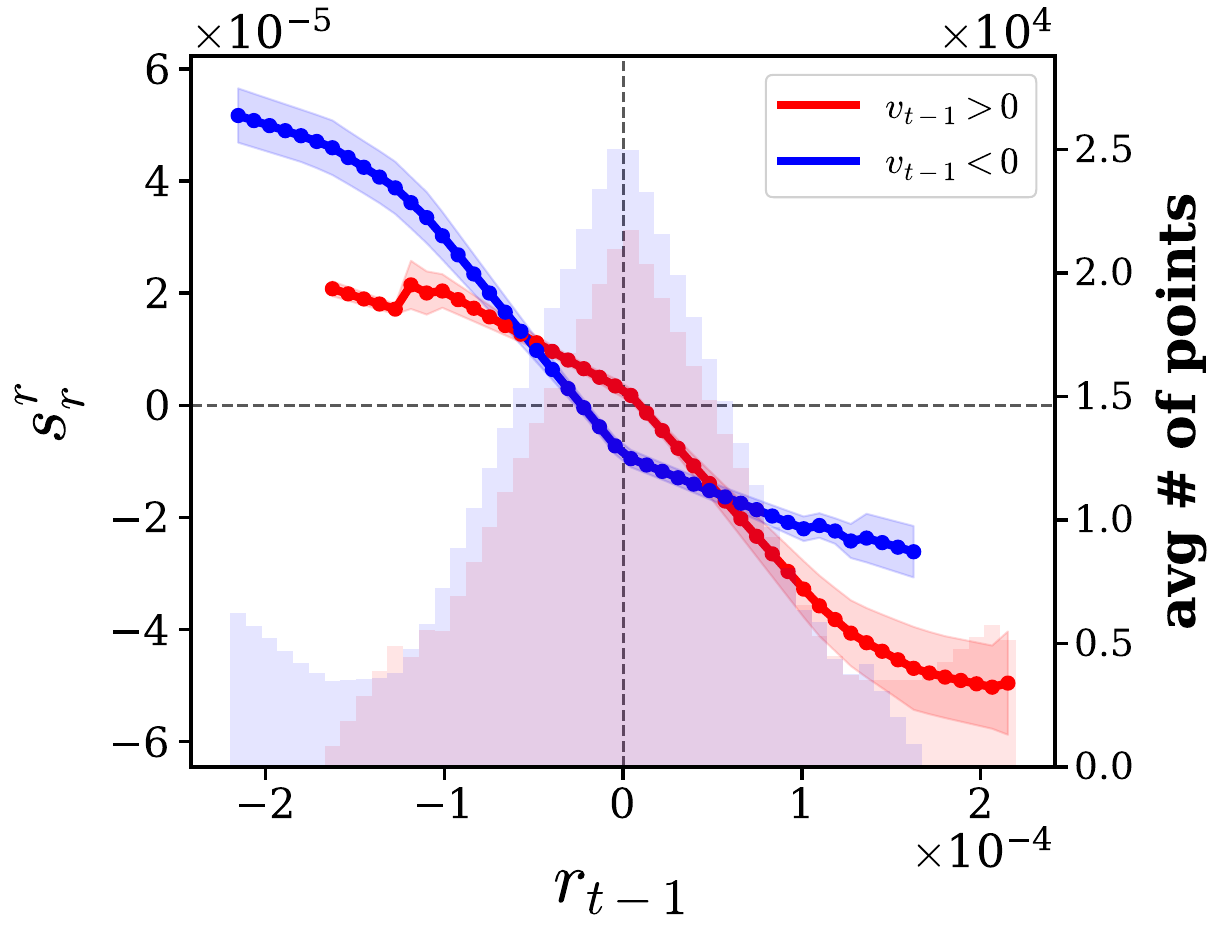}
\end{subfigure}
\begin{subfigure}{0.45\textwidth}
  \includegraphics[width=\linewidth]{figs_Final/real_shap_rt_vs_rt-1_color_vt-1_inverted_average_grid_ST.pdf}
\end{subfigure}

\vspace{0.3cm}

\begin{subfigure}{0.45\textwidth}
  \includegraphics[width=\linewidth]{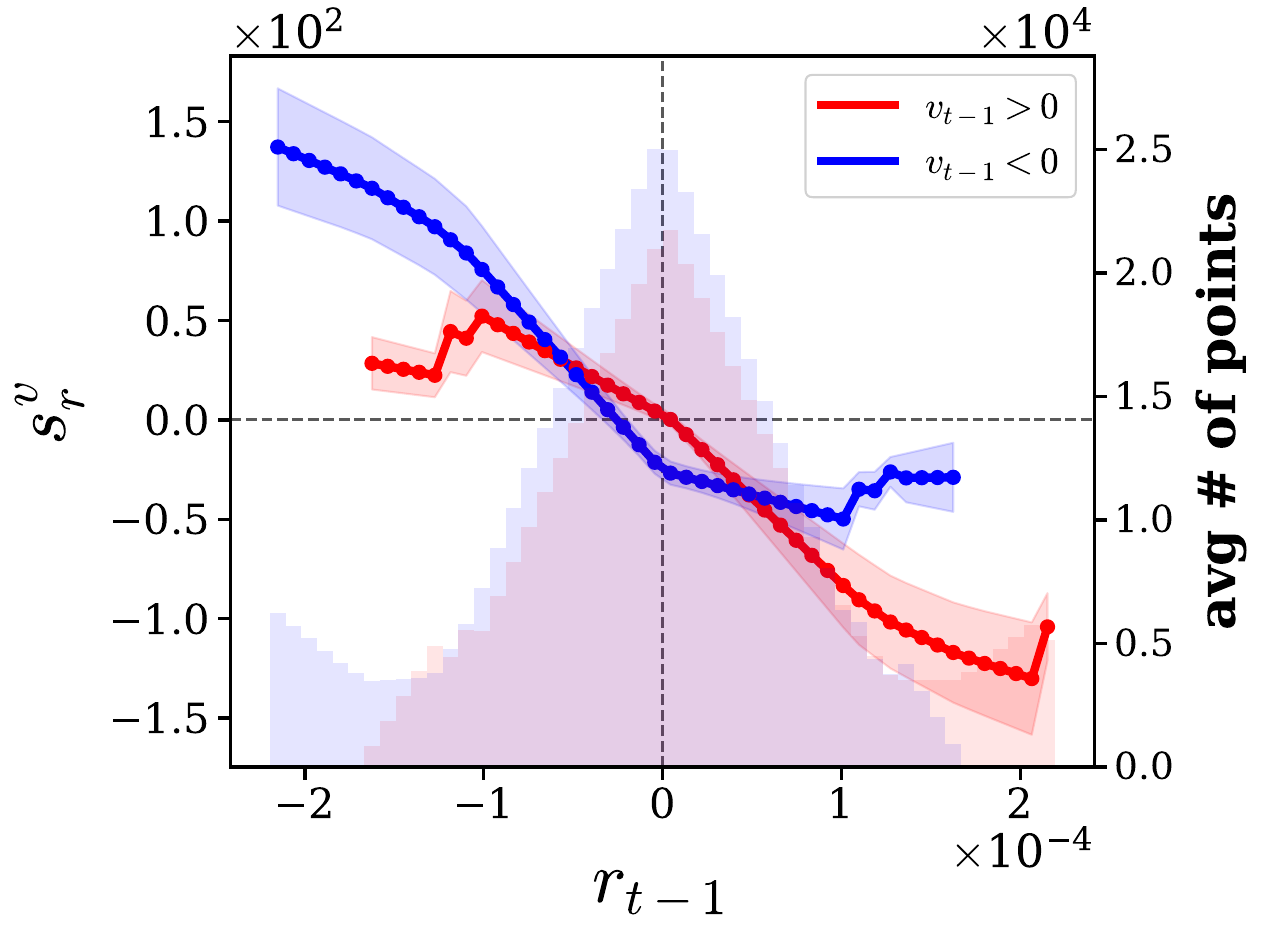}
\end{subfigure}
\begin{subfigure}{0.45\textwidth}
  \includegraphics[width=\linewidth]{figs_Final/real_shap_vt_vs_rt-1_color_vt-1_inverted_average_grid_ST.pdf}
\end{subfigure}

\caption{Average Shapley profiles for large tick (LT) stocks for the return term. Panels are organized as in Figure~\ref{fig:SHAP_star_LT}.}
\label{fig:SHAP_star_ST}
\end{figure}

In this appendix, we report the Shapley components for large-tick stocks, following the same structure used for the small-tick stocks in the main text. The corresponding profiles are shown in Figures~\ref{fig:SHAP_star_LT} and~\ref{fig:SHAP_star_ST}. Overall, the results for large-tick stocks are qualitatively similar to those obtained for small-tick stocks, although some differences are relevant and deserve to be discussed.

We start from the term associated with the contribution of \(v_{t-1}\) to the next-step prediction. The behavior is essentially analogous to the one observed for small-tick stocks. In particular, the volume term displays a sign-like dependence on \(v_{t-1}\): positive lagged signed volume gives a positive contribution, while negative lagged signed volume gives a negative contribution. Moreover, the conditional curves corresponding to non-zero values of \(r_{t-1}\) are, within the error bars, almost collapsed onto the curve associated with \(r_{t-1}=0\). Therefore, also in the large-tick case, we do not observe an additional dependence on \(r_{t-1}\) in the volume contribution. This conclusion is confirmed by the starred terms, shown in the right panels of Figures~\ref{fig:SHAP_star_LT}, which are essentially flat as functions of \(r_{t-1}\). Hence, the dependence on lagged signed volume appears to be genuinely captured by the volume Shapley term, without a relevant spurious component attributed to the return term.

We then consider the return-related contribution. Also in this case, as for small-tick stocks, a relevant part of the structure is mediated by the lagged signed volume. Looking at the starred terms in the right panels of Figures~\ref{fig:SHAP_star_ST}, we recover the same qualitative mechanism observed in the ST bucket. When the previous trade did not induce a price change, namely \(r_{t-1}=0\), the return term contributes to the next-step return and volume by preserving the sign of the previous signed volume. By contrast, when a price change occurred, \(r_{t-1}\neq 0\), the average contribution tends to act in the opposite direction, generating a pressure toward a sign reversal both in the next signed volume and in the next return.

This behavior is also visible in the non-starred return profiles. For \(r_{t-1}=0\), the contribution of the return term tends to preserve the sign of \(v_{t-1}\), as in the small-tick case. However, differently from the ST bucket, the LT profiles display a more pronounced asymmetry with respect to \(r_{t-1}\). For \(r_{t-1}>0\), the return contribution to the next-step volume and return tends to have the opposite sign, while for \(r_{t-1}<0\) the opposite pattern is observed. This suggests a stronger mean-reverting component in prices for large-tick stocks. A similar effect was already visible, although in a weaker form, for small-tick stocks in the profile \(s_r^r\), reported in the top-left panel of Figure~\ref{fig:SHAP_ST_r}.

\section{Lag dependence for the parametric multi-lag model}
\label{app:LagDependenceMulti}

\begin{figure}[t]
    \centering

    \begin{subfigure}{0.48\textwidth}
        \centering
        \includegraphics[width=\textwidth]{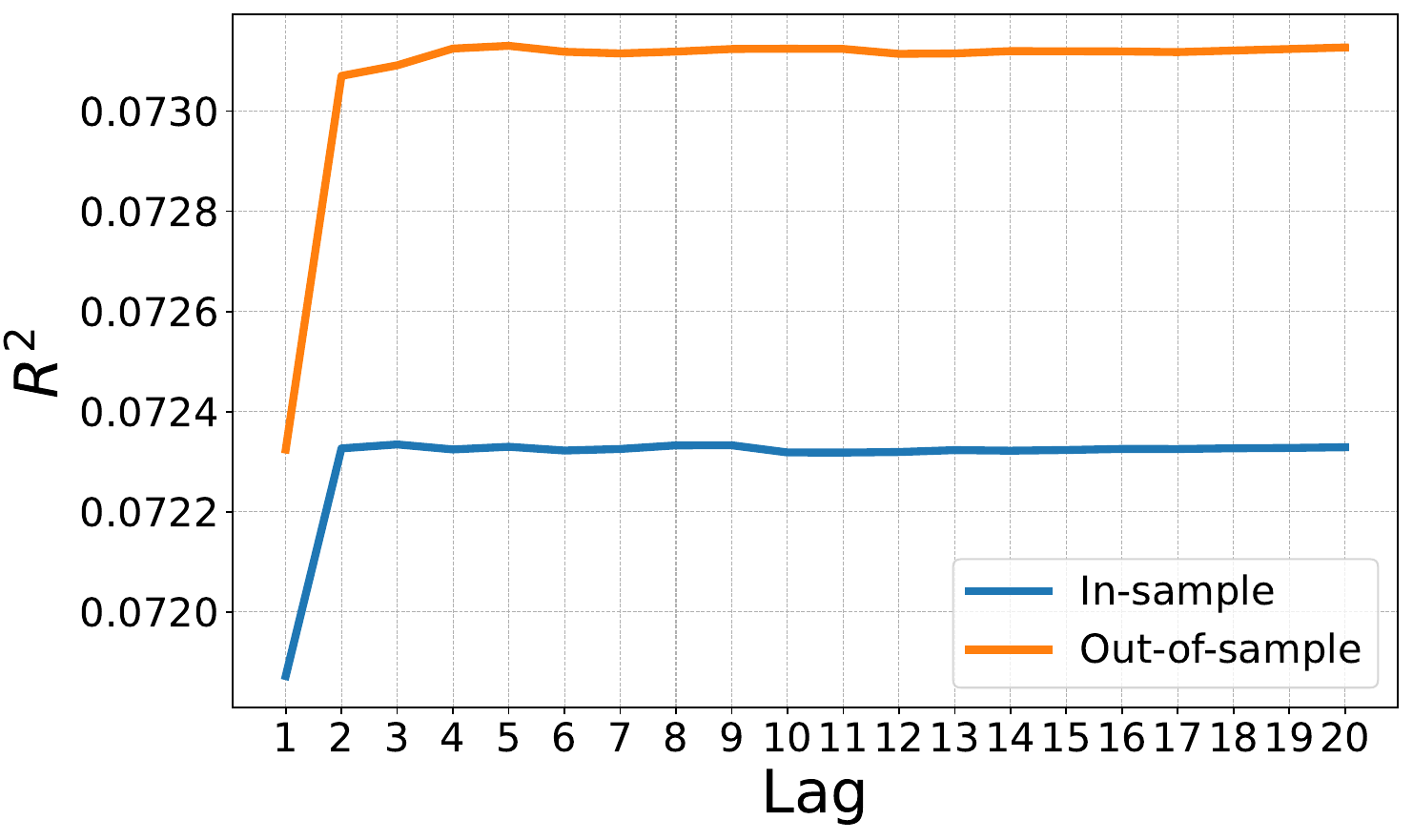}
        \caption{Returns.}
        \label{fig:parametric_GILD_lag_scan_r}
    \end{subfigure}
    \hfill
    \begin{subfigure}{0.48\textwidth}
        \centering
        \includegraphics[width=\textwidth]{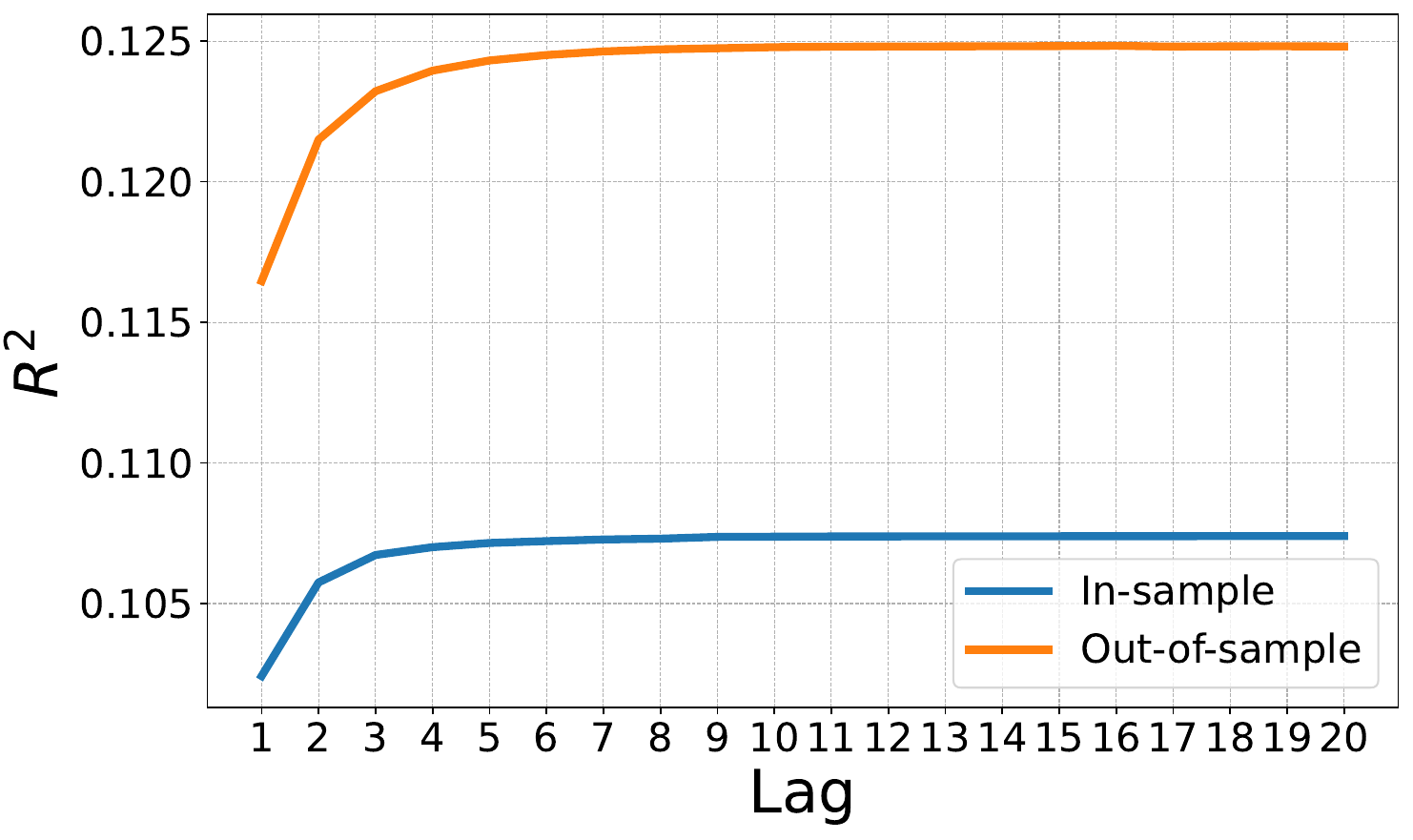}
        \caption{Signed volumes.}
        \label{fig:parametric_GILD_lag_scan_v}
    \end{subfigure}

    \caption{Evolution of the in-sample and out-of-sample \(R^2\) as a function
    of the maximum lag \(p\) for the GILD ticker. The improvement after the
    first lag is limited and the predictive performance rapidly saturates,
    consistently with a short-memory nonlinear dependence structure.}
    \label{fig:parametric_GILD_lag_scan}
\end{figure}

Figure~\ref{fig:parametric_GILD_lag_scan} reports the evolution of the
\(R^2\) as a function of the maximum lag \(p\) for the representative case of
GILD. For both returns and signed volumes, the largest improvement occurs when
moving beyond the one-lag specification, while the curves flatten as \(p\)
increases. This behaviour is consistent with the evidence obtained from the
linear VAR benchmark and with the SHAP lag-importance analysis: the process has
a short dominant nonlinear memory, whereas more distant lags contribute only
marginally. The exponential-decay parametrization therefore provides a useful
compromise between flexibility and parsimony, since it incorporates the residual
information contained in longer histories without assigning independent
nonlinear parameters to each lag.

\section{Structural shocks extraction for the parametric multi-lag model}

\begin{figure}[t]
    \centering

    \begin{subfigure}{0.48\textwidth}
        \centering
        \includegraphics[width=\textwidth]{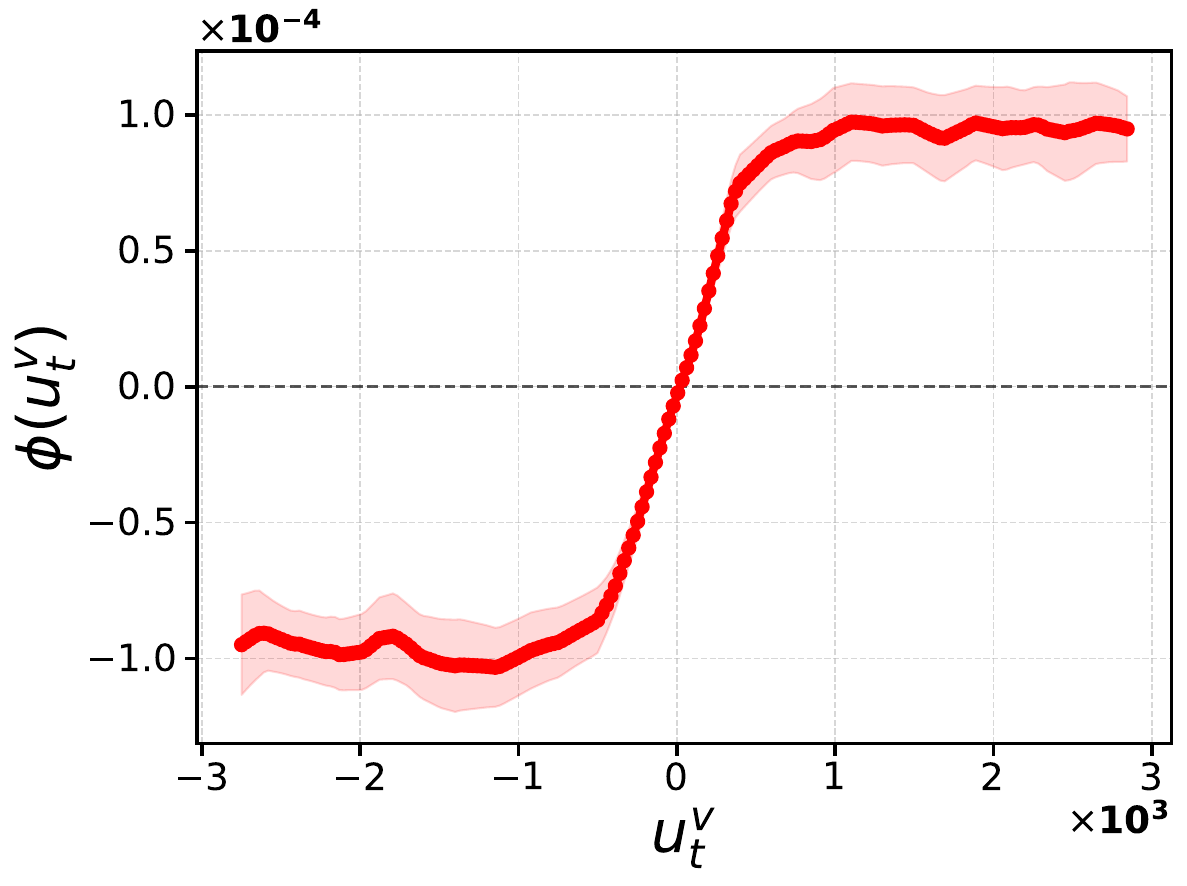}
        \caption{Large tick.}
        \label{fig:phi_parametric_LT}
    \end{subfigure}
    \hfill
    \begin{subfigure}{0.48\textwidth}
        \centering
        \includegraphics[width=\textwidth]{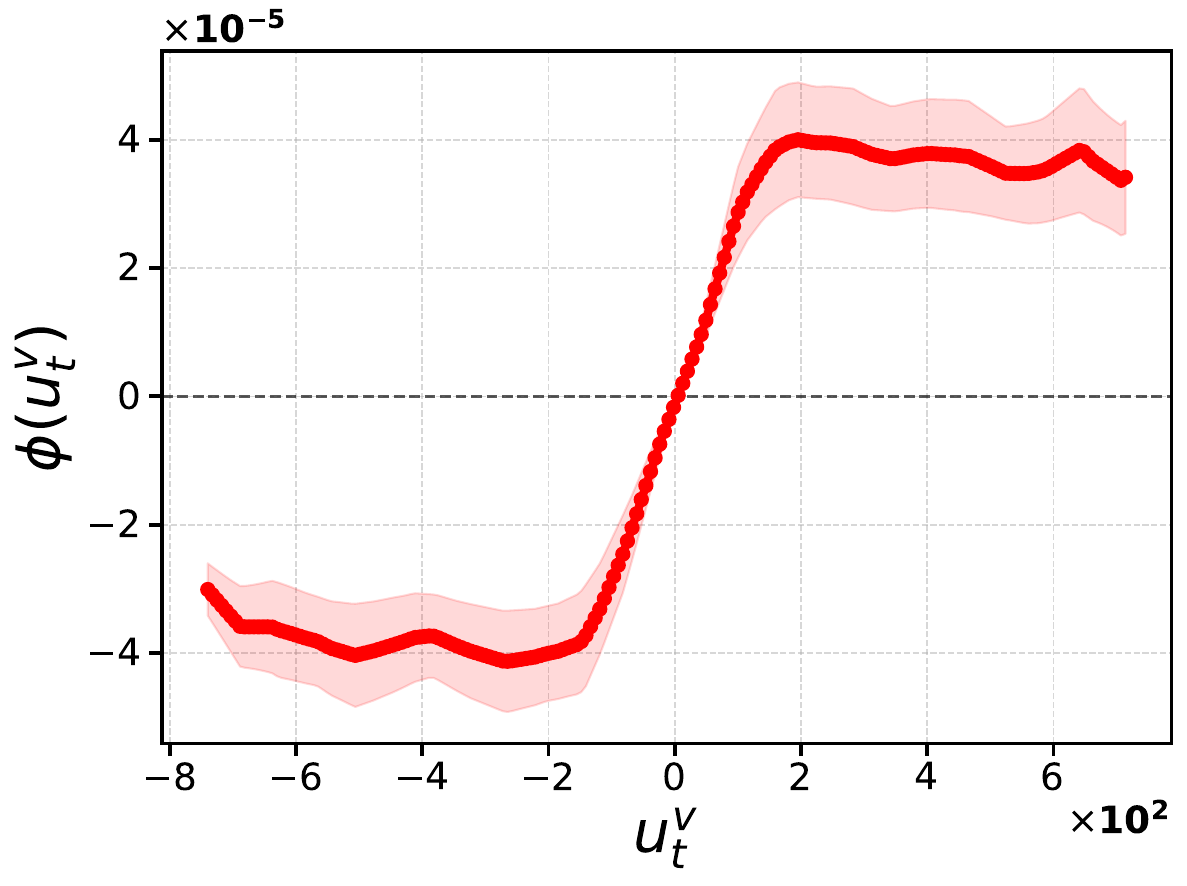}
        \caption{Small tick.}
        \label{fig:phi_parametric_ST}
    \end{subfigure}

    \caption{Estimated structural projection \(\hat\phi(u_t^v)\) for the
    multi-lag parametric model. The function is estimated through binned
    conditional averages and then averaged separately over large-tick and
    small-tick stocks.}
    \label{fig:phi_parametric_LT_ST}
\end{figure}

Figure~\ref{fig:phi_parametric_LT_ST} reports the estimated functions
\(\hat\phi(u_t^v)\), averaged separately over large-tick and small-tick stocks.
In both cases, we observe similar results as the neural network model.

The scale of the relation is larger for large-tick stocks, consistently with
the stronger discreteness of price changes and the queue-driven nature of price
formation in the large-tick regime.

Table~\ref{tab:structural_shocks_parametric} reports the residual correlations
before and after the structural correction. Before the correction, the
correlation between \(u_t^r\) and \(u_t^v\) is sizeable, with average values
around \(0.29\) for large-tick stocks and \(0.32\) for small-tick stocks. After
subtracting the fitted binned projection \(\hat\phi(u_t^v)\), the correlation is
reduced to approximately \(10^{-2}\) in both buckets. The same pattern is
observed for Spearman correlations, which fall from about \(0.37\) to values
close to zero for large-tick stocks, and from about \(0.42\) to slightly
negative values for small-tick stocks. This indicates that the binned
structural projection removes most of the contemporaneous signed dependence
between return and signed-volume residuals.

The last two columns report additional diagnostics based on nonlinear
transformations of the residuals. The correlation with squared residuals remains
very small on average, while the correlation with absolute residuals is larger
but still moderate, especially for large-tick stocks. Hence, the structural
correction removes the main directional dependence, although a residual
volatility-level dependence may remain.

\begin{table}[!htb]
\centering
\caption{Residual structural decomposition for the multi-lag parametric model
using the binned estimator of \(\phi\). The last row of each bucket reports
equal-weight averages across tickers.}
\label{tab:structural_shocks_parametric}
\begin{tabular}{llcccccc}
\toprule
Bucket 
& Ticker
& Corr. before 
& Corr. after 
& Spearman before 
& Spearman after 
& Corr. sq. 
& Corr. abs. \\
\midrule
LT & BAC   
& 0.301 & 0.0121 & 0.361 & -0.0214 &  0.0034 & 0.111 \\
LT & CMCSA 
& 0.250 & 0.0083 & 0.294 & -0.0347 & -0.0123 & 0.049 \\
LT & CSCO  
& 0.292 & 0.0109 & 0.348 & -0.0221 &  0.0075 & 0.107 \\
LT & INTC  
& 0.271 & 0.0119 & 0.323 &  0.0027 &  0.0175 & 0.126 \\
LT & PFE   
& 0.331 & 0.0135 & 0.423 &  0.0136 & -0.0152 & 0.086 \\
\midrule
LT & Average
& 0.289 & 0.0113 & 0.369 & -0.0108 &  0.0002 & 0.096 \\
\midrule
ST & AAPL  
& 0.309 & 0.0078 & 0.352 & -0.0516 &  0.0064 & 0.051 \\
ST & AMZN  
& 0.382 & 0.0118 & 0.477 & -0.0164 & -0.0196 & 0.001 \\
ST & GILD  
& 0.340 & 0.0077 & 0.422 & -0.0325 & -0.0143 & 0.020 \\
ST & NVDA  
& 0.281 & 0.0149 & 0.385 & -0.0223 &  0.0018 & 0.045 \\
ST & TSLA  
& 0.271 & 0.0100 & 0.345 & -0.0425 &  0.0054 & 0.039 \\
\midrule
ST & Average
& 0.317 & 0.0104 & 0.417 & -0.0322 & -0.0041 & 0.031 \\
\bottomrule
\end{tabular}
\end{table}

\end{document}